\documentclass[a4paper,fleqn,usenatbib,useAMS]{mnras}
\pdfoutput=1

\usepackage{times}
\usepackage{amsmath}
\usepackage{amssymb}
\usepackage{float}
\usepackage{lmodern}
\usepackage{pdflscape}
\usepackage{booktabs}
\usepackage{subcaption}
\usepackage{epsfig}
\usepackage[title]{appendix}

\def\apj{ApJ}
\def\mnras{MNRAS}
\def\nat{Nat}

\def\aap{A\&A}                   
\def\aj{AJ}                      
\def\apjs{ApJS}                  
\def\pasp{PASP}                  
\def\apjl{ApJ}                   

\def\ssr{SSRv}

\usepackage{etoolbox}
\makeatletter
\patchcmd\@combinedblfloats{\box\@outputbox}{\unvbox\@outputbox}{}{%
   \errmessage{\noexpand\@combinedblfloats could not be patched}%
}%
 \makeatother

\newcommand{\DefineRemark}[2]{%
  \expandafter\newcommand\csname rmk-#1\endcsname{#2}%
}
\newcommand{\Remark}[1]{\csname rmk-#1\endcsname}

\DefineRemark{3c120_beta}{$1.42^{+0.32}_{-0.27}$}
\DefineRemark{3c120_opening}{$27^{+10}_{-9}$}
\DefineRemark{3c120_inc}{$21.6^{+9.1}_{-7.7}$}
\DefineRemark{3c120_kappa}{$-0.17^{+0.38}_{-0.23}$}
\DefineRemark{3c120_gamma}{$2.8^{+1.5}_{-1.2}$}
\DefineRemark{3c120_eps}{$0.60^{+0.26}_{-0.30}$}
\DefineRemark{3c120_Mbh}{$7.56^{+0.33}_{-0.31}$}
\DefineRemark{3c120_fellip}{$0.62^{+0.28}_{-0.34}$}
\DefineRemark{3c120_fflow}{$0.52^{+0.33}_{-0.36}$}
\DefineRemark{3c120_thetae}{$49^{+31}_{-35}$}
\DefineRemark{3c120_rmean}{$21.74^{+6.23}_{-6.54}$}
\DefineRemark{3c120_rmin}{$3.8^{+2.4}_{-2.1}$}
\DefineRemark{3c120_taumean}{$20.3^{+6.6}_{-5.8}$}
\DefineRemark{3c120_taumedian}{$11.4^{+4.9}_{-4.1}$}
\DefineRemark{mrk335_beta}{$1.92^{+0.06}_{-0.17}$}
\DefineRemark{mrk335_opening}{$37^{+35}_{-18}$}
\DefineRemark{mrk335_inc}{$41^{+35}_{-18}$}
\DefineRemark{mrk335_kappa}{$-0.19^{+0.31}_{-0.20}$}
\DefineRemark{mrk335_gamma}{$2.9^{+1.4}_{-1.4}$}
\DefineRemark{mrk335_eps}{$0.53^{+0.33}_{-0.35}$}
\DefineRemark{mrk335_Mbh}{$6.41^{+0.41}_{-0.34}$}
\DefineRemark{mrk335_fellip}{$0.56^{+0.33}_{-0.37}$}
\DefineRemark{mrk335_fflow}{$0.66^{+0.22}_{-0.39}$}
\DefineRemark{mrk335_thetae}{$43^{+35}_{-32}$}
\DefineRemark{mrk335_rmean}{$13.6^{+4.7}_{-4.0}$}
\DefineRemark{mrk335_rmin}{$0.54^{+0.47}_{-0.30}$}
\DefineRemark{mrk335_taumean}{$13.2^{+4.9}_{-3.7}$}
\DefineRemark{mrk335_taumedian}{$3.1^{+1.8}_{-1.0}$}
\DefineRemark{mrk1501_beta}{$1.03^{+0.30}_{-0.24}$}
\DefineRemark{mrk1501_opening}{$30^{+15}_{-12}$}
\DefineRemark{mrk1501_inc}{$31^{+15}_{-12}$}
\DefineRemark{mrk1501_kappa}{$-0.07^{+0.17}_{-0.17}$}
\DefineRemark{mrk1501_gamma}{$2.4^{+1.6}_{-1.0}$}
\DefineRemark{mrk1501_eps}{$0.38^{+0.24}_{-0.25}$}
\DefineRemark{mrk1501_Mbh}{$8.06^{+0.42}_{-0.30}$}
\DefineRemark{mrk1501_fellip}{$0.42^{+0.27}_{-0.26}$}
\DefineRemark{mrk1501_fflow}{$0.25^{+0.19}_{-0.17}$}
\DefineRemark{mrk1501_thetae}{$32^{+25}_{-22}$}
\DefineRemark{mrk1501_rmean}{$55^{+12}_{-16}$}
\DefineRemark{mrk1501_rmin}{$13^{+11}_{-9}$}
\DefineRemark{mrk1501_taumean}{$48^{+12}_{-15}$}
\DefineRemark{mrk1501_taumedian}{$36^{+13}_{-12}$}
\DefineRemark{pg2130_beta}{$1.62^{+0.18}_{-0.16}$}
\DefineRemark{pg2130_opening}{$32^{+12}_{-11}$}
\DefineRemark{pg2130_inc}{$27^{+14}_{-11}$}
\DefineRemark{pg2130_kappa}{$-0.29^{+0.32}_{-0.15}$}
\DefineRemark{pg2130_gamma}{$3.2^{+1.3}_{-1.4}$}
\DefineRemark{pg2130_eps}{$0.47^{+0.33}_{-0.28}$}
\DefineRemark{pg2130_Mbh}{$7.12^{+0.36}_{-0.25}$}
\DefineRemark{pg2130_fellip}{$0.45^{+0.38}_{-0.27}$}
\DefineRemark{pg2130_fflow}{$0.48^{+0.35}_{-0.32}$}
\DefineRemark{pg2130_thetae}{$40^{+34}_{-30}$}
\DefineRemark{pg2130_rmean}{$22^{+3.0}_{-4.6}$}
\DefineRemark{pg2130_rmin}{$2.6^{+1.0}_{-1.0}$}
\DefineRemark{pg2130_taumean}{$21.5^{+3.5}_{-4.4}$}
\DefineRemark{pg2130_taumedian}{$9.6^{+2.9}_{-2.7}$}
\DefineRemark{mrk50_beta}{$1.12^{+0.22}_{-0.21}$}
\DefineRemark{mrk50_opening}{$29^{+32}_{-13}$}
\DefineRemark{mrk50_inc}{$34^{+20}_{-13}$}
\DefineRemark{mrk50_kappa}{$-0.07^{+0.17}_{-0.18}$}
\DefineRemark{mrk50_gamma}{$3.1^{+1.4}_{-1.6}$}
\DefineRemark{mrk50_eps}{$0.31^{+0.26}_{-0.21}$}
\DefineRemark{mrk50_Mbh}{$7.13^{+0.39}_{-0.36}$}
\DefineRemark{mrk50_fellip}{$0.44^{+0.23}_{-0.24}$}
\DefineRemark{mrk50_fflow}{$0.72^{+0.19}_{-0.19}$}
\DefineRemark{mrk50_thetae}{$19^{+22}_{-13}$}
\DefineRemark{mrk50_rmean}{$9.9^{+7.2}_{-4.0}$}
\DefineRemark{mrk50_rmin}{$1.9^{+1.7}_{-1.0}$}
\DefineRemark{mrk50_taumean}{$9.1^{+6.3}_{-3.6}$}
\DefineRemark{mrk50_taumedian}{$5.6^{+4.1}_{-2.2}$}
\DefineRemark{mrk141_beta}{$1.00^{+0.21}_{-0.16}$}
\DefineRemark{mrk141_opening}{$26^{+13}_{-9}$}
\DefineRemark{mrk141_inc}{$42^{+15}_{-12}$}
\DefineRemark{mrk141_kappa}{$-0.16^{+0.13}_{-0.07}$}
\DefineRemark{mrk141_gamma}{$3.4^{+1.1}_{-1.4}$}
\DefineRemark{mrk141_eps}{$0.08^{+0.17}_{-0.06}$}
\DefineRemark{mrk141_Mbh}{$7.54^{+0.31}_{-0.30}$}
\DefineRemark{mrk141_fellip}{$0.14^{+0.11}_{-0.09}$}
\DefineRemark{mrk141_fflow}{$0.73^{+0.18}_{-0.19}$}
\DefineRemark{mrk141_thetae}{$14^{+13}_{-9}$}
\DefineRemark{mrk141_rmean}{$22^{+17}_{-10}$}
\DefineRemark{mrk141_rmin}{$4.9^{+6.4}_{-2.6}$}
\DefineRemark{mrk141_taumean}{$20^{+16}_{-8}$}
\DefineRemark{mrk141_taumedian}{$13^{+12}_{-6}$}
\DefineRemark{mrk279_beta}{$1.45^{+0.41}_{-0.34}$}
\DefineRemark{mrk279_opening}{$49^{+23}_{-16}$}
\DefineRemark{mrk279_inc}{$38^{+17}_{-8}$}
\DefineRemark{mrk279_kappa}{$-0.14^{+0.56}_{-0.27}$}
\DefineRemark{mrk279_gamma}{$3.7^{+0.9}_{-1.4}$}
\DefineRemark{mrk279_eps}{$0.11^{+0.36}_{-0.08}$}
\DefineRemark{mrk279_Mbh}{$7.52^{+0.22}_{-0.20}$}
\DefineRemark{mrk279_fellip}{$0.19^{+0.51}_{-0.15}$}
\DefineRemark{mrk279_fflow}{$0.75^{+0.17}_{-0.17}$}
\DefineRemark{mrk279_thetae}{$10^{+11}_{-7}$}
\DefineRemark{mrk279_rmean}{$24^{+11}_{-7}$}
\DefineRemark{mrk279_rmin}{$10.2^{+6.6}_{-3.8}$}
\DefineRemark{mrk279_taumean}{$19.3^{+8.1}_{-6.0}$}
\DefineRemark{mrk279_taumedian}{$13.2^{+7.1}_{-4.4}$}
\DefineRemark{mrk1511_beta}{$1.02^{+0.16}_{-0.15}$}
\DefineRemark{mrk1511_opening}{$73^{+10}_{-24}$}
\DefineRemark{mrk1511_inc}{$60^{+17}_{-37}$}
\DefineRemark{mrk1511_kappa}{$0.05^{+0.29}_{-0.30}$}
\DefineRemark{mrk1511_gamma}{$2.6^{+1.6}_{-1.1}$}
\DefineRemark{mrk1511_eps}{$0.53^{+0.32}_{-0.35}$}
\DefineRemark{mrk1511_Mbh}{$7.09^{+0.30}_{-0.26}$}
\DefineRemark{mrk1511_fellip}{$0.62^{+0.21}_{-0.31}$}
\DefineRemark{mrk1511_fflow}{$0.65^{+0.24}_{-0.37}$}
\DefineRemark{mrk1511_thetae}{$36^{+33}_{-26}$}
\DefineRemark{mrk1511_rmean}{$13.0^{+9.1}_{-5.1}$}
\DefineRemark{mrk1511_rmin}{$2.6^{+2.4}_{-1.2}$}
\DefineRemark{mrk1511_taumean}{$11.1^{+6.8}_{-4.1}$}
\DefineRemark{mrk1511_taumedian}{$7.0^{+4.8}_{-2.6}$}
\DefineRemark{ngc4593_beta}{$0.72^{+0.13}_{-0.10}$}
\DefineRemark{ngc4593_opening}{$53^{+18}_{-16}$}
\DefineRemark{ngc4593_inc}{$53^{+9}_{-11}$}
\DefineRemark{ngc4593_kappa}{$-0.08^{+0.30}_{-0.17}$}
\DefineRemark{ngc4593_gamma}{$2.5^{+1.6}_{-1.1}$}
\DefineRemark{ngc4593_eps}{$0.20^{+0.29}_{-0.15}$}
\DefineRemark{ngc4593_Mbh}{$7.02^{+0.27}_{-0.23}$}
\DefineRemark{ngc4593_fellip}{$0.67^{+0.17}_{-0.29}$}
\DefineRemark{ngc4593_fflow}{$0.76^{+0.17}_{-0.19}$}
\DefineRemark{ngc4593_thetae}{$14^{+19}_{-10}$}
\DefineRemark{ngc4593_rmean}{$9.8^{+6.4}_{-3.7}$}
\DefineRemark{ngc4593_rmin}{$1.1^{+1.6}_{-0.8}$}
\DefineRemark{ngc4593_taumean}{$8.5^{+5.3}_{-3.3}$}
\DefineRemark{ngc4593_taumedian}{$6.1^{+3.9}_{-2.3}$}
\DefineRemark{pg1310_beta}{$1.28^{+0.18}_{-0.14}$}
\DefineRemark{pg1310_opening}{$63^{+18}_{-24}$}
\DefineRemark{pg1310_inc}{$50^{+26}_{-20}$}
\DefineRemark{pg1310_kappa}{$0.04^{+0.29}_{-0.37}$}
\DefineRemark{pg1310_gamma}{$2.9^{+1.4}_{-1.3}$}
\DefineRemark{pg1310_eps}{$0.56^{+0.30}_{-0.35}$}
\DefineRemark{pg1310_Mbh}{$7.16^{+0.26}_{-0.26}$}
\DefineRemark{pg1310_fellip}{$0.49^{+0.27}_{-0.30}$}
\DefineRemark{pg1310_fflow}{$0.36^{+0.39}_{-0.24}$}
\DefineRemark{pg1310_thetae}{$58^{+21}_{-37}$}
\DefineRemark{pg1310_rmean}{$18.9^{+8.2}_{-7.6}$}
\DefineRemark{pg1310_rmin}{$2.0^{+2.6}_{-1.5}$}
\DefineRemark{pg1310_taumean}{$16.0^{+7.1}_{-5.4}$}
\DefineRemark{pg1310_taumedian}{$7.9^{+4.5}_{-2.9}$}
\DefineRemark{zw229_beta}{$1.55^{+0.23}_{-0.22}$}
\DefineRemark{zw229_opening}{$34.0^{+8.4}_{-9.9}$}
\DefineRemark{zw229_inc}{$36.4^{+6.7}_{-6.4}$}
\DefineRemark{zw229_kappa}{$-0.407^{+0.070}_{-0.058}$}
\DefineRemark{zw229_gamma}{$2.9^{+1.3}_{-1.2}$}
\DefineRemark{zw229_eps}{$0.043^{+0.059}_{-0.032}$}
\DefineRemark{zw229_Mbh}{$6.79^{+0.24}_{-0.24}$}
\DefineRemark{zw229_fellip}{$0.07^{+0.10}_{-0.05}$}
\DefineRemark{zw229_fflow}{$0.75^{+0.17}_{-0.18}$}
\DefineRemark{zw229_thetae}{$8.7^{+8.5}_{-6.0}$}
\DefineRemark{zw229_rmean}{$7.3^{+4.1}_{-2.6}$}
\DefineRemark{zw229_rmin}{$2.3^{+1.5}_{-0.9}$}
\DefineRemark{zw229_taumean}{$7.0^{+4.0}_{-2.5}$}
\DefineRemark{zw229_taumedian}{$3.9^{+2.7}_{-1.4}$}

\newcommand{\Mbh}{M$_{\rm BH}$}

\newcommand{\inc}{$\theta_{i}$}
\newcommand{\opening}{$\theta_{o}$}
\newcommand{\thetae}{$\theta_{e}$}
\newcommand{\fellip}{f$_{\rm ellip}$}
\newcommand{\fflow}{f$_{\rm flow}$}
\newcommand{\taumean}{$\tau_{\rm mean}$}
\newcommand{\taumedian}{$\tau_{\rm median}$}
\newcommand{\rmean}{r$_{\rm mean}$}
\newcommand{\rmedian}{r$_{\rm median}$}
\newcommand{\rmin}{r$_{\rm min}$}
\newcommand{\RBLR}{R$_{\rm  BLR}$}

\defcitealias{raimundo19}{R19}

\begin{document}

\title[Modelling the AGN broad line region using single-epoch spectra II]{Modelling the AGN broad line region using single-epoch spectra II. Nearby AGN}
\author[Raimundo et al.]{S. I. Raimundo$^{1}$\thanks{E-mail: sandra.raimundo$@$nbi.ku.dk}, M. Vestergaard$^{1, 2}$, M. R. Goad$^{3}$, C. J. Grier$^{2}$, P. R. Williams$^{4}$, 
\newauthor B. M. Peterson$^{5,6,7}$, T. Treu$^{4}$
\\
$^{1}$ DARK, Niels Bohr Institute, University of Copenhagen, Lyngbyvej 2, 2100 Copenhagen, Denmark\\
$^{2}$ Steward Observatory, University of Arizona, 933 N. Cherry Avenue, Tucson, AZ 85721, USA\\
$^{3}$ Department of Physics and Astronomy, University of Leicester, University Road, Leicester LE1 7RH, UK\\
$^{4}$ Department of Physics and Astronomy, University of California, Los Angeles, CA 90095, USA\\
$^{5}$ Department of Astronomy, The Ohio State University, 140 West 18th Ave, Columbus, OH 43210, USA\\
$^{6}$ Center for Cosmology and AstroParticle Physics, The Ohio State University, 191 West Woodruff Avenue, Columbus, OH 43210, USA\\
$^{7}$ Space Telescope Science Institute, 3700 San Martin Drive, Baltimore, MD 21218, USA
}

\maketitle

\begin{abstract}
The structure of the broad line region (BLR) is an essential ingredient in the determination of active galactic nuclei (AGN) virial black hole masses, which in turn are important to study the role of black holes in galaxy evolution. 
Constraints on the BLR geometry and dynamics can be obtained from velocity-resolved studies using reverberation mapping data (i.e. monitoring data). However, monitoring data are observationally expensive and only available for a limited sample of AGN, mostly confined to the local Universe. Here we explore a new version of a Bayesian inference, physical model of the BLR which uses an individual spectrum and prior information on the BLR size from the radius-luminosity relation, to model the AGN BLR geometry and dynamics. We apply our model to a sample of 11 AGN, which have been previously modelled using monitoring data. Our single-epoch BLR model is able to constrain some of the BLR parameters with inferred parameter values that agree within the uncertainties with those determined from the modelling of monitoring data. We find that our model is able to derive stronger constraints on the BLR for AGN with broad emission lines that qualitatively have more substructure and more asymmetry, presumably as they contain more information to constrain the physical model. The performance of this model makes it a practical and cost-effective tool to determine some of the BLR properties of a large sample of low and high redshift AGN, for which monitoring data are not available.

\end{abstract} 


\section{Introduction}
\label{sec:introduction}

Emission lines from gas close to an actively accreting supermassive black hole can be used as a probe of the black hole's gravitational potential (e.g. \citealt{peterson&wandel99}). 
Those emission lines are typically broad, showing widths of thousands of kilometres per second caused by Doppler broadening due to bulk motions of the gas (e.g. \citealt{davidson&netzer79}). The physical region where these lines are emitted from, the broad line region (BLR), has been measured to have an average size of a few to a few 10s of light-days for Seyfert galaxies and of $\sim$ several tens to hundreds of light-days for quasars (e.g. \citealt{kaspi00}, \citealt{bentz13}, \citealt{hoormann19}), indicating that the lines are being emitted from deep in the black hole's gravitational potential.

The line emission from the BLR arises in photoionised gas, illuminated by the far/extreme ultra-violet and X-ray continuum radiation, originating in and above the accretion disc (e.g. \citealt{bahcall&kozlovsky69}, \citealt{davidson&netzer79}). 
However, the nature and structure of the BLR is still not fully understood. 
There is some suggestion that the BLR is related to the accretion disc. Theoretical models based on accretion disc outflow scenarios suggest that the broad line emission originates in a wind emerging from the accretion disc (e.g. \citealt{murray95}, \citealt{proga00}, \citealt{kollatschny03}, \citealt{elvis17}), or that the accretion disc, broad line region and obscuring medium (also known as the equatorial, dusty torus) are individual parts of a dynamical structure that changes as a function of AGN luminosity (\citealt{elitzur&shlosman06}). Other theoretical models propose that the BLR is part of a large continuous obscuring structure with the broad emission lines originating from within the dust sublimation radius (\citealt{netzer&laor93}).  

Observations of double-peaked broad emission lines in some AGN support the idea that the broad line emission arises from, or close to the outer regions of the accretion disc (e.g. \citealt{chen&halpern89}, \citealt{eracleous&halpern94}, \citealt{storchi-bergmann93}, \citealt{strateva03}, \citealt{storchi-bergmann17}), although there are alternative models to explain double-peaked lines \citep{goad&wanders96}. 
The broad line emission appears to originate in a continuous distribution of gas (e.g. \citealt{murray95}) as opposed to individual gas clouds (e.g. \citealt{laor06}), however in some cases the presence of discrete clouds is a viable explanation to variable obscuration in the BLR (e.g. \citealt{risaliti09}, \citealt{elvis17}).

Knowledge of the BLR structure is a key component in the determination of virial black hole masses and to understand the systematic uncertainties associated with reverberation-mapping and single-epoch mass measurements (e.g. \citealt{horne04}, \citealt{vestergaard19}). Reverberation mapping uses the delayed response of the broad emission lines to changes in the continuum flux to infer the geometry and kinematics of the BLR (\citealt{blandford&mckee82}, \citealt{peterson93}, \citealt{peterson14}). 
The time delay of the emission line response times the speed of light ($\tau c$) is used as a proxy for a size scale of the BLR, namely the radius of the BLR (\RBLR). By assuming that the emission line velocity width ($\Delta V$) reflects virial motion (e.g. \citealt{peterson&wandel99}), the black hole mass (\Mbh) can be calculated as: \Mbh $= f R_{\rm BLR} \Delta V^{2}/G$ (e.g. \citealt{peterson14}). The dimensionless $f$ factor is of the order of unity and encompasses the unknown BLR geometry, dynamics and orientation that can vary for each emission line and for each AGN \citep{goadkorista&ruff12}. 

Our limited information on the BLR structure has a potential strong impact on the current knowledge of black hole and AGN evolution, as the sample of AGN with reverberation mapping measurements is used to calibrate the zero-point of black hole and galaxy scaling relations (e.g. \citealt{vestergaard02}, \citealt{peterson04}, \citealt{onken04}, \citealt{vestergaard&osmer09}, \citealt{grier13b}). These scaling relations are in turn used to infer most of the supermassive black hole masses at low and high redshift (e.g.  \citealt{vestergaard04}, \citealt{vestergaard&peterson06}, \citealt{trump11}, \citealt{shen11} \citealt{jun15}, \citealt{kozlowski17}, \citealt{banados18}, \citealt{shen19}).

Further necessary constraints on the BLR geometry and dynamics 
can be obtained from velocity-resolved reverberation mapping studies. These studies analyse the response of the broad emission line profile to continuum changes as a function of the line of sight velocity and the time delay between continuum and emission line changes (e.g. \citealt{bahcall72}, \citealt{welsh&horne91}, \citealt{horne04}). This method has been successful at determining the geometry and dynamics of the BLR for a small sub-sample of AGN with reverberation mapping data available (e.g. \citealt{kollatschny03}, \citealt{bentz09}, \citealt{denney10}, \citealt{bentz10}, \citealt{barth11}, \citealt{grier13a}, \citealt{du16}, \citealt{pei17}). While the analyses described above are mostly model-independent, the same velocity-resolved datasets have also been modelled using an underlying physical model that allows for quantitative constraints to be obtained on the BLR geometry and dynamics parameters in the context of that model (\citealt{pancoast11}, \citealt{brewer11}, \citealt{li13}, \citealt{pancoast14b}, \citealt{grier17}, \citealt{pancoast18}, \citealt{williams18}). The model-independent and model-dependent velocity-resolved studies referred to above, have found a large variety of geometry and dynamics for the BLR: e.g. gas in near-circular elliptical orbits, signatures of inflowing or outflowing gas, with thick disk-like or spherical geometry distributions.

The small number of AGN for which velocity-resolved analysis has been carried out is somewhat limited by observational requirements: the need to have monitoring data suitable for reverberation mapping studies, i.e. spectroscopic, fast cadence, multi-epoch monitoring over a sufficiently long time span. Although single-epoch spectra per definition do not contain any timing information and hence BLR size information, by virtue of being an account of the collective emission from across the entire BLR, these spectra also contain information on the velocity distribution in the BLR gas. Additionally, they are more readily available than monitoring data. 

Several works have investigated the BLR geometry and kinematics by modelling the shape of the broad emission lines as seen in individual spectra. They find that while not all line profiles can constrain the full BLR structure, the modelling of some profiles allow us to establish meaningful constraints on the geometry or velocity of the BLR gas (e.g. \citealt{capriotti80}, \citealt{kwan&carroll82}, \citealt{eracleous94}, \citealt{rosenblatt94}, \citealt{robinson95}, \citealt{kollatschny&zetzl13}, \citealt{storchi-bergmann17}). 

Recent progress has been made to constrain the geometry and dynamical parameters of the BLR from single-epoch spectra. In a previous work \citep[][hereafter \citetalias{raimundo19}]{raimundo19}, we presented a modified version of the BLR geometry and dynamics Bayesian inference model of \cite{pancoast14a} and \cite{pancoast18}. While previous versions of the model used reverberation mapping monitoring data to constrain the BLR geometry and dynamical parameters, our model is tailored to use single-epoch spectra. \citetalias{raimundo19} validate and test our model with simulated data and with observed data on Arp 151, which allowed us to compare the single-epoch modelling with monitoring data modelling. \citetalias{raimundo19} find that single-epoch spectral modelling can constrain some of the BLR geometry and dynamics parameters to a reasonable degree. Notably, the single-epoch results agree with the original monitoring data modelling results within the 68\% confidence range.
The fact that this new modelling effort requires only a single AGN spectrum, brings significant power to this method since it may be applicable to much larger subsets of the broad-line AGN population (e.g. in the Sloan Digital Sky Survey, e.g. \citealt{paris18}) than those accessed by monitoring studies alone.

In this work we apply our single-epoch model to a more extended sample of AGN that were targets of monitoring campaigns and modelling of the BLR (\citealt{grier17}, \citealt{williams18}). These AGN display a large range of black-hole masses and accretion rates. Our goal is to carry out the first sample test of our model using broad emission lines of different shapes, to evaluate to what degree the model can be applied to AGN with the purpose of constraining some of the parameters characterising the BLR geometry and dynamics.
We note that the original model, on which our present work is based, has significant limitations. In its current version, the model is not designed to represent or infer the full physical properties of the BLR. The goal of the model is to provide a flexible albeit simplified approach to predict emission line shapes based on an idealized distribution of point particles in the gravitational potential of a supermassive black hole, in order to derive the black hole mass and a description of the BLR geometry and kinematics. Despite its limitations the model has been helpful to determine that in some cases gas inflow trajectories are observed in the BLR and that there is a dependence of the f-factor with inclination. In addition, the model has also been helpful in outlining that the distribution of the gas responding to the central ionizing radiation is a thick disk like structure. These results are particularly helpful given the limited knowledge about the BLR structure and kinematics that we currently have. The aim of applying this rudimentary approach to single-epoch spectra is simply to identify which of the structure and kinematics parameters, determined by the original code (i.e., applied to monitoring data), can be constrained using a single spectrum and at which confidence level. The hope is that this will be insightful for understanding the AGN population typical BLR structure and kinematics.
In Section~\ref{sec:model}, we describe our single-epoch spectra model and describe the limitations of this model approach and the upgrades currently being implemented to improve the model's BLR physical representation in Section~\ref{sec:model_limitations}. In Section~\ref{sec:procedure}, we describe our AGN sample, the observational data available and the modelling procedure employed to model single-epoch spectra of low-redshift AGN. In Section~\ref{sec:results} we present the results of the model for each AGN and discuss them in detail in Section~\ref{sec:discussion}. The conclusions and future applications of the model are presented in Section~\ref{sec:conclusions}.
 
We use the following cosmological parameters to calculate luminosity distances from redshifts measurements: H$_{0} = 70$ km s$^{-1}$ Mpc$^{-1}$, $\Omega_{M} = 0.3$ and $\Omega_{\lambda} = 0.7$.

\section{Model description}
\label{sec:model}
In this work we model the structure of the BLR of a sample of low redshift AGN using the model presented by \citetalias{raimundo19}. This model is a modified version of the BLR model of \cite{pancoast18}, first presented by \cite{pancoast11}. Our model uses single-epoch spectra (as opposed to monitoring data) to constrain the geometry and dynamics parameters of the BLR. In the following section, we describe the model, its underlying physical prescription and the associated parameters.

\subsection{A single-epoch broad line region model}
\label{sec:single_epoch}
The single-epoch BLR model was described in detail by \citetalias{raimundo19}. Here, we summarise the main features of the model and refer the reader to \citetalias{raimundo19} and  \cite{pancoast14a} for further details.

We use an underlying physical model to define the BLR. With the physical model and a set of associated free parameters, we are able to generate a variety of possible geometry and dynamical configurations for the BLR. The physical model consists of a set of point particles representing the BLR gas, around a central emitting source (i.e., accretion disc). The accretion disc is defined as a point source emitting isotropically. It is assumed that the point particles instantaneously reprocess the incoming accretion disc continuum flux into emission-line flux. This is a reasonable assumption given that the light travel time of the continuum photons is of several days compared to the much shorter recombination time scale of dense gas ($\sim$ 0.1h) (\citealt{osterbrock&ferland06}, \citealt{peterson&horne06}).
The point particles are spatially distributed according to an array of geometry parameters that control the position of each particle with respect to the central source. 
As defined by \cite{pancoast14a}, the radial distribution of point particles is parametrised as a Gamma distribution with the following form:
\begin{equation}
p(r|\alpha) \propto r^{\alpha-1} e^{\left( -\frac{r}{\theta}\right)}
\label{eq:gamma}
\end{equation}
where $r$ is the radius and $\alpha$ and $\theta$ are the standard parameters (shape parameter and scale parameter, respectively). The Gamma distribution is shifted from $r =0 $ by a value equal to the Schwarzschild radius ($R_{S} = 2GM_{BH}/c^{2}$), plus a minimum radius of the BLR (r$_{\rm min}$). A variable change is performed to convert Eq.~\ref{eq:gamma} from standard units to the units of the mean radius of the BLR radial profile ($\mu$) using:
\begin{equation}
\mu = r_{\rm min} + \alpha\theta \hspace{1.0cm} \beta = \frac{1}{\sqrt{\alpha}} \hspace{1.0cm} F = \frac{r_{\rm min}}{r_{\rm min}+\alpha\theta}
\end{equation}

\noindent with the three free parameters: $\beta$, $\mu$ and $F$.
The parameter $\mu$ is the mean radius of the Gamma distribution (therefore the mean radius of the BLR radial distribution) and $F$ is the minimum radius of the BLR ($r_{\rm min}$) in units of the mean radius. The shape parameter $\beta$, controls if the radial distribution of particles is a narrow Gaussian-like distribution ($\beta < 1$), an exponential profile ($\beta = 1$) or a radial distribution that is steeper than an exponential profile ($\beta > 1$). The radial distribution is truncated at a fixed maximum radius r$_{\rm out} = \Delta T \times c/2$ where $\Delta T$ is the total duration of the continuum light curve that is also simulated by the model (see details below). The assumed values for $r_{\rm out}$ are shown in the Appendix in Table~\ref{table_results}. The spatial distribution of point particles is inclined by an inclination angle $\theta_{i}$ with respect to the observer$'$s line-of-sight, with the inclination angle being the angle between the normal to the BLR mid-plane and the observer's line of sight. A face-on structure will have $\theta_{i} = 0^{\circ}$ while an edge-on structure will have $\theta_{i} = 90^{\circ}$. The opening angle, $\theta_{o}$, defines the angular half-thickness of the BLR as seen from the central emitting source and defined from the mid-plane of the BLR. A spherical distribution of point particles will have $\theta_{o} = 90^{\circ}$ while a thin disc will approach $\theta_{o} \rightarrow 0^{\circ}$.

\begin{table*}
\centering
\begin{tabular}{l | c | c | c | c | c | c | c}
\hline
\hline
Object & z & f$_{\lambda, \rm AGN}$ & Reference & R$_{\rm BLR}$ & log(M$_{\rm BH} [M_{\odot}]$) & Reference & $L_{\rm bol}/L_{\rm Edd}$\\
$[1]$ & $[2]$ & $[3]$ & $[4]$ & $[5]$ & $[6]$ & $[7]$ & $[8]$\\
\hline
\multicolumn{6}{ c }{\textbf{AGN10}}\\[0.1cm]
3C 120 & 0.0330 & 2.71 $\pm$ 0.10 & a, b & 28.7 &  7.745$^{+0.038}_{-0.040}$ & d & 0.10\\[0.1cm]
Mrk 335 & 0.0258 & 5.84 $\pm$ 0.29 & a, b & 22.7 & 7.230$^{+0.042}_{-0.044}$ & d & 0.22\\[0.1cm]
Mrk 1501  & 0.0893 & 2.06 $^{ + 0.25}_{ - 0.22}$ $*$ & a & 49.8 & 8.067$^{+0.119}_{-0.165}$ & d & 0.14\\[0.1cm]
PG 2130+099 & 0.0630 & 2.53 $\pm$ 0.09 & a, b & 40.0 & 7.433$^{+0.055}_{-0.063}$ & d & 0.40\\[0.2cm]
\multicolumn{6}{ c }{\textbf{LAMP2011}}\\[0.1cm]
Mrk 50 & 0.0234 & 1.20 $\pm$ 0.20 & c & 8.5 & 7.422$^{+0.057}_{-0.068}$ & d & 0.02\\[0.1cm]
Mrk 141 & 0.0417 & 1.09 $\pm$ 0.16 & c & 15.2 & 7.46$^{+0.15}_{-0.21}$ & e & 0.06\\[0.1cm]
Mrk 279 & 0.0305 & 3.98 $\pm$ 0.53 & b & 21.6 & 7.435$^{+0.099}_{-0.133}$ & d & 0.13\\[0.1cm]
Mrk 1511 & 0.0339 & 0.86 $\pm$ 0.09 & c & 10.7 & 7.11$^{+0.20}_{-0.17}$ & e & 0.07\\[0.1cm]
NGC 4593 & 0.0090 &  8.02 $\pm$ 0.90 & b & 8.4 & 6.882$^{+0.084}_{-0.104}$ & d & 0.08\\[0.1cm]
PG 1310$-$108 & 0.0343 & 1.66 $\pm$ 0.17& c & 15.4 & 6.48$^{+0.21}_{-0.18}$ & e & 0.60\\[0.1cm]
Zw 229$-$015 & 0.0279 & 0.56 $\pm$ 0.06& c & 6.9 & 6.913$^{+0.075}_{-0.119}$ & d & 0.05\\[0.2cm]

\hline
\end{tabular}
\caption{List of objects and their properties. $[1]$ AGN name. $[2]$ Redshift from the NASA/IPAC Extragalactic Database. $[3]$ AGN continuum flux density at the rest-frame wavelength of $5100$ \AA\ and in units of $10^{-15}$ erg s$^{-1}$ cm$^{-2}$ \AA$^{-1}$. $[4]$ Reference for the AGN flux density measurement: (a) \citealt{grier12}; (b) \citealt{bentz13}; (c) \citealt{barth15} $[5]$ BLR rest-frame radius in units of light days inferred from the flux in column $[3]$ and the R$-$L relation of \citealt{bentz13}. $[6]$ Logarithm of the black hole mass in units of solar masses. $[7]$ Reference for the M$_{\rm BH}$ measurement: (d) reverberation mapping measurement quoted in the AGN database \citep{bentz&katz15}; (e) BLR modelling using monitoring data \citep{williams18}. $[8]$ Indicative Eddington ratio calculated from the monochromatic luminosity $L_{5100} = \lambda L_{\lambda}$ (5100 \AA) (based on column $[3]$ and the cosmology used in this paper), and an approximate average bolometric correction factor of $f_{\rm bol} = L_{\rm bol}/L_{5100}$ = 10 (e.g. \citealt{castello-mor16}, \citealt{kilerci-eser18}). $*$ Flux value is the total (galaxy $+$ AGN flux) because imaging data were not available for the \citealt{grier12} analysis.}
\label{object_table}
\end{table*}

Three other parameters ($\kappa$, $\xi$ and $\gamma$) control the relative emission weights of specific groups of point particles in the BLR. The parameter $\kappa$ is defined based on the angle between the observer's line of sight to the central source and the point particle line of sight to the central source. It can be used to give different relative weights to point particles that are on the far side or the near side of the BLR (with respect to the observer). For $\kappa \rightarrow - 0.5$, the particles on the far side contribute with more line emission while for $\kappa \rightarrow 0.5$, the particles on the near side contribute with more line emission. Uniform weighting corresponds to $\kappa = 0$.
The parameter $\xi$ controls the transparency of the mid-plane of the BLR, here defined as the x$-$y plane in Fig. 9 of \cite{pancoast11}. If $\xi \rightarrow 0$ the mid-plane is opaque, and only point particles in the foreground of the mid-plane will be observed. For $\xi \rightarrow 1$ the mid-plane is transparent. The parameter $\gamma$ determines the angular displacement of the point particles and can have values between 1 and 5. This angle is measured from the mid-plane of the BLR to the opening angle. The parameter $\gamma$ can be used to transform a uniform distribution of particles into a distribution of particles that are more concentrated towards the opening angle, i.e. if one imagines a BLR as a thick disc, the particles would be more concentrated towards the outer faces of the disc. A value of $\gamma = 1$ corresponds to a uniform angular distribution of point particles while $\gamma > 1$ will result in the particles being more concentrated towards the opening angle value. 

We mentioned earlier that the point particles reprocess the continuum radiation from the accretion disc as line emission. The wavelength of the radiation emitted by each point particle will depend on the particle's velocity, which is defined by the dynamical state of the BLR and the corresponding dynamical parameters.
The point particles are assumed to move in Keplerian orbits in the gravitational potential of the black hole, with a black hole mass defined by the $M_{\rm BH}$ parameter. A fraction of point particles ($f_{\rm ellip}$) will have their velocities drawn from a distribution of velocities centred around the values corresponding to circular orbits. The remaining point particles (1 - $f_{\rm ellip}$) have velocities drawn from the distribution of inflowing or outflowing escape velocities. The parameter $f_{\rm flow}$ is a binary parameter that determines if the velocities are drawn from the inflowing ($f_{\rm flow} < 0.5$) or outflowing ($f_{\rm flow} > 0.5$) distributions. The centre of the inflowing and outflowing velocity distributions can be rotated by $\theta_{\rm e}$ along an ellipse towards the circular orbit velocity, so that an increasingly larger fraction of particles are in bound orbits as $\theta_{\rm e} \rightarrow 90$ degrees.

The model requires a simulated continuum light-curve determined at random instants to be able to model the BLR. Although we only model one epoch, the emission line profile will be the sum of the line emission from different portions of the BLR. Each portion of the BLR is located at a specific distance from the ionising continuum and hence associated with a specific time delay between the continuum and reprocessed emission. This requires different instants of the continuum light-curve to be probed, even for the modelling of a single emission line spectrum. To determine the continuum flux at each time instant, the model generates continuum light curves (i.e. continuum flux as a function of time) using several free parameters that we will refer to as the continuum hyper-parameters. In previous versions of the model, the generated continuum light curve was constrained based on the observed continuum light curve as determined from monitoring campaigns (e.g. \citealt{pancoast14b}, \citealt{grier17}, \citealt{pancoast18}, \citealt{williams18}). 
In our version of the model, where we use only a single spectrum instead of monitoring data, we do not have an observed light curve to analyse \citepalias{raimundo19}. Instead, the continuum light curves in our model are generated based on the literature-quoted range of possible continuum light curves observed for AGN. A description of the continuum light-curve generation can be found in Section~\ref{sec:continuum}.

Defining the geometry and dynamics parameters result in a specific structure for the BLR, which is then used to infer the time delay between the continuum emission from the accretion disc and the reprocessed radiation for each of the point particles. These time delays are used to match the emission line flux for each point particle with the previously emitted continuum flux drawn from the generated continuum light curve. Two additional parameters, an additive ($C_{\rm add}$) and a multiplicative ($C_{\rm mult}$) constant are used to normalise the continuum such that the modelling is not dependent on the absolute flux level of the continuum or of the emission line.

The model then generates a simulated spectrum based on the parameterised model of the spatial and kinematical structure of the BLR to be compared with the observed spectrum. In the previous version of the model, a simulated spectrum is generated for each epoch of the emission line light curve, i.e. for every epoch of the monitoring campaign. In our version of the model this is done only once and for one epoch. Since we do not use monitoring data to constrain our model, we do not have timing information from the observed light curves to constrain the time delays. Without a time delay constraint, there is limited information to establish the absolute physical scale of the BLR (and hence to independently constrain \Mbh).
Instead we provide this information to the model in the form of a Gaussian-like prior probability distribution for the mean time delay. The model's mean time delay is the flux-weighted mean time delay for the whole BLR. A detailed description of why the mean time delay is used and the effect of choosing such a prior is described by \citetalias{raimundo19}. 

The model parameters are constrained via Bayesian inference using DNest3, a Diffusive Nested Sampling algorithm (\citealt{brewer09}, \citealt{brewer_dnest}). DNest3 is a Markov Chain Monte Carlo method based on Nested Sampling, an algorithm for Bayesian computation \citep{skilling06}. 
Using a Bayesian formalism we define a prior probability distribution for the parameters (see Table 1 of \citetalias{raimundo19} for specific assumptions) and update it to the posterior probability distribution taking into account the observed spectrum. DNest3 is used to explore the parameter space and is able to handle strong correlations between the parameters. DNest3 is particularly suited for cases where the posterior probability distribution has a complex or unknown shape. A detailed description of how DNest3 is implemented in the model can be found in Appendix A of \citetalias{raimundo19}.

The output of the model will be a multi-dimensional posterior probability distribution from which we extract the posterior probability distribution for each of the BLR parameters. The posterior probability distribution for each parameter is obtained after marginalising over all the other parameters (e.g. \citealt{sivia&skilling06}). The inferred parameter values and their associated uncertainties are obtained from calculating the median value of the parameter posterior probability distribution and the 68\% confidence intervals around the median. 
\subsection{Limitations of the modelling approach}
\label{sec:model_limitations}
Due to the simplified nature of the model, there are significant limitations as to its physical interpretation. In the following we describe the main limitations of the model and the upgrades planned and currently under development to improve it.
The model does not aim to conserve the information on the absolute energy scale, in the sense that the continuum and line emission flux are rescaled. The main focus and goal of the model is to reproduce the shape of the emission line and learn about the geometry and kinematics of the part of the BLR from which we receive emission. The advantage of rescaling the flux is that this goal can be pursued without the more challenging task of having to deal with the complex physics that regulates the production, scattering, and absorption of photons. 
The approach of the original model (e.g. \citealt{pancoast18}) is not meant to describe the absolute physical size of the gas within the BLR, but just to give a description of where the observed emission is coming from.  Furthermore, in this work, since we spatially scale our model based on the empirical R-L relation, we do not attempt to constrain even the size of the emissivity of the BLR, but focus on finding the possible BLR emissivity configurations that can reproduce the observed emission line shape.
 
Another known limitation of the model is a simplified representation of emission line sources as point sources. While the large number of point sources used in the model can easily reproduce a variety of possible line emitting gas geometries and their kinematics, the absence of a physical size associated with individual `particles' in the model implies that the model cannot describe physical covering factors and local surface emissivities for the gas in the BLR. In the absence of a radial covering factor dependence, the model cannot take into account shadowing effects from particles at smaller radii on particles at larger radii. Thus, the model cannot predict if for a certain BLR structure and ionizing flux, there are enough ionising photons to ionize gas at all radii. In any case, the model's inferred distribution of particles and their kinematics will always be limited to the part of the BLR that emits in response to the continuum emission at a particular time.
An additional important general caveat is that the original model only accounts for the gravitational force of the black hole and neglects all effects of radiation pressure. Thus, model-based results for high Eddington ratio AGN should be interpreted with this in mind.
A related limitation is that in the model, the ionizing flux from the continuum source is assumed to not change as a function of radius. We know that the ionizing photon flux decreases with increasing distance from the source and this is essential to consider in the implementation of photoionisatin physics. At present efforts are underway to implement in the model: a) a radial profile of the ionizing flux based on physical assumptions and b) energy conservation between the continuum and total line flux (Williams et al. in prep.). These modifications will be available in the short-term and will allow us to test and apply the model under more physical assumptions. However, one should keep in mind that all these modifications involve additional uncertainties, and that obtaining a fully self-consistent description of the BLR remains a coveted but still distant goal owing to the complexity of the problem.

\section{Modelling the broad-line region of low-redshift AGN}
\label{sec:procedure}
In this work we apply our single epoch BLR modelling \citepalias{raimundo19} to a sample of AGN that were the target of recent reverberation mapping campaigns. Here we model 11 AGN spanning a wide range of black hole mass and accretion rate.
In the previous section we describe the general setup and parameters of our BLR model. In this section we describe the data used to constrain the model and the specific assumptions we make to model the AGN sample. This includes the selection of the spectral epoch, the prior probability distribution we assume for the mean time delay and how the continuum light curve is generated.
\subsection{Data}
\label{sec:data}
We select a sample of 11 AGN: 4 AGN from the AGN10 monitoring campaign (\citealt{grier12}, \citealt{grier17}) and 7 AGN from the Lick AGN Monitoring Project 2011 (LAMP 2011; \citealt{barth15}). The main properties of the sample are shown in Table~\ref{object_table}. These AGN were selected because their BLRs have recently been modelled by \cite{grier17} (AGN10) and \cite{williams18} (LAMP2011) using the original model as implemented by \cite{pancoast18} and the full monitoring dataset. This allows for a direct comparison with our results. 

Henceforth, the BLR modelling using the monitoring dataset as input will be referred to as the `full light-curve modelling'. The full light-curve modelling approach uses: i) the continuum and integrated H$\beta$ broad emission line flux light curves and ii) monitoring spectra (i.e. multi-epoch spectra) of the H$\beta$ broad emission line to constrain the BLR parameters. Our approach will be referred to as single-epoch modelling, as we use a single spectrum to constrain the model parameters.
The BLR parameters inferred from the full light-curve modelling are presented by \cite{grier17} and \cite{williams18} and will be used as a comparison to our modelling results using single-epoch spectra. We use the same original datasets as \cite{grier17} and \cite{williams18} as a starting point, including the same spectral decompositions. The narrow emission line component is not subtracted before modelling the data, to avoid introducing uncertainties. For all the objects in our sample the narrow emission line is modelled with a narrow Gaussian function, with the total narrow line flux and systematic central wavelength as free parameters in the model. This approach is the same as that adopted by \cite{pancoast14a} and \cite{pancoast18} and a more detailed description can be found in Section 3.3 of \cite{pancoast18}. In general we find narrow line fluxes from the single-epoch modelling that agree within the 68\% confidence range with what was found from modelling of monitoring data. The exceptions are Mrk 1511 and PG 1310$-$108 that agree within the 95\% confidence range and Mrk 279 that agree within the 99.7\% range. Below we describe the datasets and monitoring campaigns in more detail. 

\subsubsection{AGN10}
\label{sec:agn10}
We model 4 AGN that were part of the AGN10 reverberation mapping campaign: 3C 120, Mrk 335, Mrk 1501 and PG 2130+099, all classified as Seyfert 1 galaxies. The black holes in these AGN are in the upper range of black hole masses and luminosities compared with the other low-redshift AGN in our sample. 

The monitoring data consist of photometric and spectroscopic measurements in the optical range, taken at several epochs within a time span of a few months. 
For each object there are $\sim 70$ epochs of optical spectra covering the wavelength range of the H$\beta$ broad emission line, and $\sim 100 - 200$ epochs of continuum flux measurements (at a wavelength of 5100 \AA). Before modelling the broad emission line, the spectra are decomposed to account for all the emission components that are not due to broad H$\beta$ line emission, such as the AGN continuum, the host galaxy light and the narrow line emission (\citealt{barth15}, \citealt{grier17}). More details on the observations and the data analysis are provided by \cite{grier12} and \cite{grier17}. BLR full light-curve modelling (i.e. using the monitoring dataset) was carried out by \cite{grier17}. From the initial dataset we select a fixed epoch and use the spectrum obtained at that epoch as input to our model, disregarding all the remaining data.

\subsubsection{LAMP2011}
We model 7 AGN that were part of the LAMP2011 campaign: Mrk 50, Mrk 141, Mrk 279, Mrk 1511, NGC 4593, PG 1310$-$108 and Zw 229$-$015. 
The monitoring data consist of photometric and spectroscopic measurements in the optical range, taken at several epochs within a time span of a few months. 
For each object there are $\sim 30-50$ epochs of optical spectra covering the wavelength range of the H$\beta$ broad emission line, and $\sim 60 - 170$ epochs of continuum flux measurements, depending on the AGN. Before modelling the broad emission line, the spectra are decomposed as described above. More details can be found in the work by \cite{barth15} and \cite{williams18}. 

Modelling of the full light curve was carried out by \cite{williams18}. They use three different spectral decompositions that differ only in the form of Fe II template used. They model the result of each spectral decomposition prescription independently and present the posterior probability distributions for each. Additionally they also show the result of combining the posterior probability distributions of all the spectral decomposition prescriptions. We model the data that were spectrally decomposed using the Fe II template of \cite{kovacevic10}, which due to its larger number of normalisation parameters provides the most flexible spectral modelling \citep{williams18}. Our results are however compared with both the full light-curve posterior probability distribution using the \cite{kovacevic10} spectral decomposition and the combined posterior probability distribution shown by \cite{williams18}. Similar to the AGN10 modelling described in \ref{sec:agn10}, we select a fixed epoch from the initial dataset and use the spectrum from that epoch as input to our model, disregarding all the remaining data.
\subsection{Epoch selection}
\label{sec:epoch_selection}
As all of our objects have been the target of a reverberation mapping campaign, we have access to the full continuum lightcurves and monitoring spectra. For each object we select the epoch which shows the highest signal-to-noise ratio (S/N) in the H$\beta$ broad emission line. The signal-to-noise ratio for each epoch is measured as the mean signal-to-noise ratio per pixel of the spectrum across the H$\beta$ broad emission line profile.

In our previous work \citepalias{raimundo19}, we selected the epochs based on their continuum flux level with respect to the full light-curve, to sample low, medium and high continuum flux levels. We proposed that the specific line shape at that epoch and not the continuum flux level was responsible for most of the constraints on the BLR parameters that we obtained in the modelling. Here we select the individual epochs based on the S/N of the line flux in the spectrum to generalise this approach to most of the AGN in the literature, where one does not have information on the historical light-curve but for which a single-epoch spectrum is available. 

In Fig.~\ref{light-curves} we show the V-band continuum and H$\beta$ emission-line light curves for the full monitoring dataset of each source. 
The green stars show the epochs that we select for our analysis. Each epoch in the emission line light curve identifies a specific (observed) spectrum. Since the V-band continuum epochs may not be exactly the same as the spectral epoch, we select the (observed) continuum epoch that is closest in time to the spectral epoch. The single-epoch H$\beta$ profiles that correspond to the selected epochs are shown in Fig.~\ref{epoch_spec}. We note that the epoch with the highest S/N ratio typically has high line flux, but does not necessarily correspond to the epoch of highest line flux level, as can be seen in Fig.~\ref{light-curves}.

In addition to our main analysis with the highest S/N epoch we also carry out a test using a second epoch for each AGN. This second epoch was chosen randomly from the subset of epochs that show an average S/N $\sim$ 20 across the line profile. The exceptions were Zw 229-015 and Mrk 141 that have lower S/N for all epochs and do not reach S/N $\sim$ 20. For these two cases we selected random epochs from the subset that showed an average S/N $\sim$ 9 across the line profile. Since the results from the second epoch are fully consistent with those of the highest S/N epoch we focus our detailed discussion on the highest S/N epoch results and give an overview of the second epoch results in Section~\ref{sec:full_sample}.

\begin{figure*}
\centering
\includegraphics[width=0.4\textwidth]{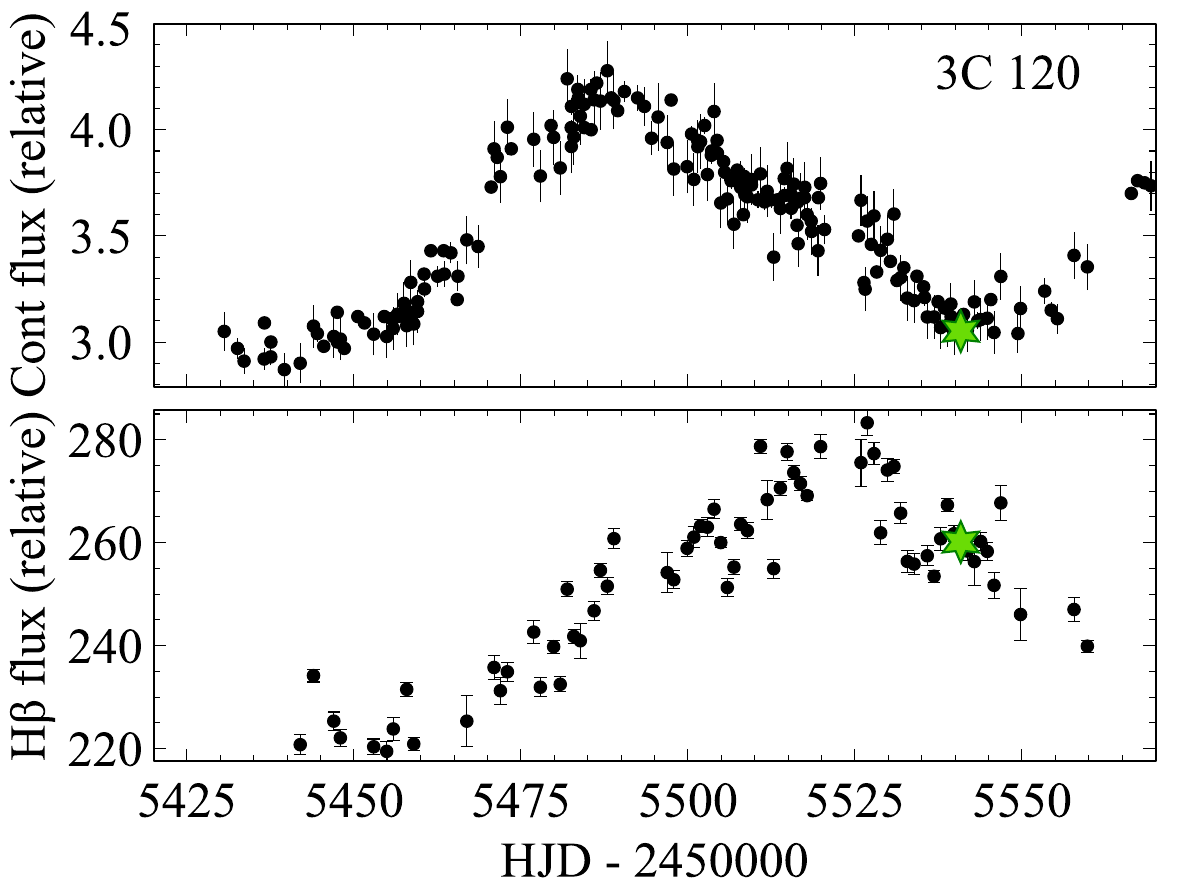}\hspace{0.8cm}
\includegraphics[width=0.4\textwidth]{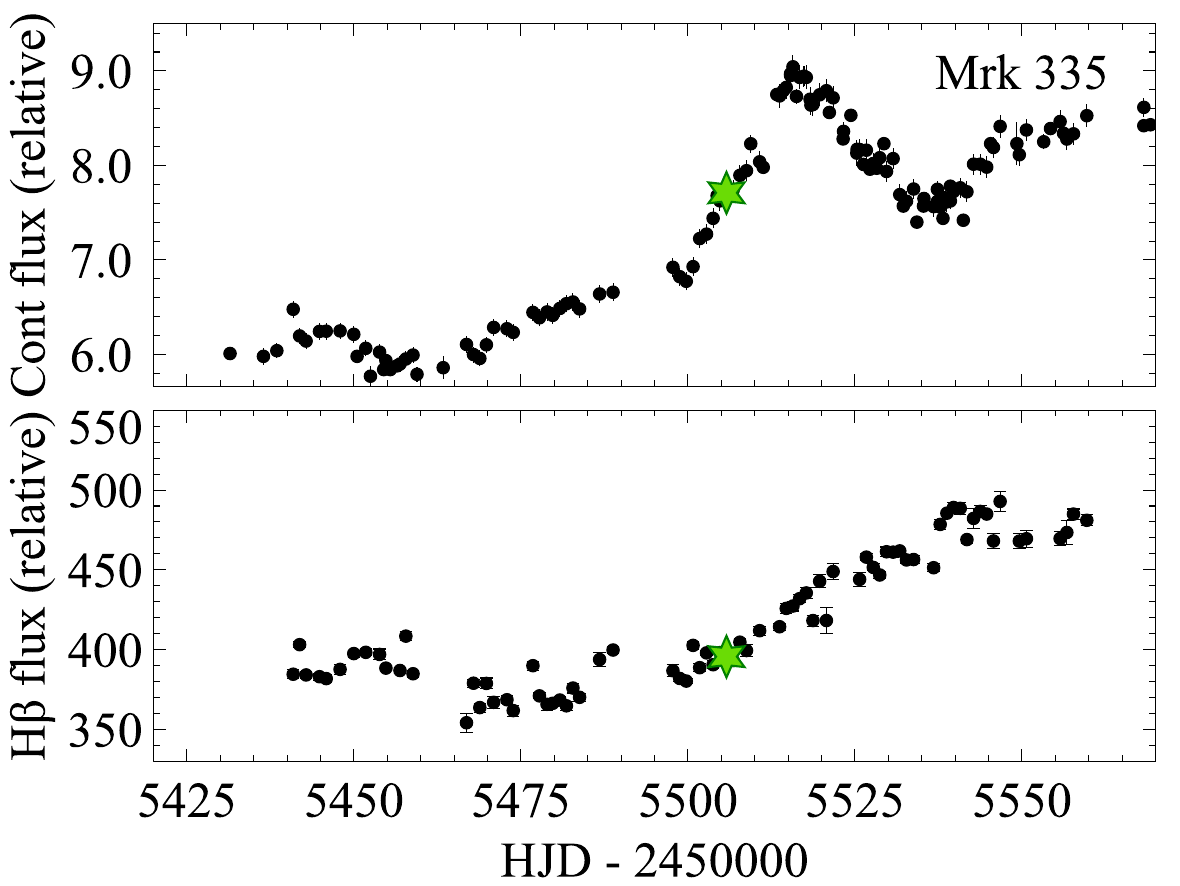}\\[0.2cm]
\includegraphics[width=0.4\textwidth]{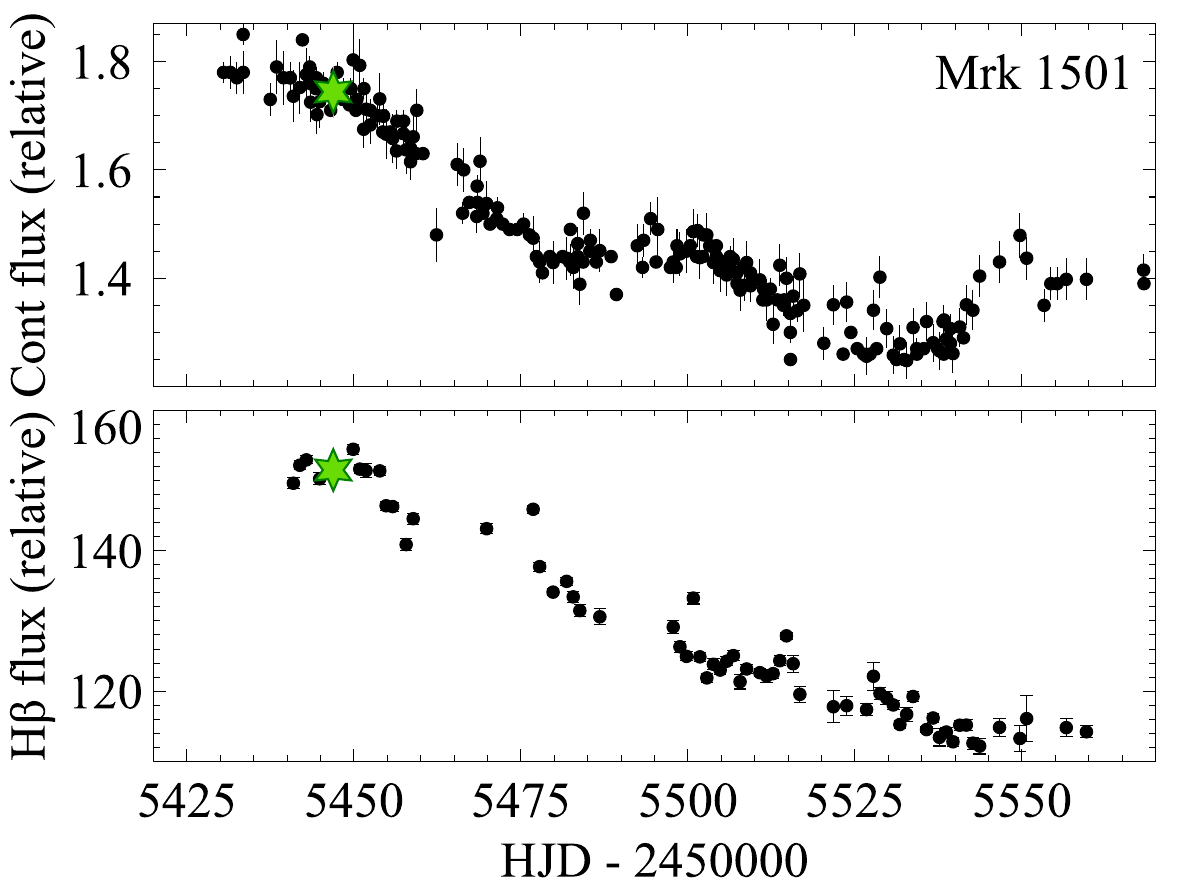}\hspace{0.8cm}
\includegraphics[width=0.4\textwidth]{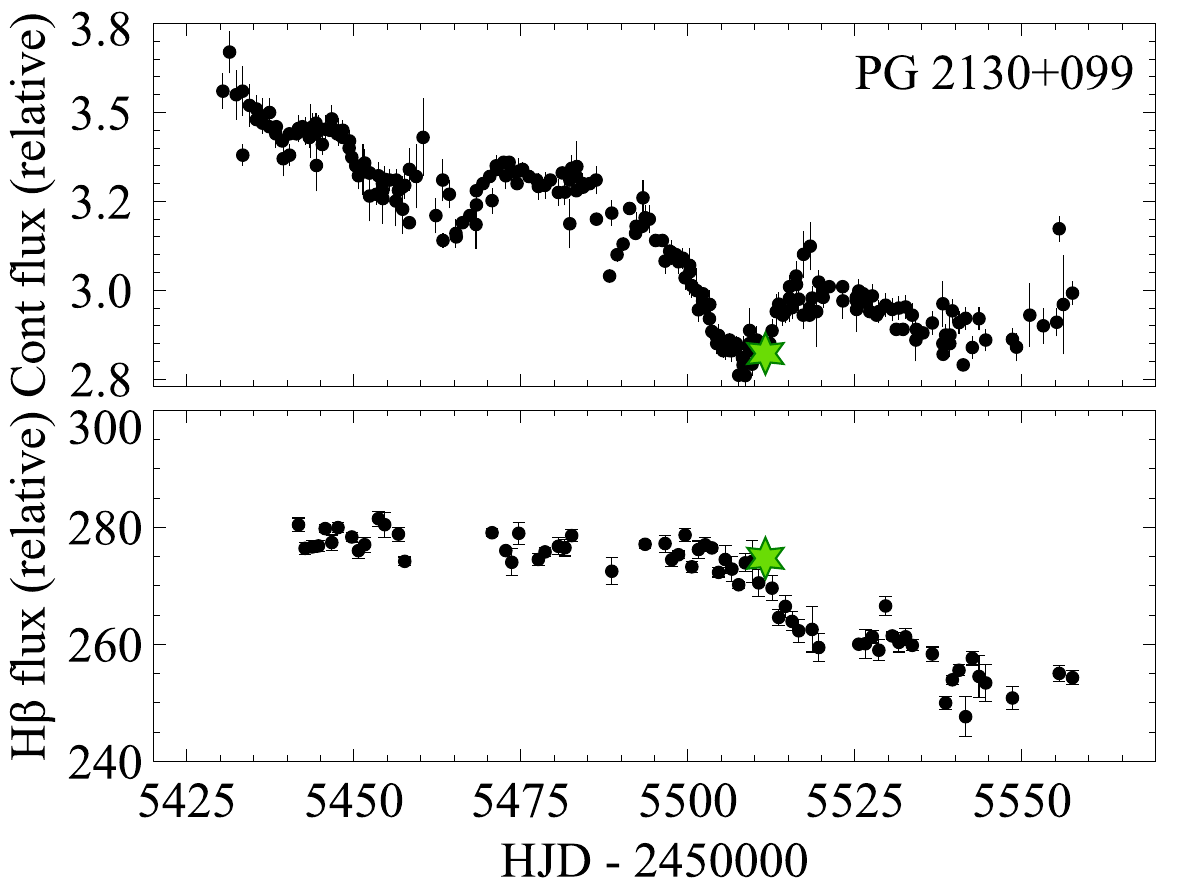}\\[0.2cm]
\includegraphics[width=0.4\textwidth]{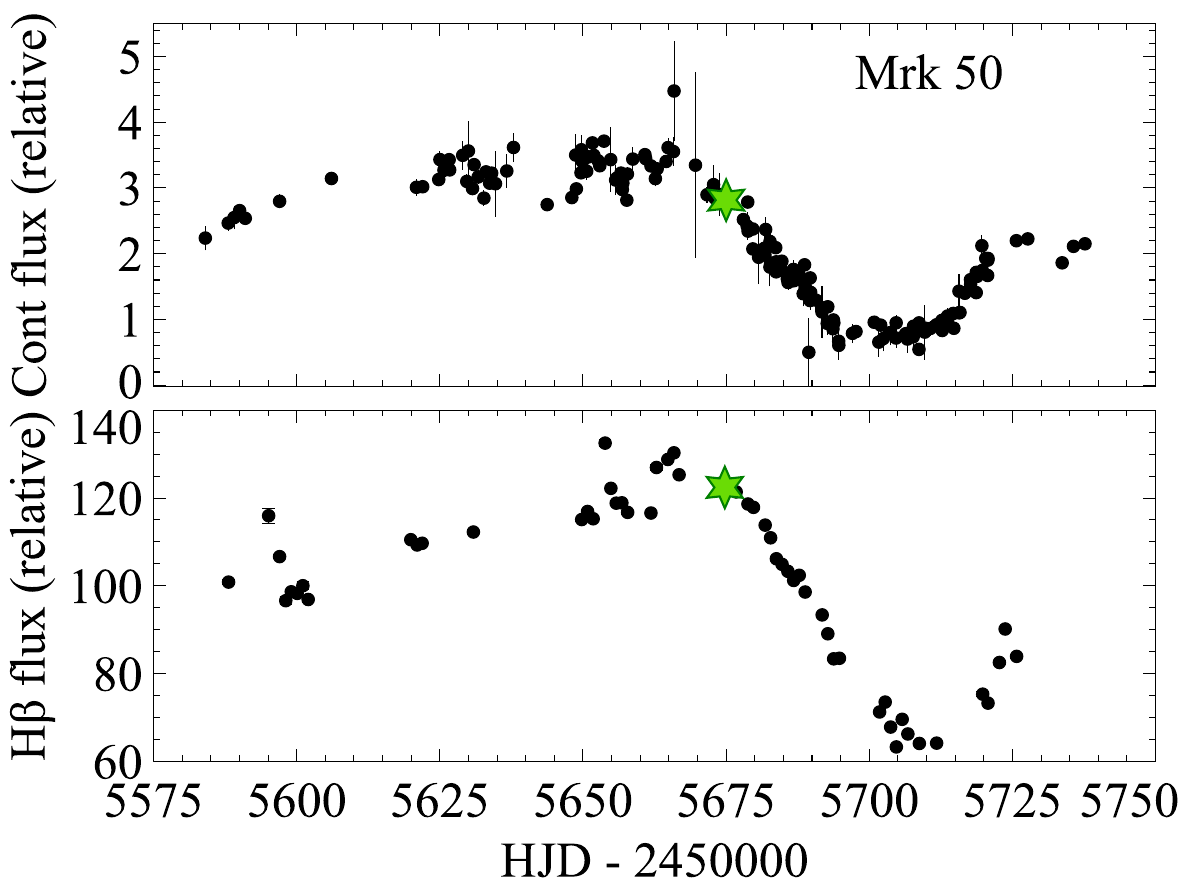}\hspace{0.8cm}
\includegraphics[width=0.4\textwidth]{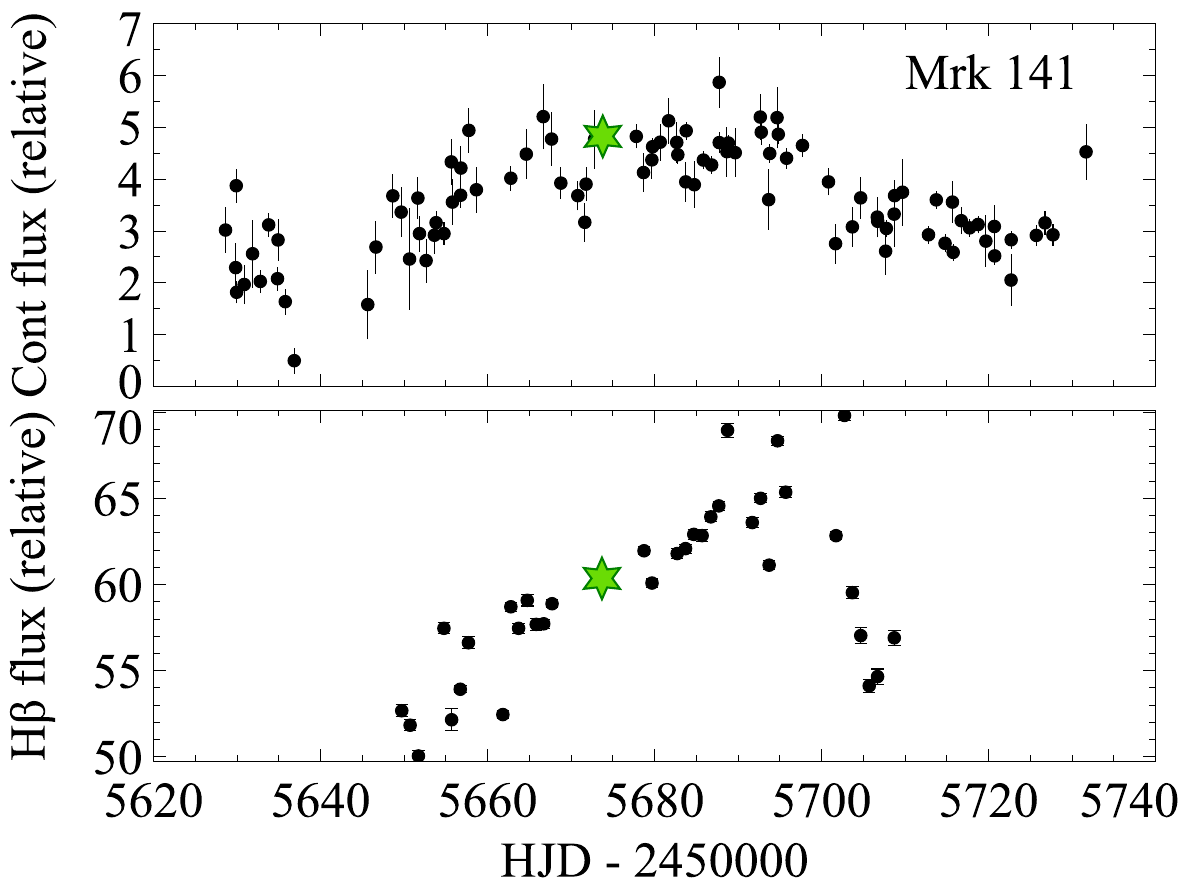}\\[0.2cm]
\includegraphics[width=0.4\textwidth]{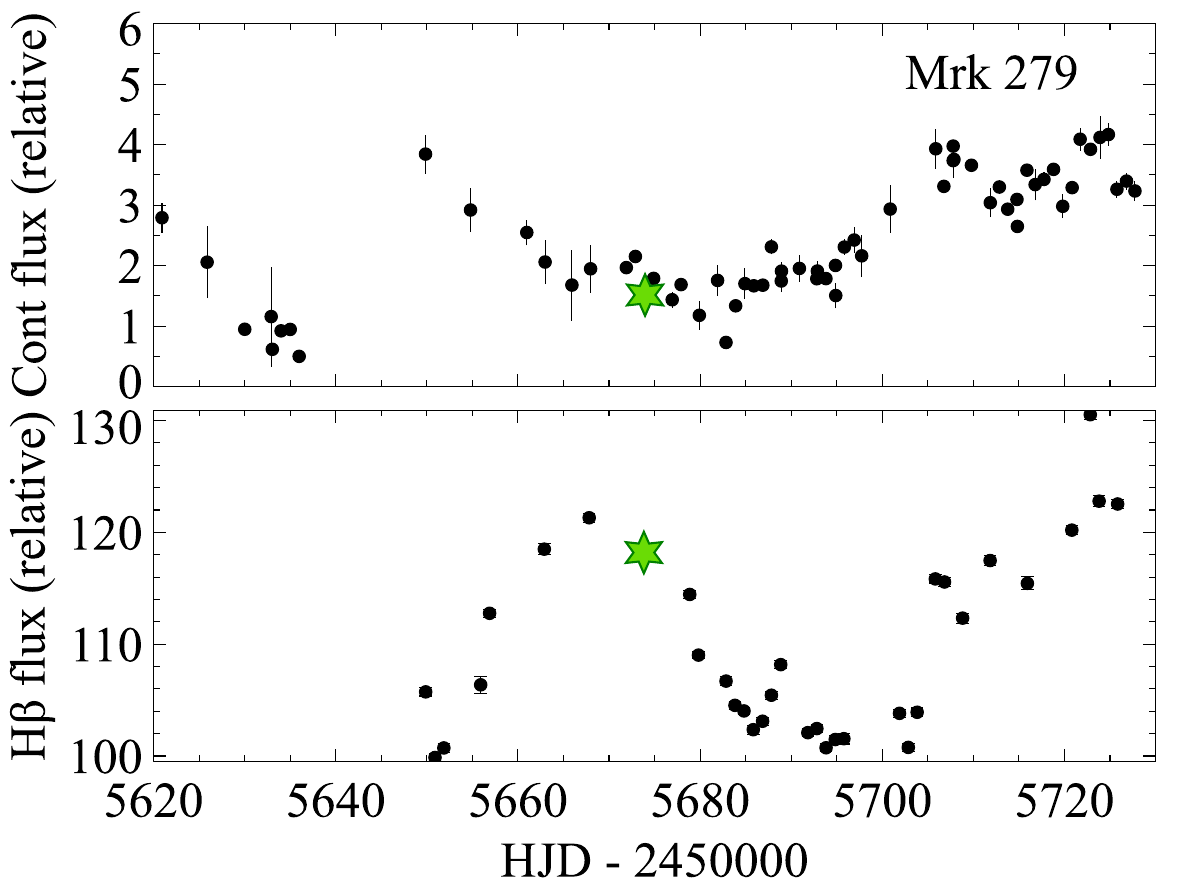}\hspace{0.8cm}
\includegraphics[width=0.4\textwidth]{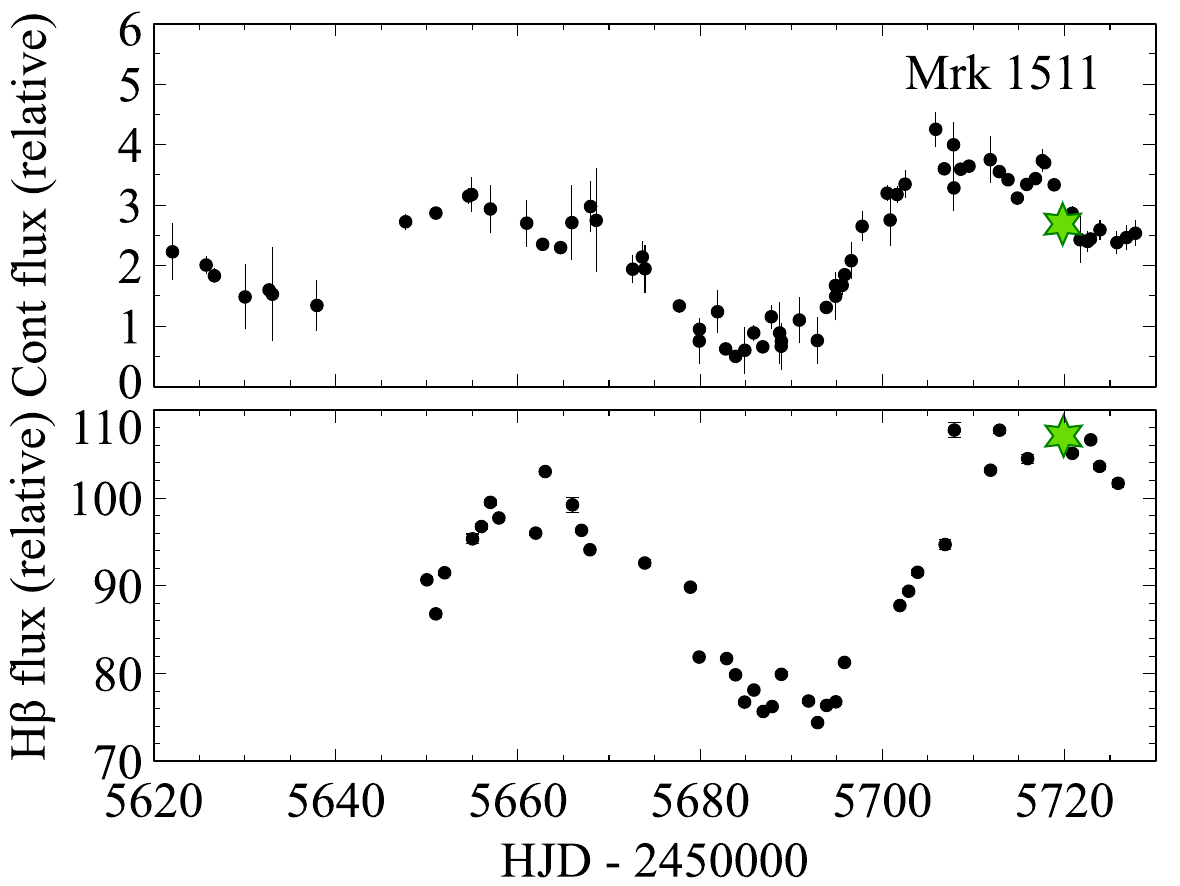}
\caption{V-band continuum and integrated H$\beta$ line flux lightcurves for each object in our sample. The lightcurves shown by the black filled circles are part of the monitoring dataset described by \citealt{grier17} and \citealt{williams18}. For each object the green star symbol indicates the epoch selected for our single-epoch modelling analysis. The single-epoch spectra extracted from these specific epochs are shown in Fig.~\ref{epoch_spec}.}
\end{figure*}
\begin{figure*}\ContinuedFloat
\centering
\includegraphics[width=0.4\textwidth]{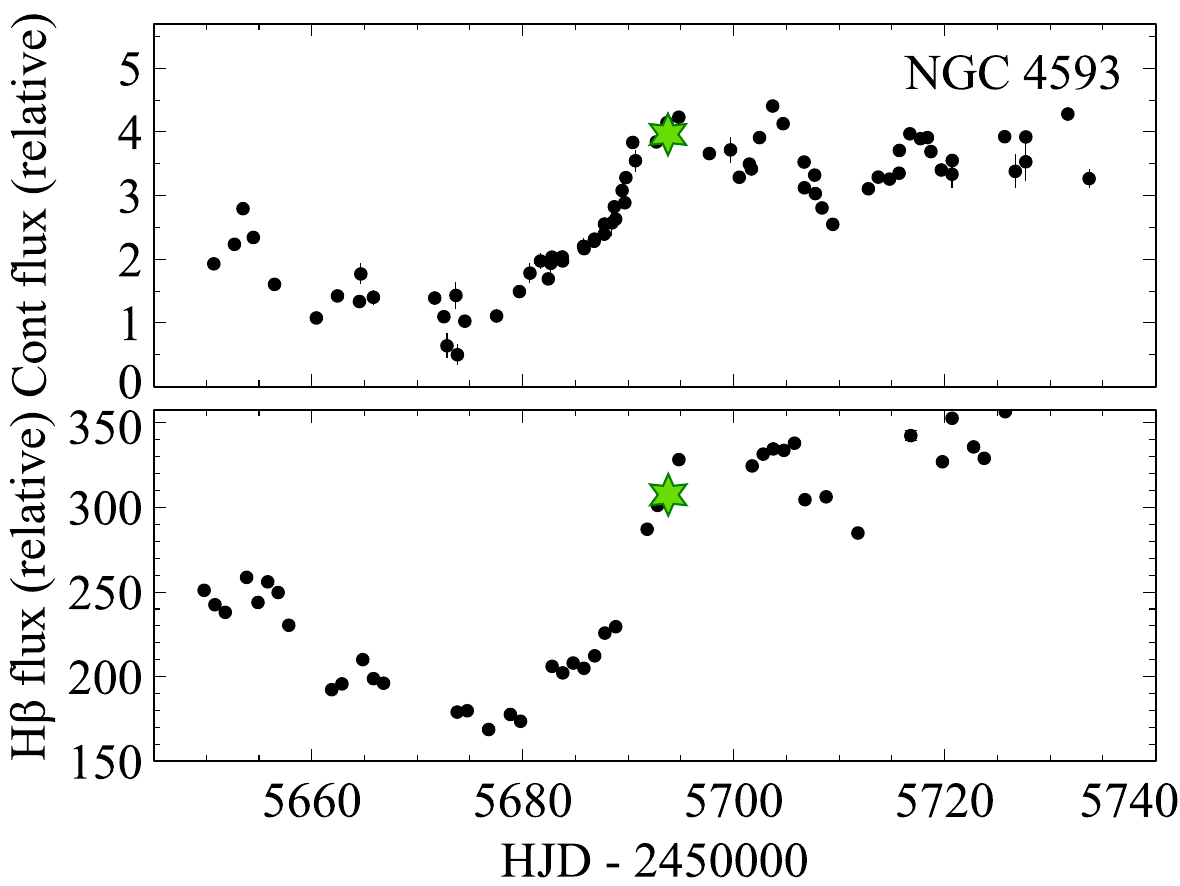}\hspace{0.8cm}
\includegraphics[width=0.4\textwidth]{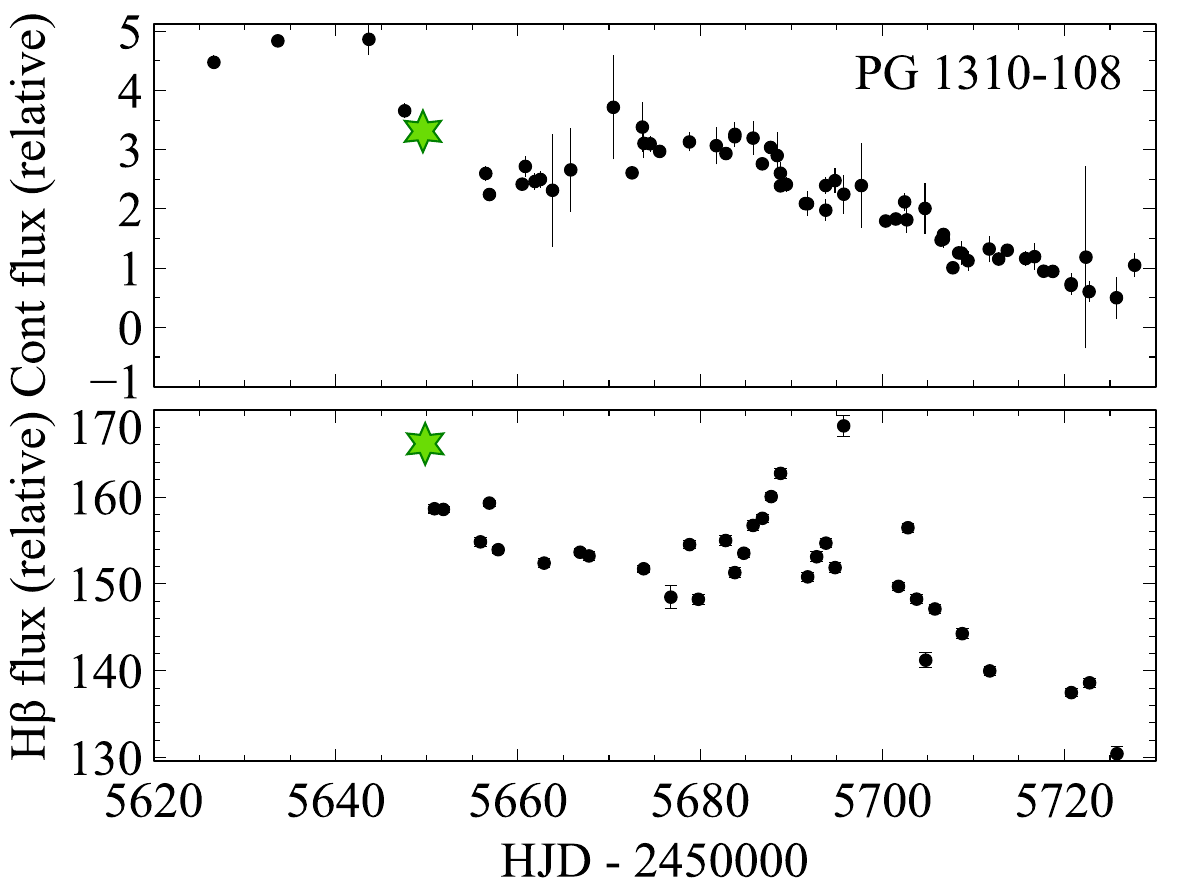}\\[0.2cm]
\includegraphics[width=0.4\textwidth]{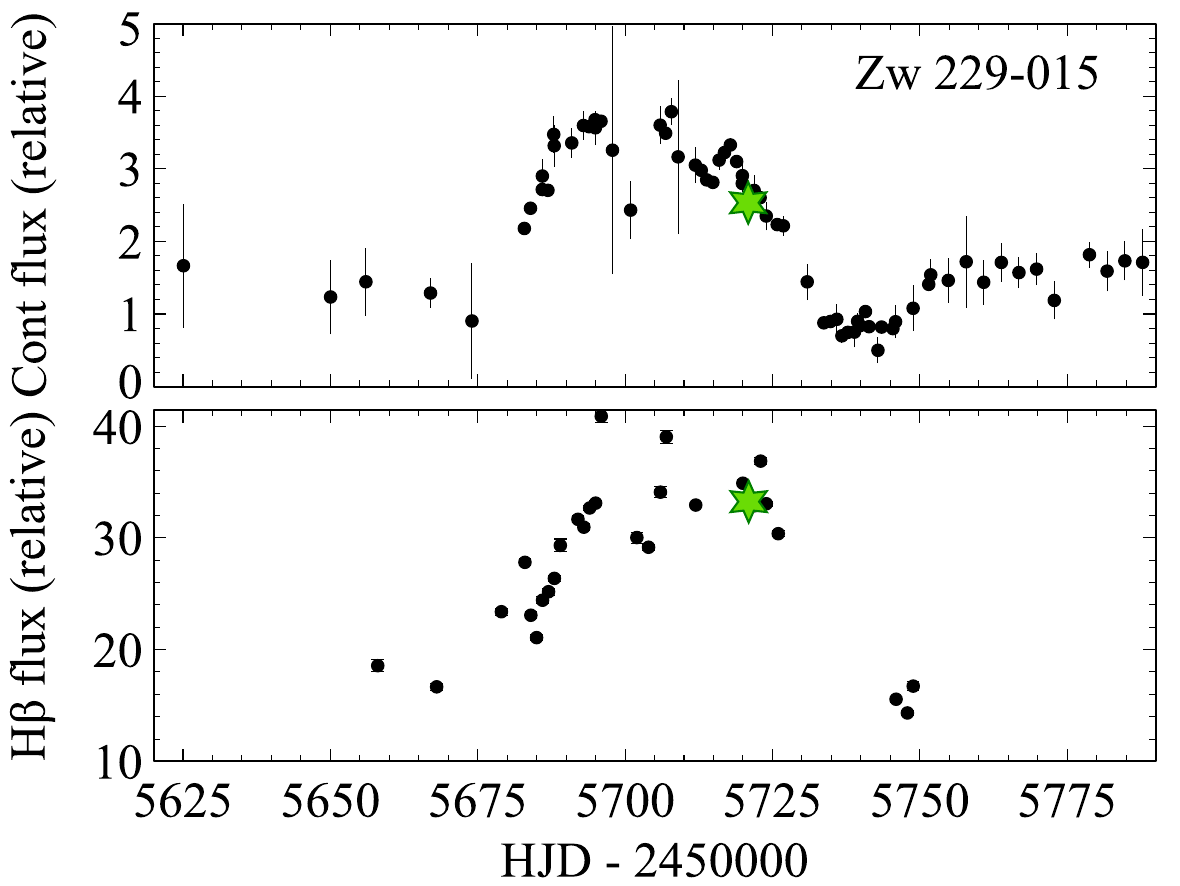}
\caption{Continued.}
\label{light-curves}
\end{figure*}

\begin{figure*}
\centering
\includegraphics[width=0.3\textwidth]{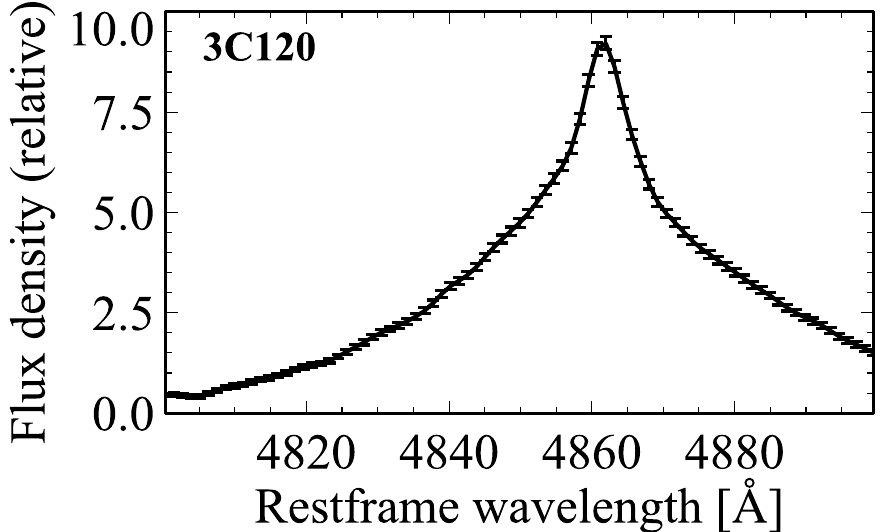}
\includegraphics[width=0.3\textwidth]{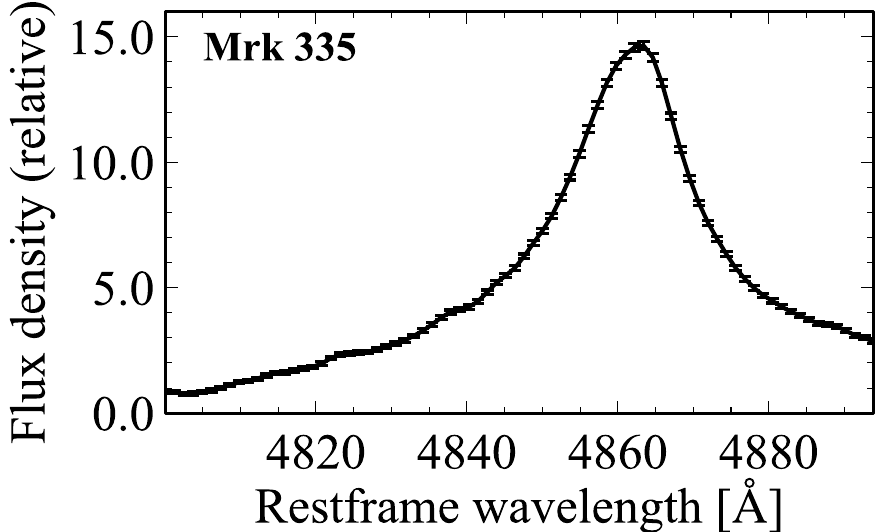}
\includegraphics[width=0.3\textwidth]{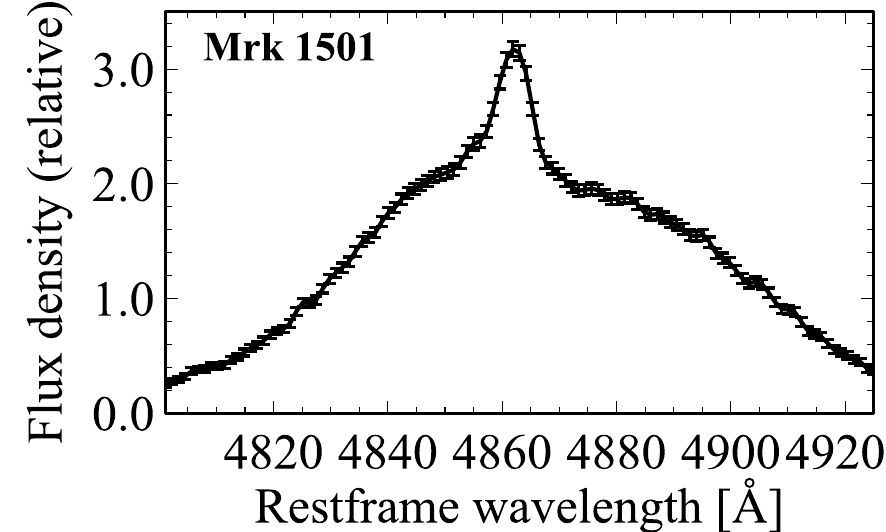}\\[0.2cm]
\includegraphics[width=0.3\textwidth]{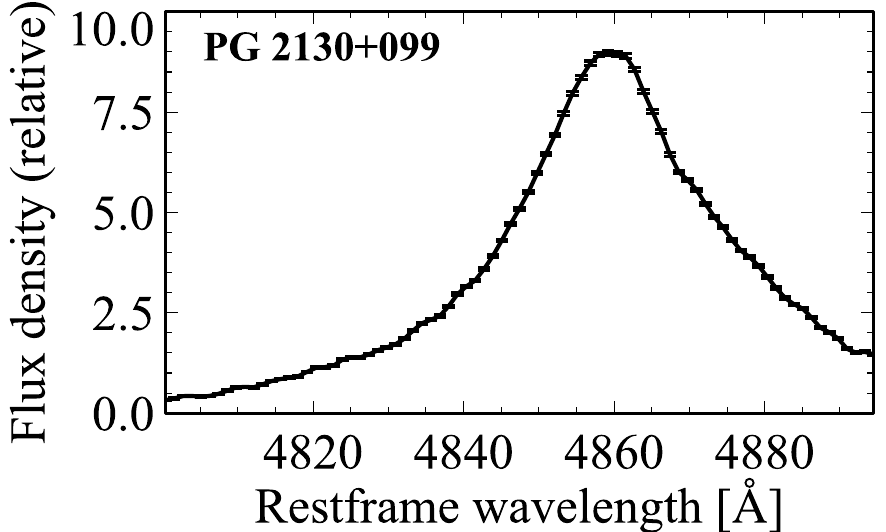}
\includegraphics[width=0.3\textwidth]{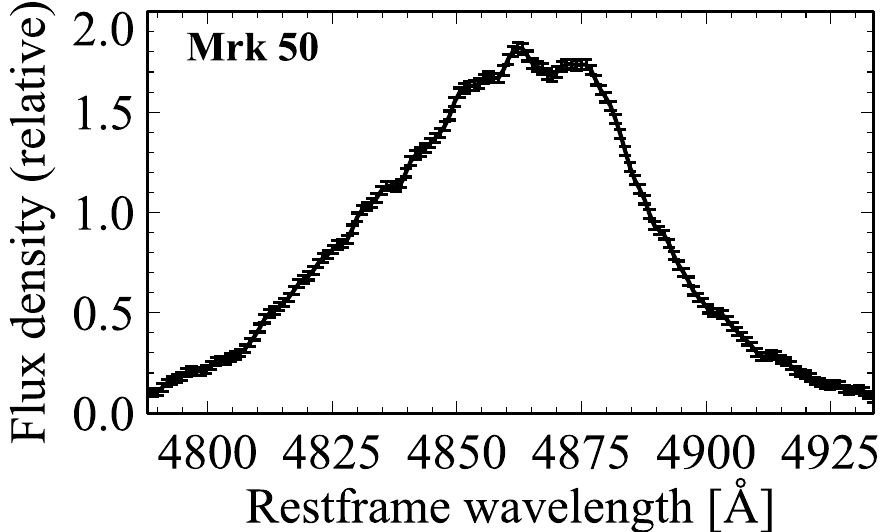}
\includegraphics[width=0.3\textwidth]{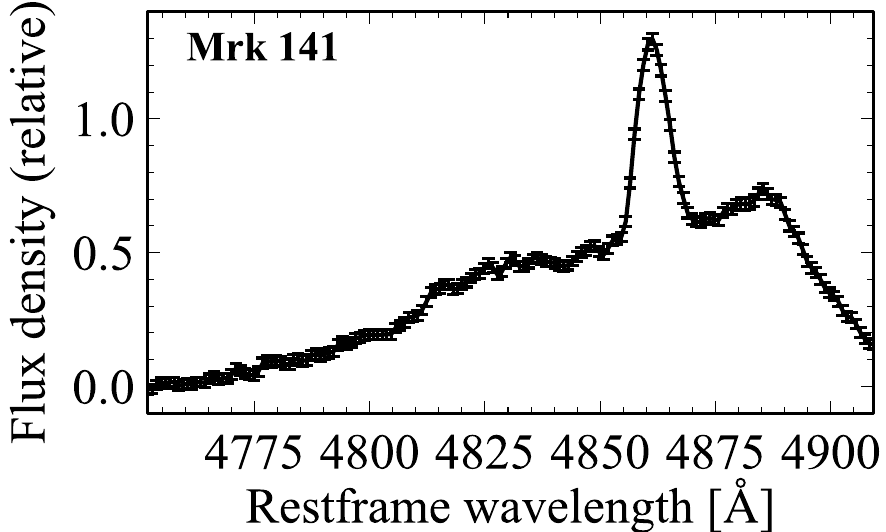}\\[0.2cm]
\includegraphics[width=0.3\textwidth]{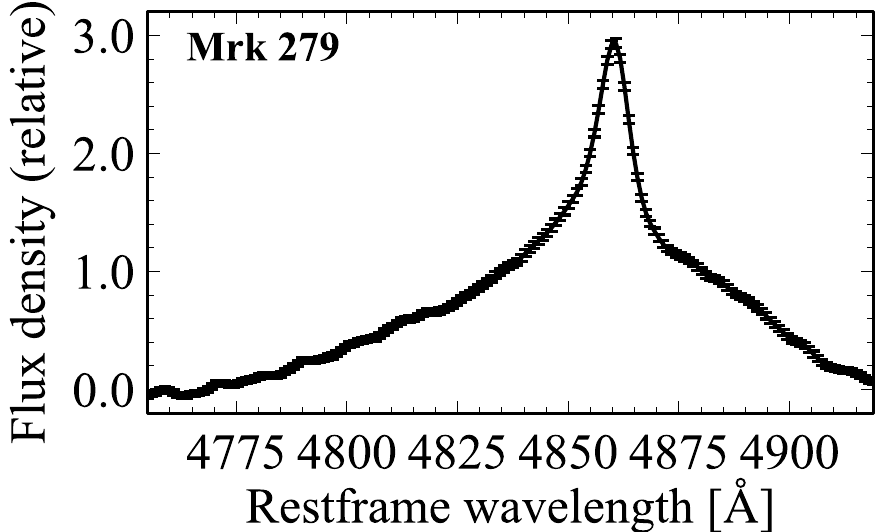}
\includegraphics[width=0.3\textwidth]{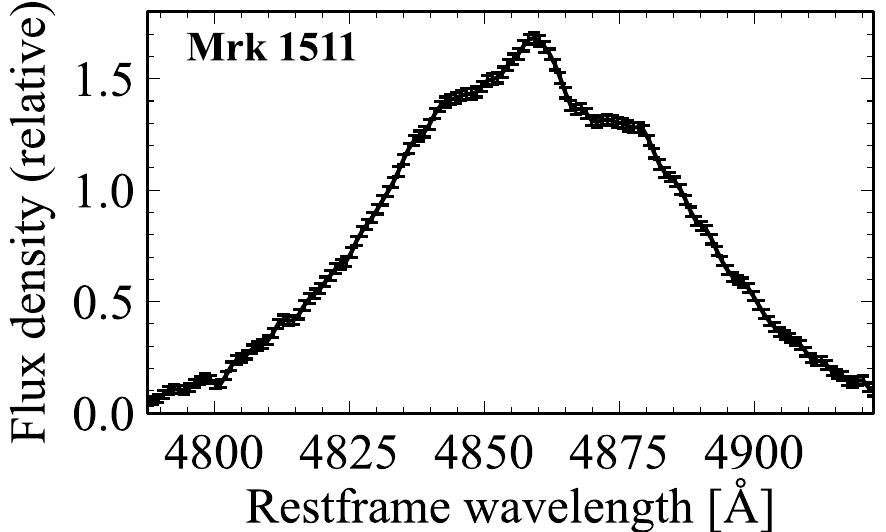}
\includegraphics[width=0.3\textwidth]{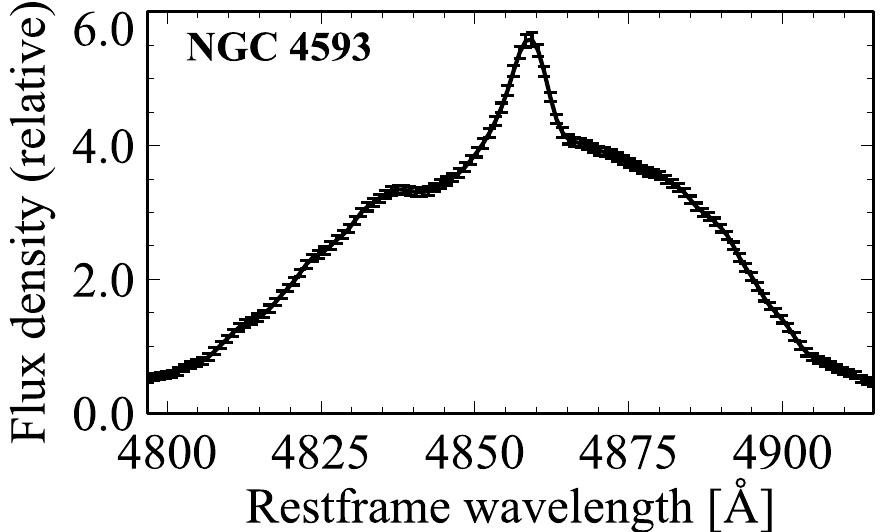}\\[0.2cm]
\includegraphics[width=0.3\textwidth]{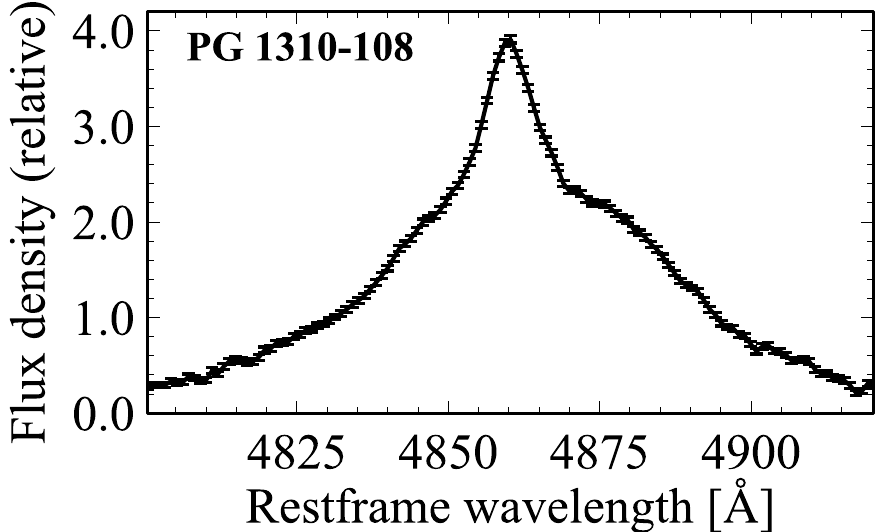}
\includegraphics[width=0.3\textwidth]{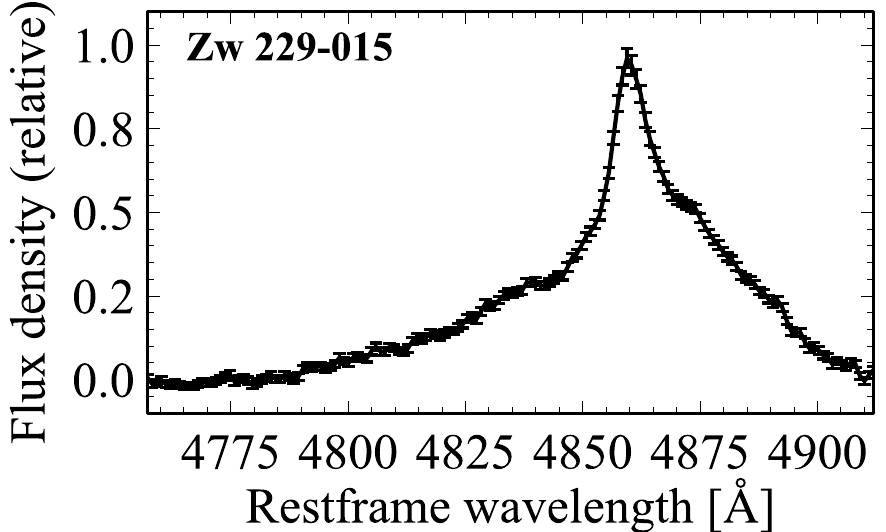}
\caption{Single-epoch spectra used in the BLR modelling. For each object, the spectrum shown corresponds to the epoch highlighted by a green star in Fig.~\ref{light-curves}. The spectra show the broad and narrow H$\beta$ emission line as a function of rest-frame wavelength, measured above the continuum flux level. The H$\beta$ line profile shown is the result of eliminating the other line and continuum contributions as defined by the spectral decomposition (see Section~\ref{sec:data} for details).}
\label{epoch_spec}
\end{figure*}

\subsection{Prior on the mean time delay}
Our model requires the use of a prior probability distribution for the mean time delay ($\tau_{\rm mean}$), as mentioned in Section~\ref{sec:single_epoch}. We adopt a log-normal prior distribution for $\tau_{\rm mean}$, with a mean and width based on the radius luminosity relation (R$-$L relation) and its scatter distribution as determined by \cite{bentz13}. The AGN luminosity for each object of our sample is calculated from the rest-frame AGN monochromatic flux density at 5100 \AA. We use AGN flux density values from \cite{grier12} and \cite{bentz13}, which were obtained by modelling and subtracting the host galaxy flux from \emph{Hubble Space Telescope} images. For objects where HST images are not available, we use the AGN 5100\AA\ rest-frame flux density measurements of \cite{barth15} based on spectral decompositions (see column 4 of Table~\ref{object_table}). The AGN flux density is then converted to an AGN monochromatic luminosity, L$_{\lambda}$ (5100\AA), using the luminosity distance for each object and the cosmological parameters listed in Section~\ref{sec:introduction}. Finally, the BLR radius (R$_{\rm BLR}$) is inferred using the R$-$L relation of \cite{bentz13} and L$_{\lambda}$ (5100\AA). The calculated R$_{\rm BLR}$ can be found in Table~\ref{object_table}. We also calculate and list in Table~\ref{object_table} the Eddington ratio for our targets ($\lambda = L_{\rm bol}/L_{\rm Edd}$, where $L_{\rm Edd}$ is the Eddington luminosity) as a measure of the mass accretion rate level of each AGN. We note that the Eddington ratio is only an approximate estimate as we use an average bolometric correction in the calculation.

In our previous work \citepalias{raimundo19}, we show that regardless of the width of the prior probability distribution, the model is able to derive similar constraints on the majority of the geometry and dynamics parameters of the BLR (with the exception of the M$_{\rm BH}$ and the mean and median time delays). Here we assume a width of 0.2 dex for the prior probability distribution of $\tau_{\rm mean}$, which represents the typical dispersion in the R$-$L relation \citep{bentz13}. This choice comes with the knowledge that the inferred M$_{\rm BH}$ will be strongly affected by the assumptions on the $\tau_{\rm mean}$ prior and should not be taken as an independent estimate of the intrinsic M$_{\rm BH}$ value. This choice is reasonable since our main goal is to constrain the remaining geometry and dynamics parameters of the BLR.

For all the remaining parameters, the prior probability distributions are the same as those shown by \citetalias{raimundo19} and can be found in Table~1 of that paper.

\subsection{Simulated continuum lightcurves}
\label{sec:continuum}
As discussed in Section~\ref{sec:model}, the model requires a simulated continuum light-curve determined at random instants to be able to model the BLR, even for the single-epoch modelling we implement.  

In our previous work \citepalias{raimundo19} we tested several assumptions for the statistical properties of the simulated continuum lightcurves and found that they did not affect the inferred geometry and dynamics parameters. This is in part due to the fact that the model renormalises the absolute continuum and spectral flux by means of two rescaling free parameters included in the model, $C_{\rm add}$ and $C_{\rm mult}$. Additionally, to infer the geometry and dynamics parameters we marginalise over the continuum light-curve parameters, meaning that we take into account all the possible lightcurves that could generate the observed line profile. 

The continuum light-curve parameters, hereafter the continuum hyper-parameters, are not fixed in the model but defined with flat (i.e. uninformative) prior probability distributions within a fixed wide range. The three main continuum hyper-parameters are $\tau_{\rm cont}$, the typical time-scale for variations; $\mu_{\rm cont}$, the long-term mean flux value of the light-curve and $\sigma_{\rm cont}$, the long-term standard deviation of the light-curve, as in Equations~5 and 6 of \cite{pancoast14a}. For consistency we use upper and lower limits for the priors that encompass the broad range of values found from studies of AGN variability (e.g. \citealt{kelly09}, \citealt{kozlowski10}, \citealt{macleod10}). We assume $\tau_{\rm cont}$ between 1 and 10$^{4}$ days, $\mu_{\rm cont}$ variations of 40\% around the mean flux value (i.e. around our single-epoch continuum flux value) and $\sigma_{\rm cont}$ is defined via the parameter $\hat{\sigma}_{\rm cont} = \sigma_{\rm cont} \sqrt{2/\tau_{\rm cont}}$ within the range $\hat{\sigma}_{\rm cont} \sim [8\times10^{-4} - 0.3]$ mag/day$^{1/2}$. We simulate a continuum light curve with a duration of 1.8$\times$10$^{7}$s for all AGN except Mrk 1501. For Mrk 1501 we simulate a light curve of 2.7$\times$10$^{7}$s to be conservative, since there is evidence from the R$-$L relation that the BLR is larger for this AGN than for the other AGN in our sample. Some of the simulated light curves are similar to the real continuum light curves for the AGN studied, but in general our simulated light curves are different since we are exploring a broad parameter space in continuum parameters. An example of simulated lightcurves is shown in the appendix.

The continuum light-curve parameters are all defined in the rest-frame of the AGN, therefore, our simulated continuum light-curve is the continuum light-curve in the rest-frame as well. All the remaining parameters are also defined in the rest-frame of the AGN.

\section{Results}
\label{sec:results}
\begin{figure*}
\centering
\includegraphics[width=0.95\textwidth]{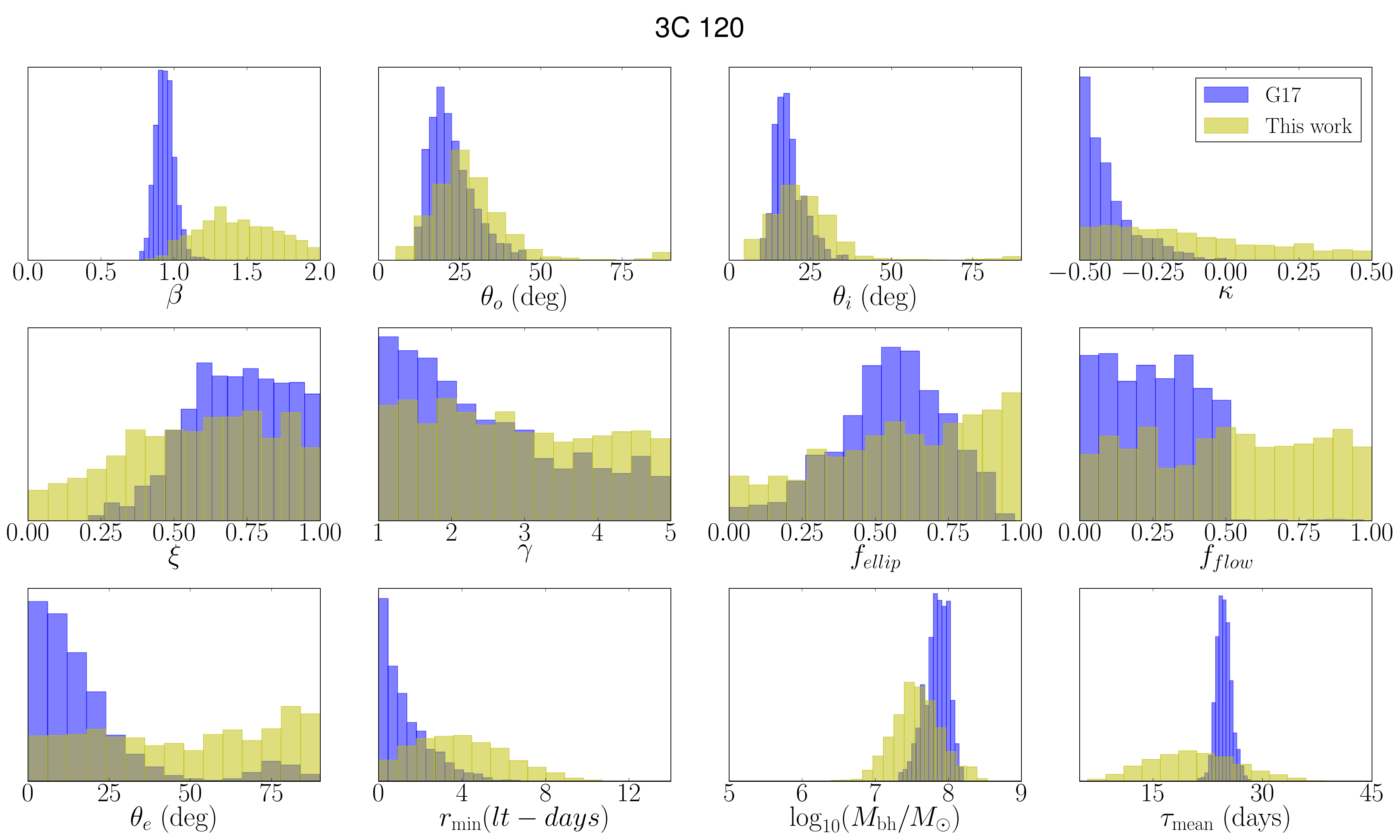}
\caption{Histograms showing the posterior probability distributions for the parameters in the single-epoch BLR modelling of 3C 120. Results from our single-epoch model are shown in yellow and compared with the full light-curve modelling from previous work. For the AGN10 objects our results are compared with those of \citealt{grier17} (marked as G17 in the figures) shown as the blue histograms. For the LAMP 2011 objects we compare our results with the \citealt{williams18} results using two different prescriptions: in blue we show the \citealt{williams18} combined posterior probability distribution (marked as W18 - combined in the figures) and in red the posterior probability distribution for the modelling using the \citealt{kovacevic10} spectral decomposition (marked as W18 - K10 in the figures).}
\label{posterior_3c120}
\end{figure*}

In this section we show the results from the single-epoch BLR modelling for all objects of Table~\ref{object_table}.  Figs.~\ref{posterior_3c120} to \ref{posterior_zw229} in the main text and Figs.~\ref{posterior_mrk335} to \ref{posterior_iisz10} in the Appendix show the final posterior probability distributions for the BLR parameters (yellow histograms), obtained by modelling the single-epoch line profiles of Fig.~\ref{epoch_spec}. Overlaid in the figures are the previous results obtained by modelling the full light-curve (\citealt{grier17} and \citealt{williams18}) shown as blue and red histograms. The blue histograms for the LAMP 2011 objects refer to the combined posterior probability distribution, while the red histograms are the posterior probability distributions obtained using the \cite{kovacevic10} Fe II templates in the spectral decomposition (see \citealt{williams18} for more details).
The meaning of each BLR parameter is described in detail in Section~\ref{sec:model}. In addition to those parameters, we show the flux-weighted median time delay ($\tau_{\rm median}$), the flux-weighted mean radius of the BLR ($r_{\rm mean}$) and the flux-weighted median radius of the BLR ($r_{\rm median}$), which are calculated by the model and related to the mean time delay. A list of the inferred parameter values and their uncertainties can be found in Table~\ref{table_results}. Examples of the line profiles generated by the model can be found in Fig.~\ref{spec_fit}. The profiles in this figure correspond to a solution found by the model and randomly extracted from the posterior probability distribution.

Fig.~\ref{inferred_parameters} shows a visual comparison between the 68\% confidence ranges of the inferred parameters in our work (shown by the solid lines) and the confidence ranges found using the full light-curve modelling (shown by the dashed lines). The median values of the posteriors are indicated by the blue filled circles for the results in this work and by the black filled stars for the full light-curve modelling result. We consider that a parameter cannot be constrained when the 68\% confidence range covers more than 50\% of the parameter space.

In agreement with our previous work on Arp 151 \citepalias{raimundo19}, we find that several of the BLR parameters can be constrained using a single-epoch model. The degree to which a parameter can be constrained and which parameters are constrained depends on the AGN analysed and likely on the shape of the broad emission line. 
We find that when the BLR parameters can be constrained, the inferred values for those parameters tend to agree within the 68\% confidence intervals with the inferred values from the full light-curve modelling. Below we describe the results for each individual object using Figures~\ref{posterior_3c120} to \ref{inferred_parameters} and Figs.~\ref{posterior_mrk335} to \ref{posterior_iisz10} as reference. We discuss all the parameters here in this section, except \Mbh\, and $\beta$, which will be discussed in detail in Section~\ref{sec:individual_parameters}.

\begin{figure*}
\centering
\includegraphics[width=0.95\textwidth]{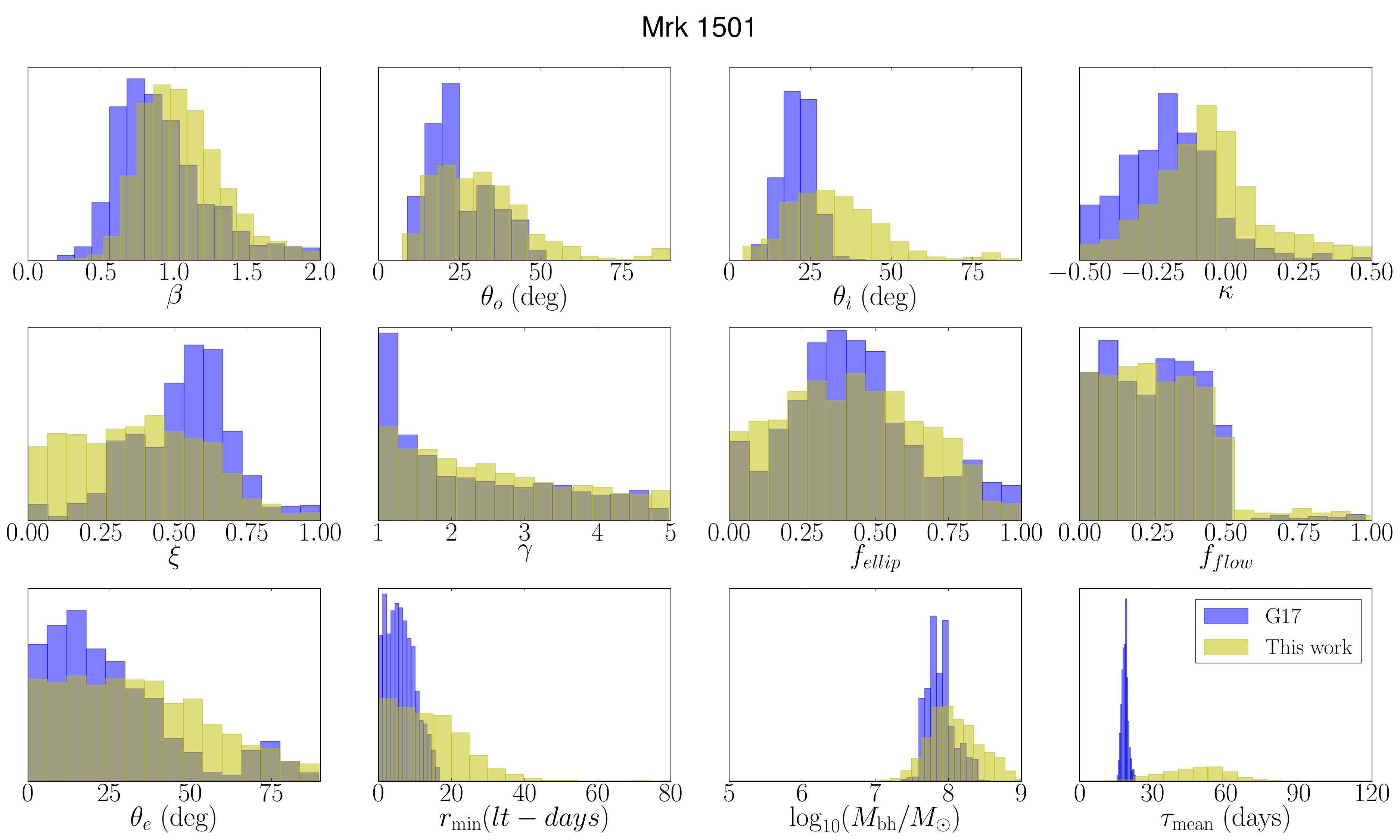}
\caption{Same as Fig.~\ref{posterior_3c120} but for Mrk 1501.}
\label{posterior_mrk1501}
\end{figure*}

\begin{figure*}
\centering
\includegraphics[width=0.95\textwidth]{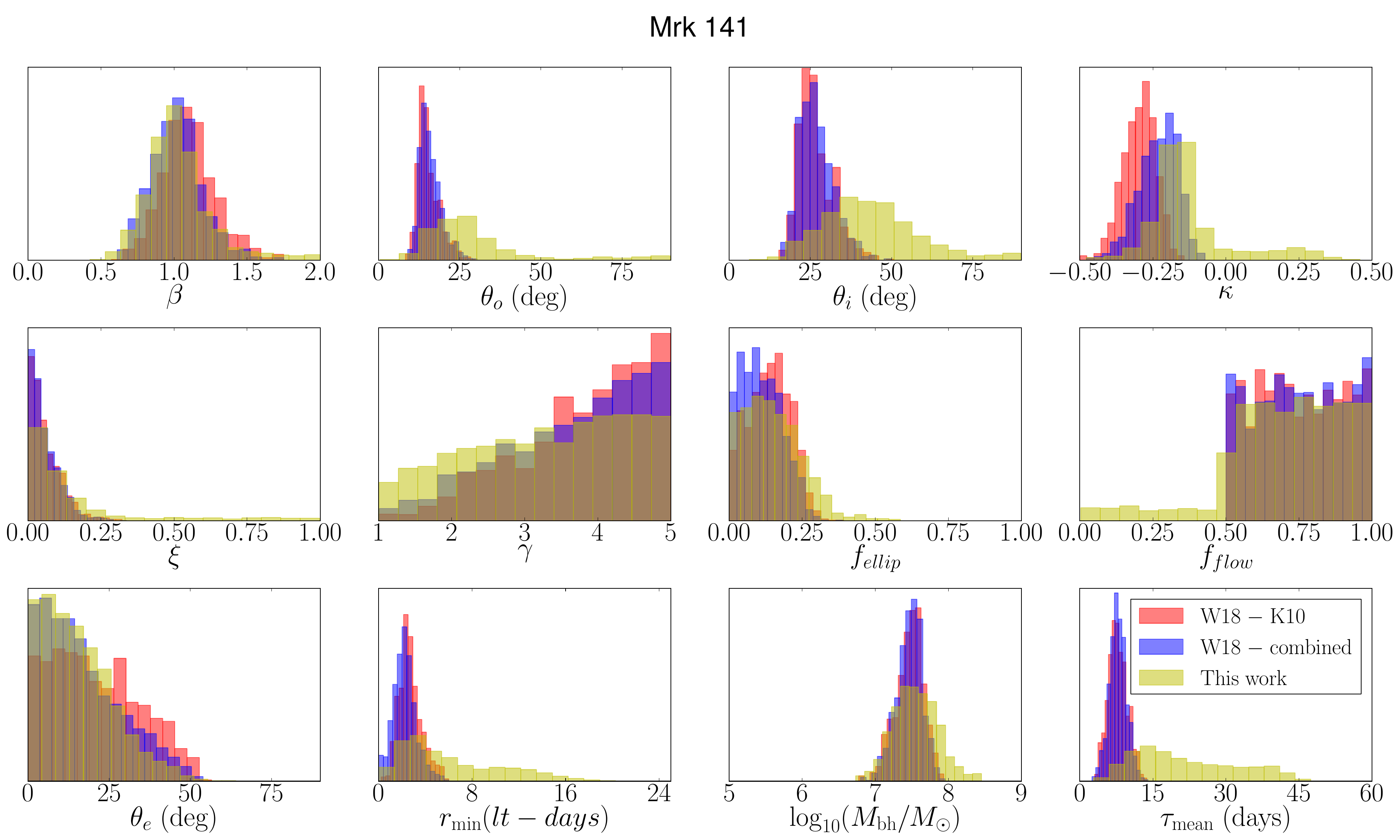}
\caption{Same as Fig.~\ref{posterior_3c120} but for Mrk 141.}
\label{posterior_mrk141}
\end{figure*}

\begin{figure*}
\centering
\includegraphics[width=0.95\textwidth]{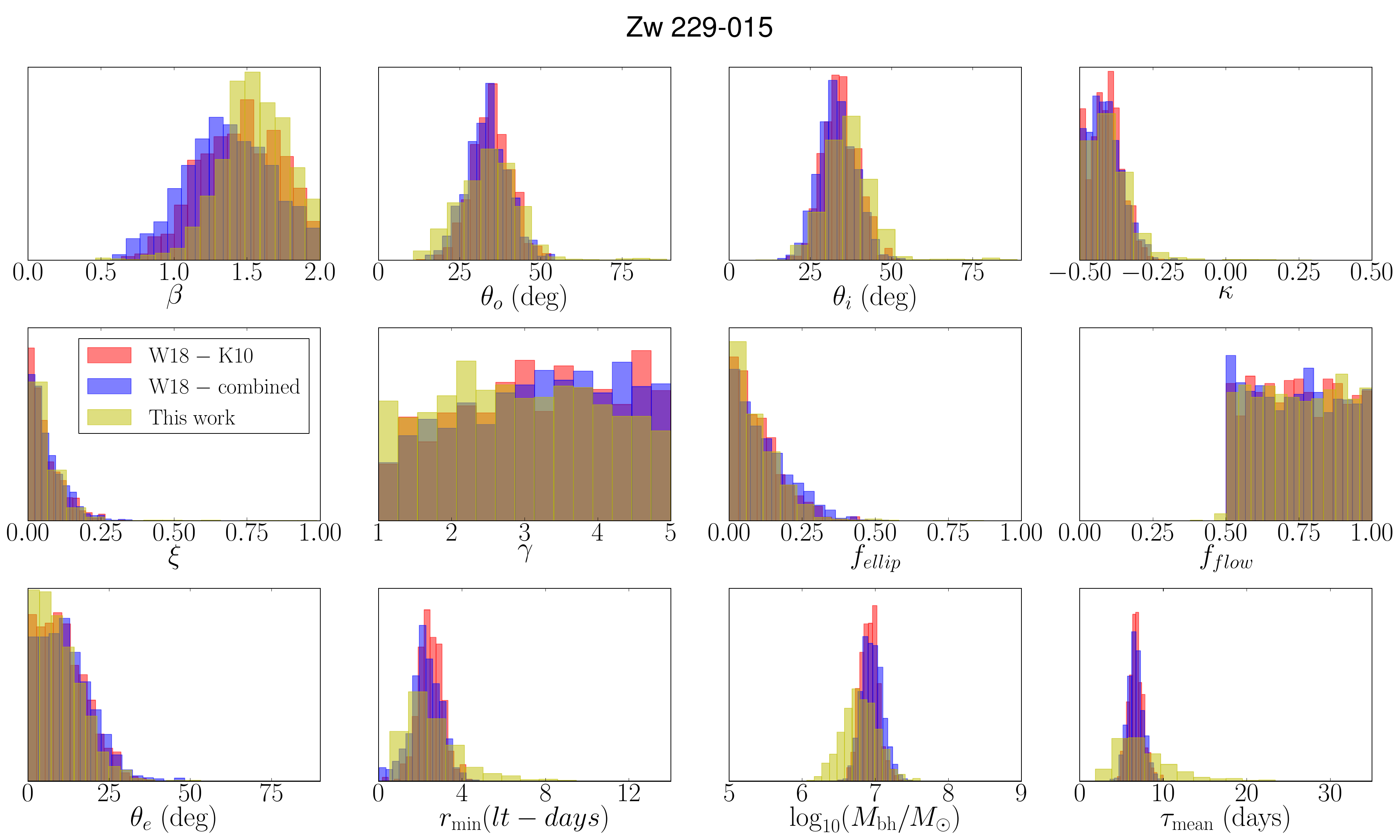}
\caption{Same as Fig.~\ref{posterior_3c120} but for Zw 229-015.}
\label{posterior_zw229}
\end{figure*}

\subsection{Results on individual objects}

\subsubsection{3C 120}
Some of the parameters can be constrained for this AGN and agree within the errors with the full light-curve result: we find \opening $=$ \Remark{3c120_opening} degrees, \inc $=$ \Remark{3c120_inc} degrees and \rmin $=$ \Remark{3c120_rmin} light-days. 
The remaining parameters have very broad posterior probability distributions (Fig.~\ref{posterior_3c120}), which indicates that they cannot be accurately constrained from the single-epoch spectrum.

\subsubsection{Mrk 335}
The posterior probability distributions for most of the parameters of Mrk 335 are very broad (Fig.~\ref{posterior_mrk335}), indicating that the BLR parameters cannot be constrained from the single spectrum. The only exception is \rmin $=$ \Remark{mrk335_rmin} light-days which is constrained to a value similar to what has been found from the full light-curve model.

\subsubsection{Mrk 1501}
Some of the parameters can be constrained and agree within the errors with the full light-curve result: we find \opening $=$ \Remark{mrk1501_opening} degrees, \inc $=$ \Remark{mrk1501_inc} degrees and $\kappa =$ \Remark{mrk1501_kappa}. \fflow\ is also constrained and found to be $< 0.5$ (i.e. signature of inflowing trajectories), indicating that even though \fellip is not tightly constrained, the model is able to identify that some of the orbits have velocities close to the radial inflowing escape velocity.
The parameters \fellip, \thetae\ and $\gamma$ are not constrained, but we note that the full light-curve model was also not able to constrain these parameters significantly. The parameter $\xi$ is somewhat constrained to $\xi =$ \Remark{mrk1501_eps}, indicating that the mid plane of the BLR is not completely opaque nor completely transparent, although the posterior probability distribution is broad, similar to what was found for the full light-curve modelling (Fig.~\ref{posterior_mrk1501}).
\subsubsection{PG 2130+099}
Some of the parameters can be constrained: we find \opening $=$ \Remark{pg2130_opening} degrees, \inc $=$ \Remark{pg2130_inc} degrees and \rmin $=$ \Remark{pg2130_rmin} light-days, consistent within the uncertainties with the full light-curve result. The parameter $\kappa$ is found to be lower than zero, favouring scenarios in which the particles on the far side of the BLR contribute with more emission. However, this is not a strong constraint as the posterior probability distribution for this parameter has a tail extending to positive $\kappa$ values, indicating possible degeneracies. However, the posterior distribution for $\kappa$ from the full light-curve modelling is also broad. The remaining parameters have very broad posterior probability distributions (Fig.~\ref{posterior_pg2130}), and cannot be accurately constrained.

\subsubsection{Mrk 50}
Some of the parameters can be constrained and agree within the errors with the full light-curve result: we find \inc $=$ \Remark{mrk50_inc} degrees, $\kappa =$ \Remark{mrk50_kappa}, \fflow\ is constrained to be $> 0.5$ i.e. signatures of outflowing trajectories, \thetae $=$  \Remark{mrk50_thetae} degrees and \rmin $=$ \Remark{mrk50_rmin} light-days. The parameter \fellip\ is somewhat constrained to \Remark{mrk50_fellip}. Although the posterior probability distribution for \fellip\ is broad, it indicates that scenarios where \fellip $\rightarrow$ 1.0, i.e. no gas is inflowing/outflowing, are not favoured. The parameter $\xi$ is also marginally constrained to $\xi = $\Remark{mrk50_eps} with a broad posterior probability distribution. Both \fellip and $\xi$ nevertheless agree with the full light-curve results within the uncertainties. The remaining parameters have very broad posterior probability distributions (Fig.~\ref{posterior_mrk50}), and cannot be accurately constrained.

\subsubsection{Mrk 141}
Mrk 141 is a remarkable case since most of its BLR parameters can be constrained using a single spectrum. As discussed in Section~\ref{sec:discussion} this is likely due to the significant asymmetry and structure of the line profile in this AGN, which provides more information for the model to constrain the BLR parameters. We find \opening $=$ \Remark{mrk141_opening} degrees, \inc $=$ \Remark{mrk141_inc} degrees $\kappa =$ \Remark{mrk141_kappa}, $\xi =$  \Remark{mrk141_eps}, \fellip $=$ \Remark{mrk141_fellip},  \fflow\ $> 0.5$ i.e. signatures of outflowing trajectories, \thetae $=$  \Remark{mrk141_thetae} degrees and \rmin $=$ \Remark{mrk141_rmin} light-days, all agreeing within the uncertainties with the full light-curve result (Fig.~\ref{posterior_mrk141}). The only exception is the parameter $\gamma$, which cannot be constrained. However we note that an unconstrained $\gamma$ is a common feature for all our objects.

\subsubsection{Mrk 279}
Some of the parameters can be constrained: we find \inc $=$ \Remark{mrk279_inc} degrees, \fflow\ is constrained to be $> 0.5$ i.e. signatures of outflowing trajectories, indicating that even though \fellip is not tightly constrained, the model is able to identify that some of the orbits have velocities close to the radial outflowing escape velocity. We also find \rmin $=$ \Remark{mrk279_rmin} light-days and \thetae $=$  \Remark{mrk279_thetae} degrees indicating that not all of the outflowing orbits are bound.
The opening angle is somewhat constrained: \opening $=$ \Remark{mrk279_opening} degrees, indicating that very thin disc configurations are not favoured. A similar result is obtained for $\xi = $ \Remark{mrk279_eps}, indicating that the mid-plane of the BLR is not completely transparent. Both the constrained and marginally constrained parameters agree within the uncertainties with the result from the full light-curve modelling (Fig.~\ref{posterior_mrk279}). 

\subsubsection{Mrk 1511}
\label{sec:mrk1511}
Our model is not able to constrain most of the parameters for Mrk 1511 (Fig.~\ref{posterior_mrk1511}). The only exception is \rmin $=$ \Remark{mrk1511_rmin} light-days which is constrained and agrees within the uncertainties with what was found for the full light-curve.
The model marginally constrains \opening $=$ \Remark{mrk1511_opening} degrees. This parameter is one of the few cases where the single-epoch estimate is not consistent within the uncertainties (68\% confidence range) with what was found from the full light-curve modelling. 
Since we used the total (AGN + host galaxy) flux to calculate the mean time delay from the R$-$L relation, we assume a mean value for the prior on $\tau_{\rm mean}$ which is significantly higher than what was found in the full light-curve modelling. This results in the time delay parameters ($\tau_{\rm mean}$, $\tau_{\rm median}$, $r_{\rm mean}$ and $r_{\rm median}$) to be overestimated. The inferred M$_{\rm BH}$ from single-epoch modelling agrees with the full light-curve result. This indicates that independent information on the H$\beta$ line profile helps to constrain M$_{\rm BH}$ beyond the approximate value suggested by the R$-$L relation. As observed by \cite{williams18}, it is likely that Fe II has a strong contribution to the spectra of Mrk 1511. \cite{williams18} found that due to the strong Fe II contribution, different spectral decompositions resulted in different values for the BLR parameters. A residual Fe II contribution may be one of the reasons why the single-epoch spectrum of Mrk 1511 does not provide meaningful constraints.

\subsubsection{NGC 4593}
Some of the parameters can be constrained: we find \opening $=$ \Remark{ngc4593_opening} degrees, \inc $=$ \Remark{ngc4593_inc} degrees, \fflow\ is constrained to be $> 0.5$ i.e. signatures of outflowing trajectories, indicating that even though \fellip\ is only marginally constrained (\fellip $=$ \Remark{ngc4593_fellip}), the model is able to identify that some of the orbits have velocities close to the radial outflowing escape velocity. We also find \thetae $=$  \Remark{ngc4593_thetae} degrees and \rmin $=$ \Remark{ngc4593_rmin} light-days.
The parameter $\kappa$ is somewhat constrained $\kappa =$ \Remark{ngc4593_kappa}, indicating that there is no significant asymmetry between the emission strength from the near or far side of the BLR. A weak constraint is also found for  $\xi = $ \Remark{ngc4593_eps}, indicating that the mid-plane of the BLR is not completely transparent. All the constrained and marginally constrained parameters mentioned above agree within the uncertainties with the full light-curve result.
We note that the 68\% confidence ranges for the parameters determined from the full light-curve modelling are also wide and comparable with what we find from the single-epoch modelling (Fig.~\ref{posterior_ngc4593}). This is in part due to the fact that the full light-curve result takes into account the three different spectral decompositions of \cite{williams18} which broadens the 68\% confidence intervals.

\subsubsection{PG 1310-108}
The parameter \rmin $=$ \Remark{pg1310_rmin} light-days is constrained and agrees with the full light-curve result within the uncertainties. Additionally there is one parameter that can be marginally constrained for PG 1310-108, the opening angle: \opening $=$ \Remark{pg1310_opening} degrees. Looking at the posterior probability distributions for \opening\ and \inc\ in Fig~\ref{posterior_iisz10} one can see that the probability distribution appears to have two peaks, which translates to broad 68\% confidence ranges and poorly constrained angles. These two peaks are also present in the full light-curve results, indicating that it is difficult to constrain the angles for PG 1310-108, even when using monitoring data.
All the remaining parameters have broad posterior probability distributions and therefore cannot be significantly constrained from the single-epoch spectrum.

\subsubsection{Zw 229-015}
The parameters for this object are remarkably well constrained and agree with the full light-curve result within the uncertainties. As for Mrk 141, this is likely due to the information associated with the significant asymmetry and structure of the line profile in this AGN. The only exception is $\gamma$, which is not constrained for Zw 229-015 or any of the AGN in our sample. We find \opening $=$ \Remark{zw229_opening} degrees, \inc $=$ \Remark{zw229_inc} degrees, $\kappa =$ \Remark{zw229_kappa}, $\xi = $  \Remark{zw229_eps}, \fellip $=$ \Remark{zw229_fellip}, \fflow\ is constrained to be $> 0.5$ i.e. signatures of outflowing trajectories,  \thetae $=$  \Remark{zw229_thetae} degrees and \rmin $=$ \Remark{zw229_rmin} light-days. The posterior probability distributions is shown in Fig.~\ref{posterior_zw229}.

\section{Discussion}
\label{sec:discussion}

We have information on the full light-curve modelling results for all the AGN we modelled. For the purpose of this work, we assume that the full light-curve inferred BLR parameters are representative of the intrinsic parameters of the BLR. Our exercise in this paper is to determine what BLR parameters can be constrained and to what degree of confidence, using single-epoch spectra modelling of a significant sample of AGN. This is necessary prior to applying our single-epoch model to AGN for which no monitoring data nor information on the BLR structure are available, which is our ultimate goal.

Below we discuss the overall model performance, describe in detail our findings for each of the BLR parameters, present a summary of the overall findings for the whole sample of AGN and discuss the caveats associated with our model.

\subsection{Model performance}
\label{sec:model_performance}
There are two important questions that our work addresses: 1) Is the single-epoch modelling able to constrain the BLR parameters? That is, are the posterior probability distributions narrow enough that useful information can be obtained? 2) Are the inferred parameters from single-epoch spectra in agreement with the full light-curve results, within the uncertainties?

The results from Section~\ref{sec:results} show that some of the BLR parameters can be constrained using single-epoch spectra, as the posterior probability distributions are significantly narrower than the prior probability distributions. More importantly, when the parameters can be constrained, the large majority of the inferred values agree with the full light-curve modelling results, within the uncertainties. For the case where the parameter posterior probability distribution is wide, this indicates that the model is not able to constrain the parameter. The model is successful at a) inferring correctly the BLR parameters, and b) at returning a broad posterior probability distribution when no sufficient constraints are found for a specific parameter.
After the performance tests we did in this work, combined with the results from \citetalias{raimundo19}, we are confident that the model can be used to derive quantitative constraints on the BLR parameters and be applied to AGN without prior knowledge of the BLR structure.

Notably the model is able to constrain most of the BLR parameters for two AGN: Mrk 141 and Zw 229-015. From Fig.~\ref{epoch_spec}, we can see that the line profiles for these two AGN have a significant substructure, which may be an advantage in the amount of information provided to the model, as suggested for Arp 151 \citepalias{raimundo19}. Fairly symmetrical and featureless profiles, such as for 3C120, Mrk 335 and PG 2130+099 tend to result in fewer constraints on the BLR parameters. This is not always the case, and there may be other factors contributing to how well the BLR parameters can be determined from a spectrum. Mrk 1511 for example, has some features and asymmetry in the profile but may have Fe II residuals in the spectrum, causing the model to not be able to constrain the BLR parameters, as discussed in Section~\ref{sec:mrk1511}. 

\begin{figure*}
\centering
\includegraphics[width=0.33\textwidth]{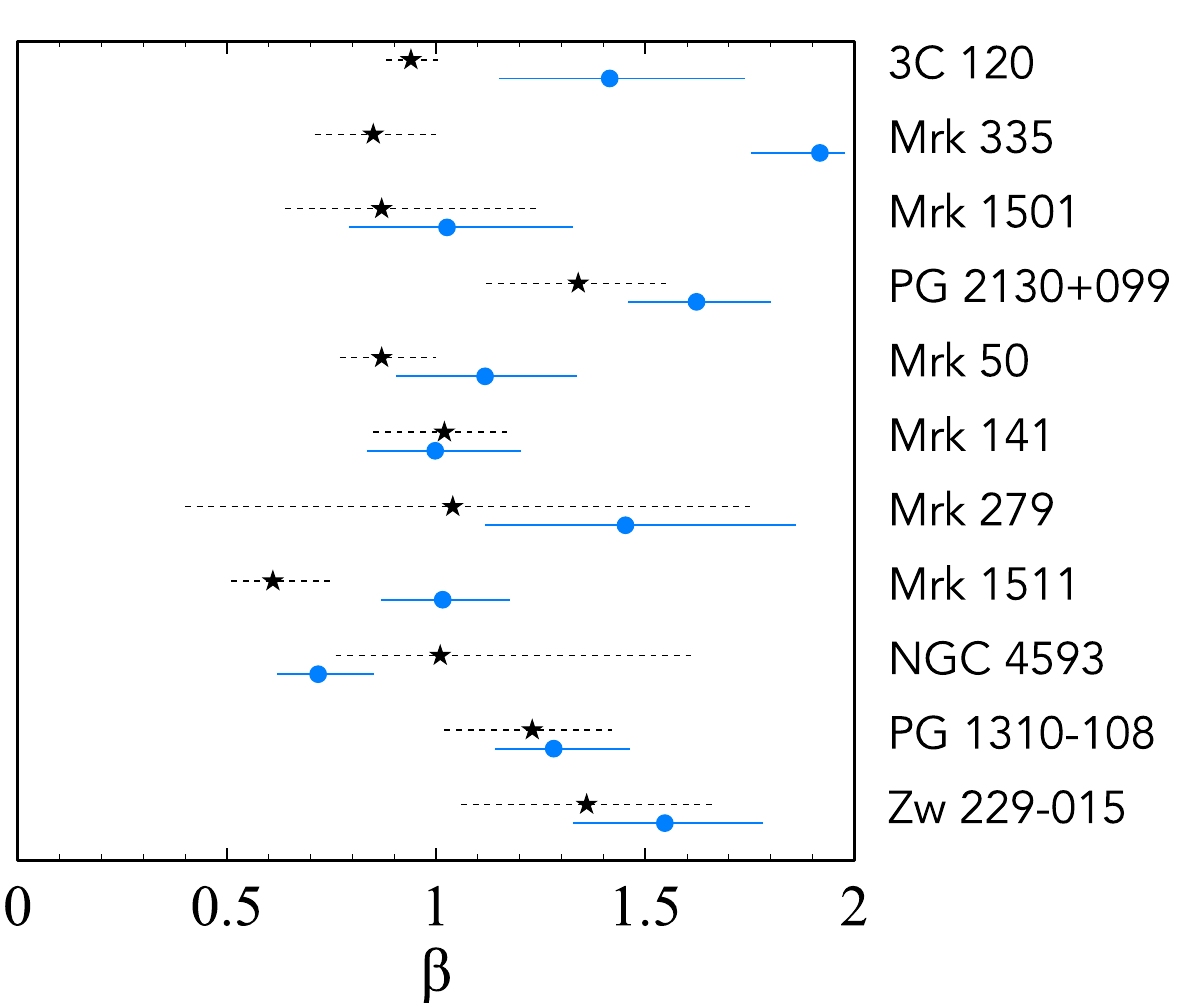}
\includegraphics[width=0.33\textwidth]{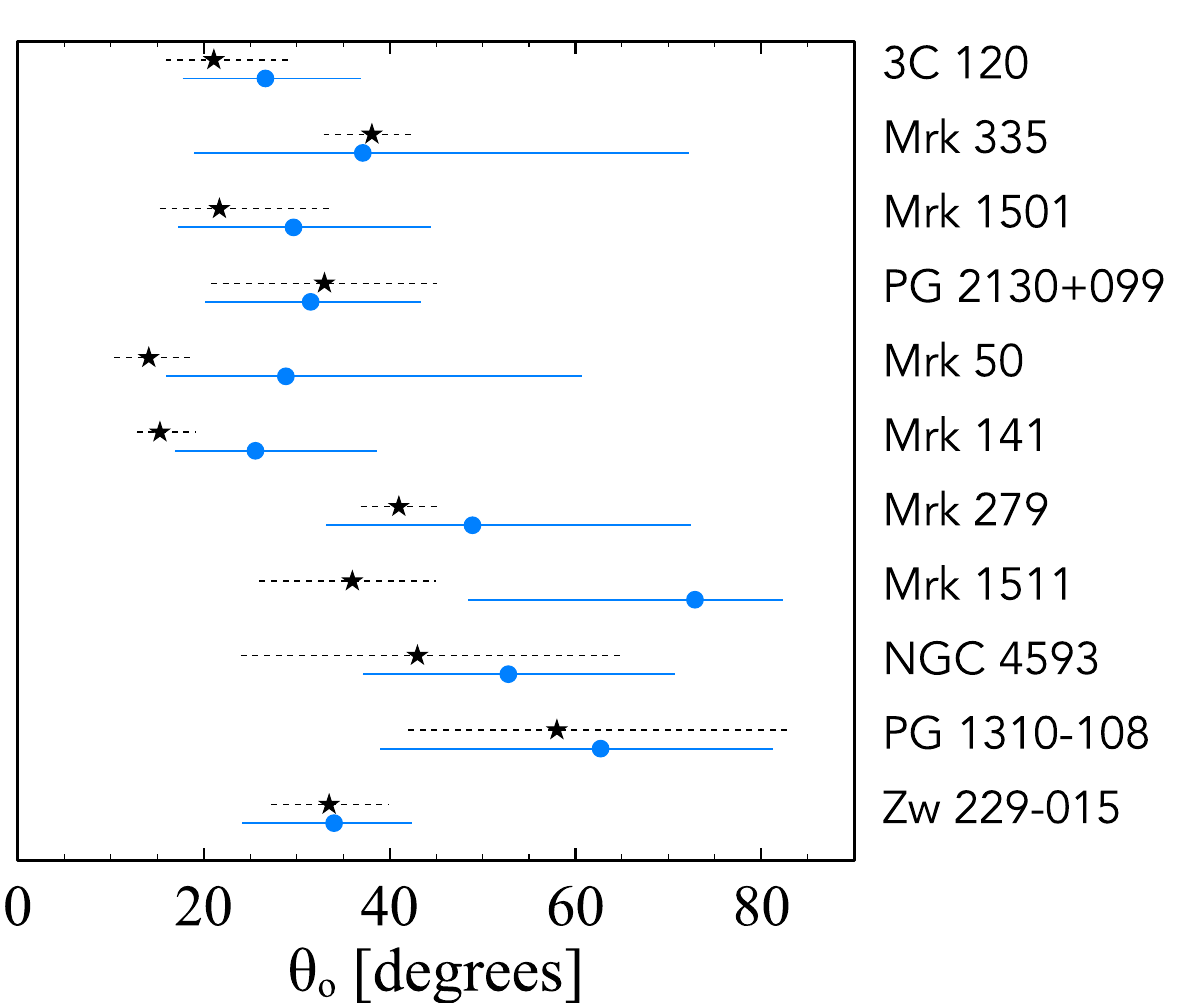}
\includegraphics[width=0.33\textwidth]{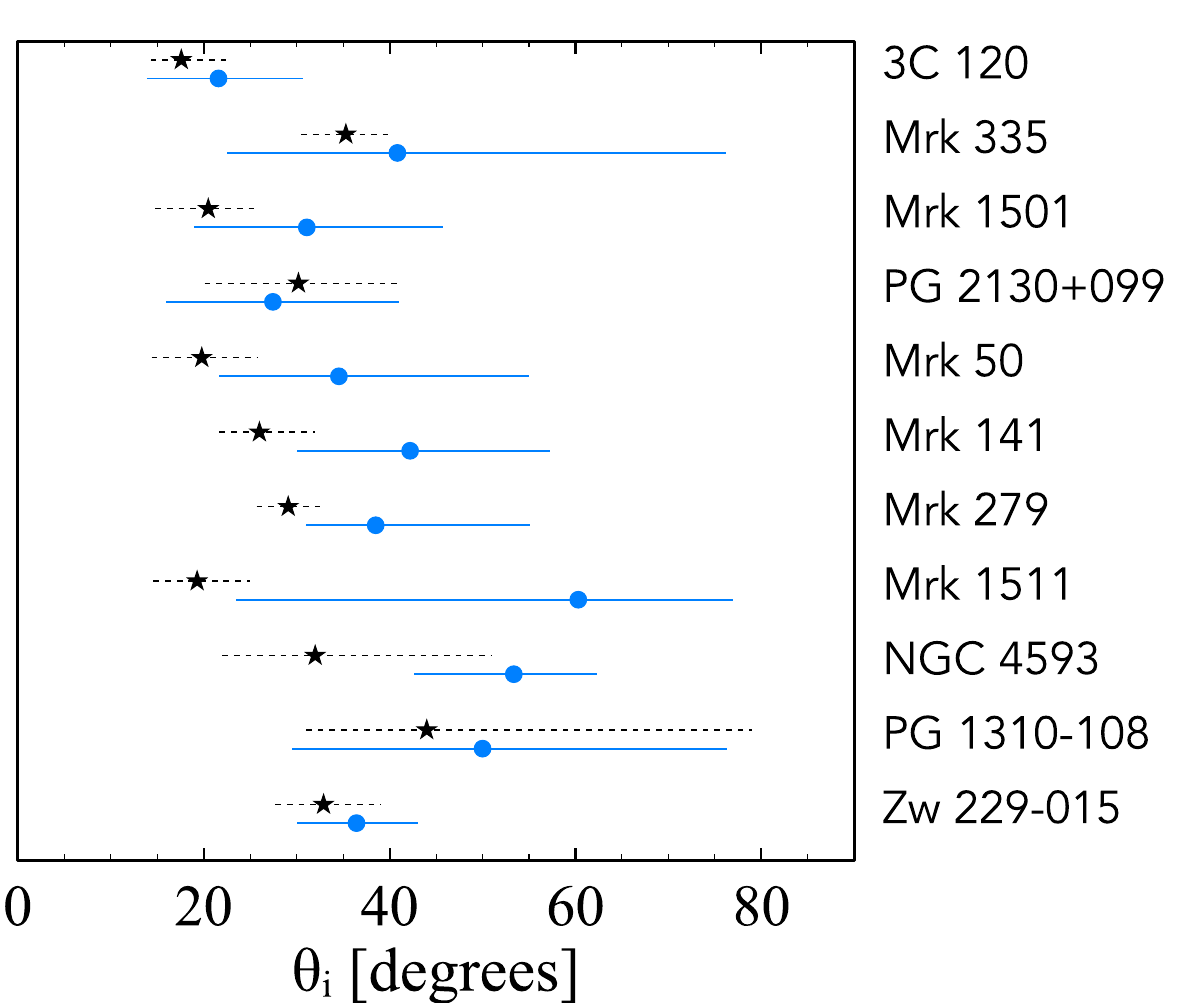}\\[0.1cm]
\includegraphics[width=0.33\textwidth]{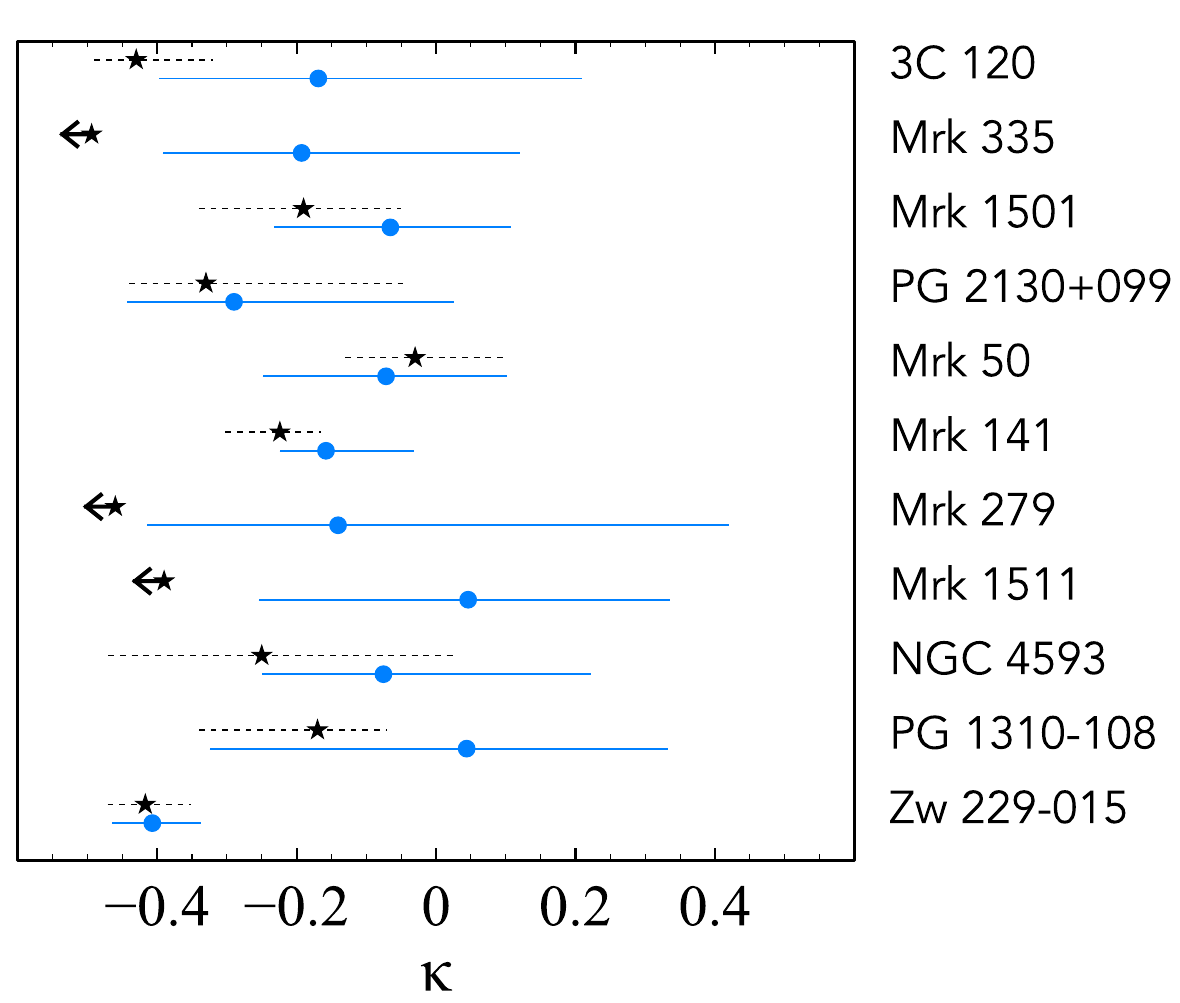}
\includegraphics[width=0.33\textwidth]{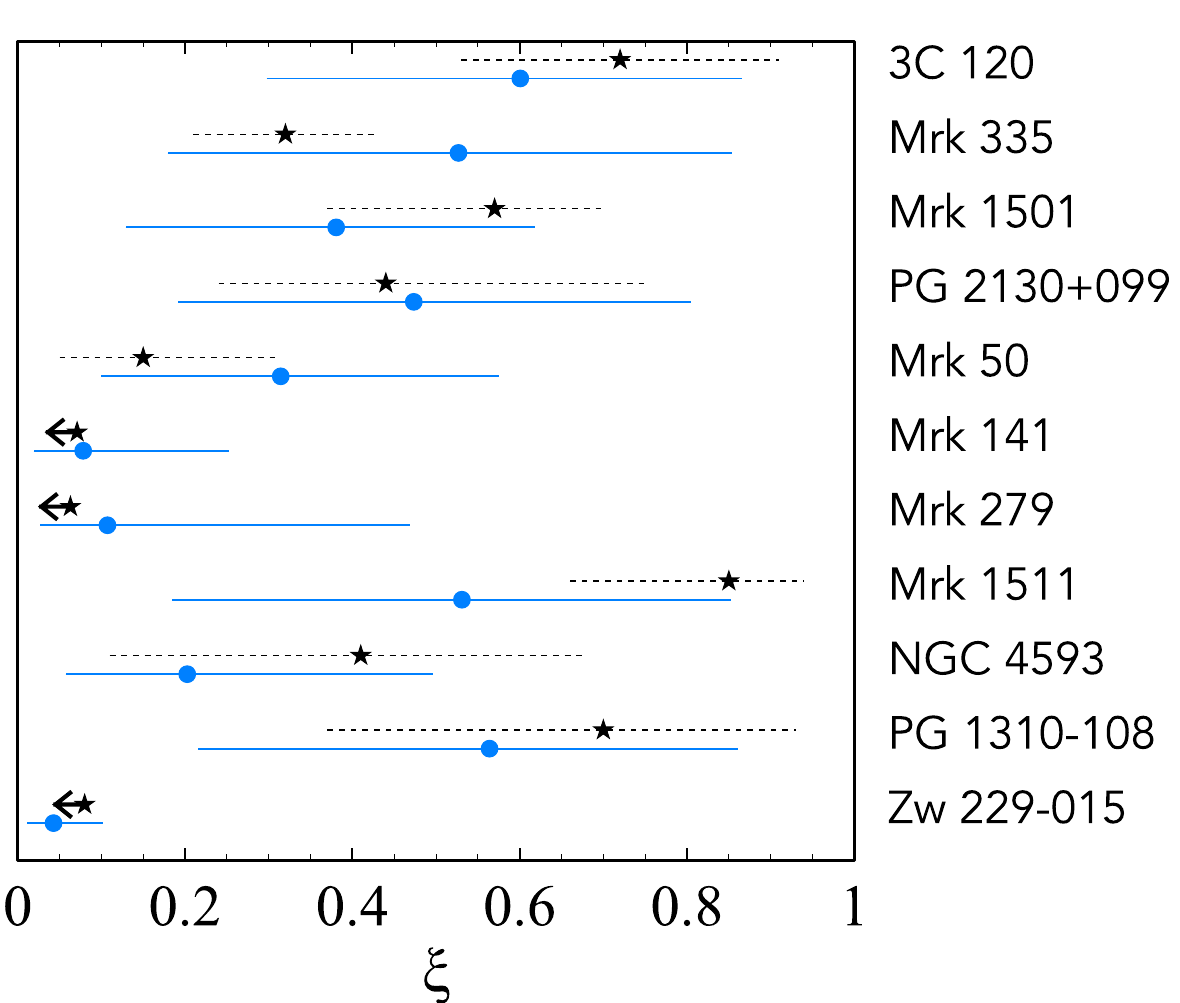}
\includegraphics[width=0.33\textwidth]{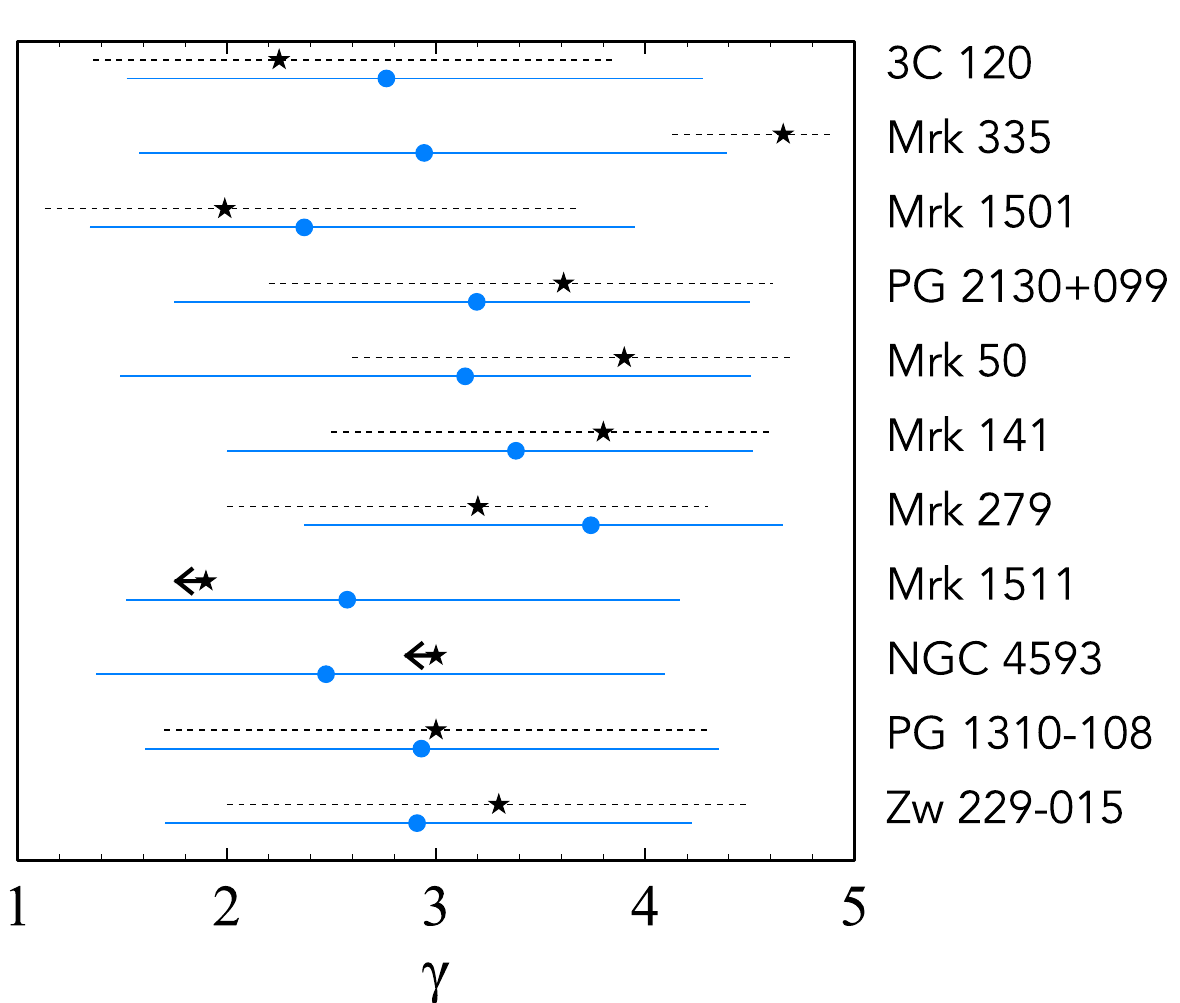}\\[0.1cm]
\includegraphics[width=0.33\textwidth]{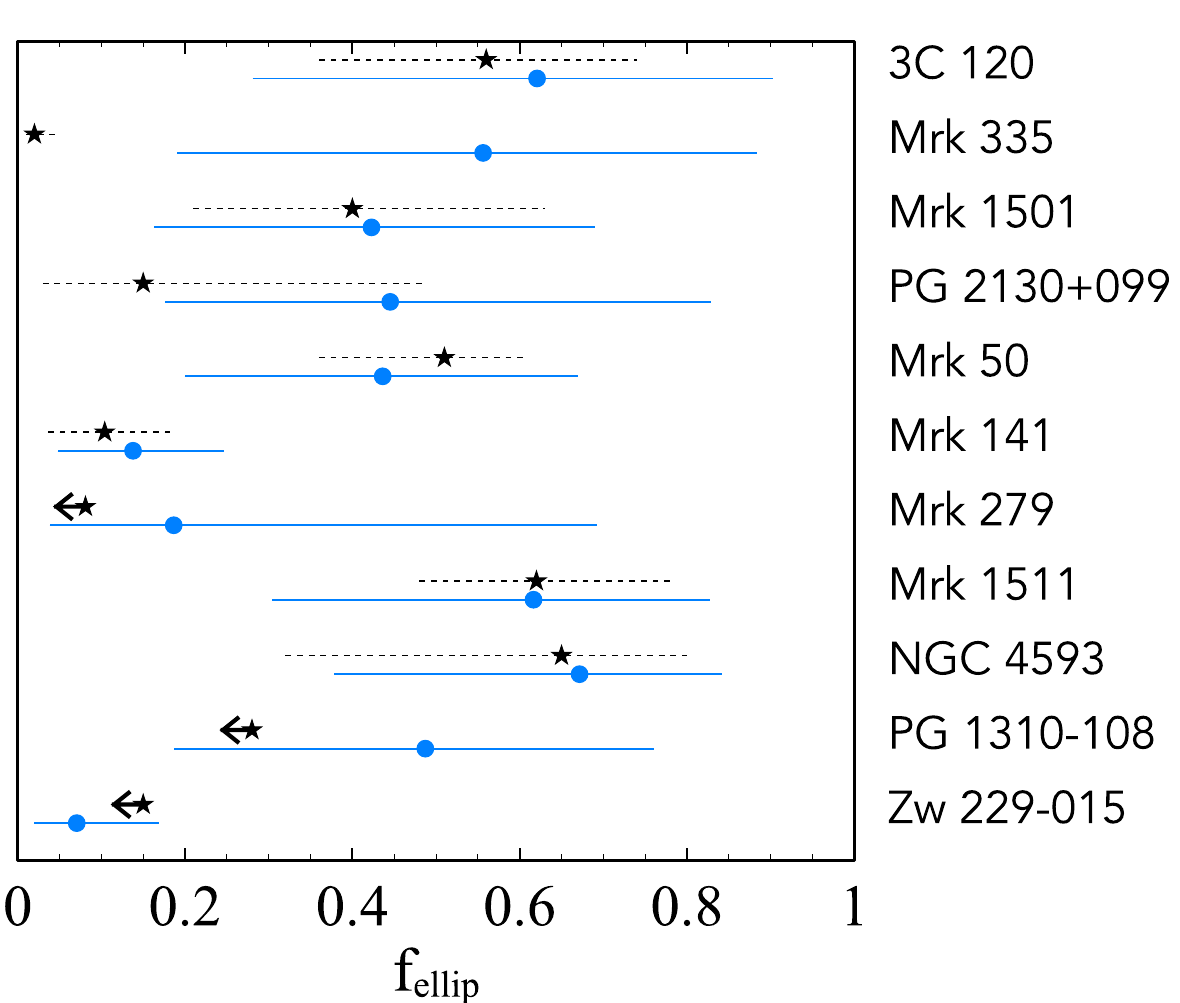}
\includegraphics[width=0.33\textwidth]{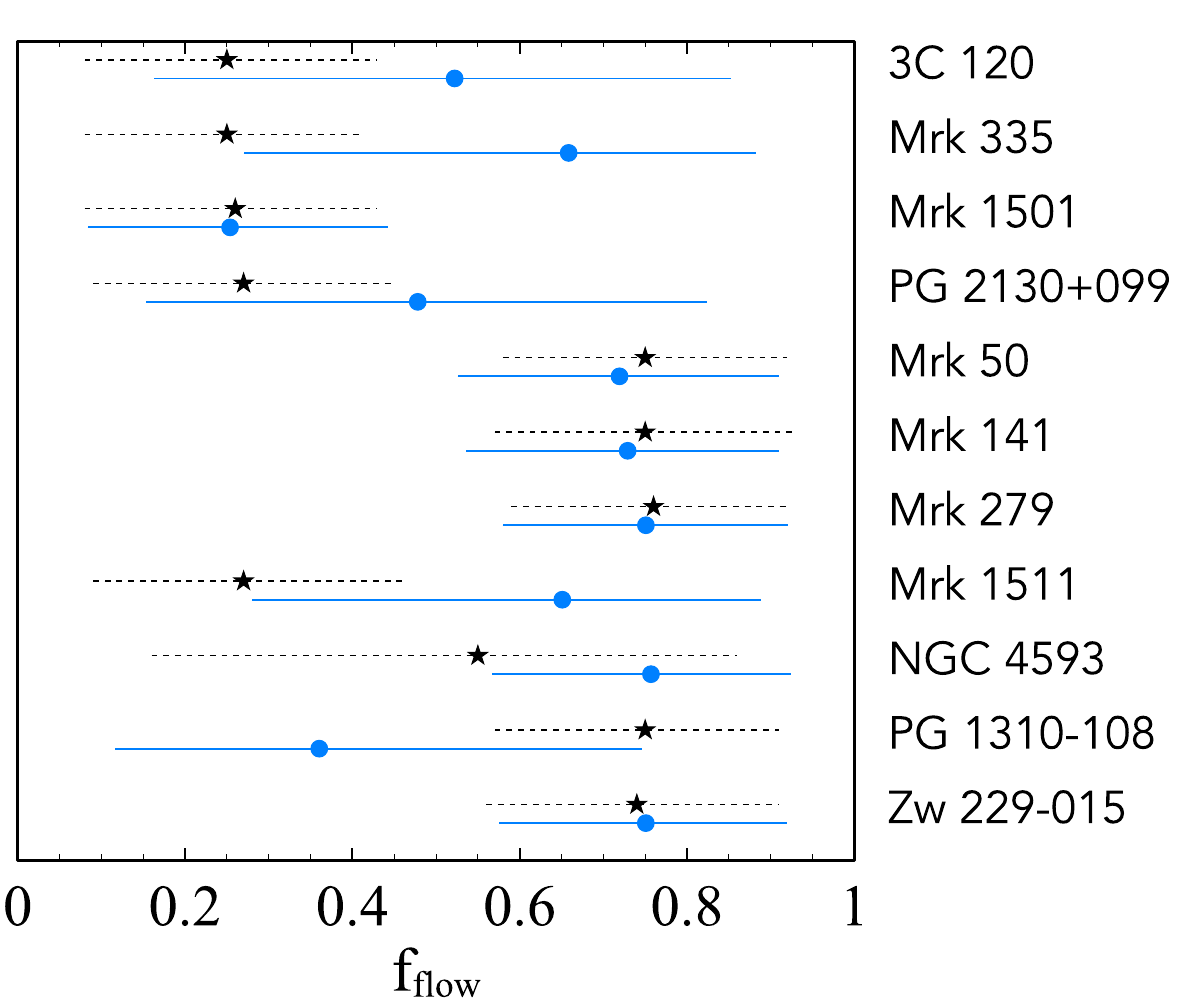}
\includegraphics[width=0.33\textwidth]{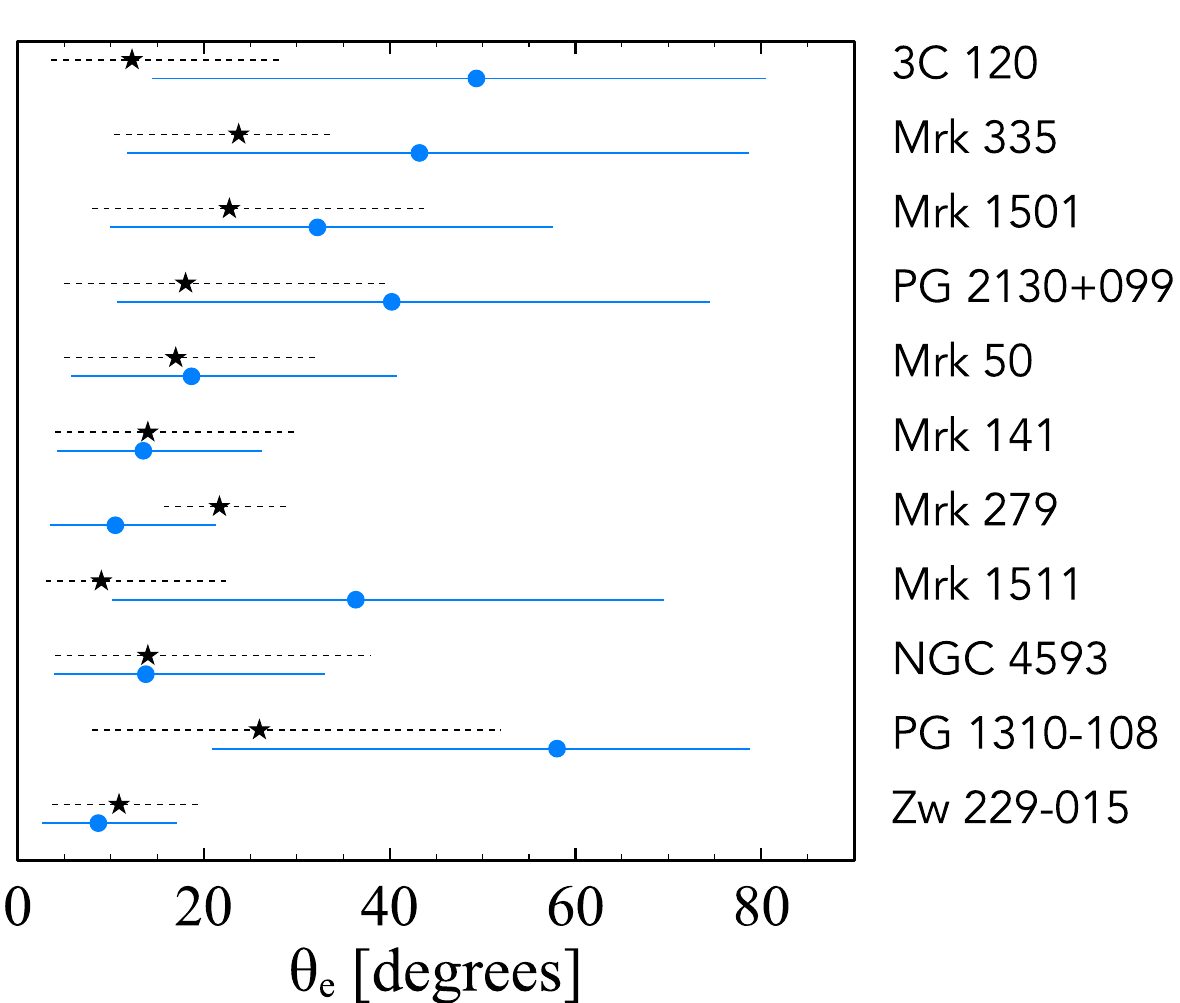}\\[0.1cm]
\includegraphics[width=0.33\textwidth]{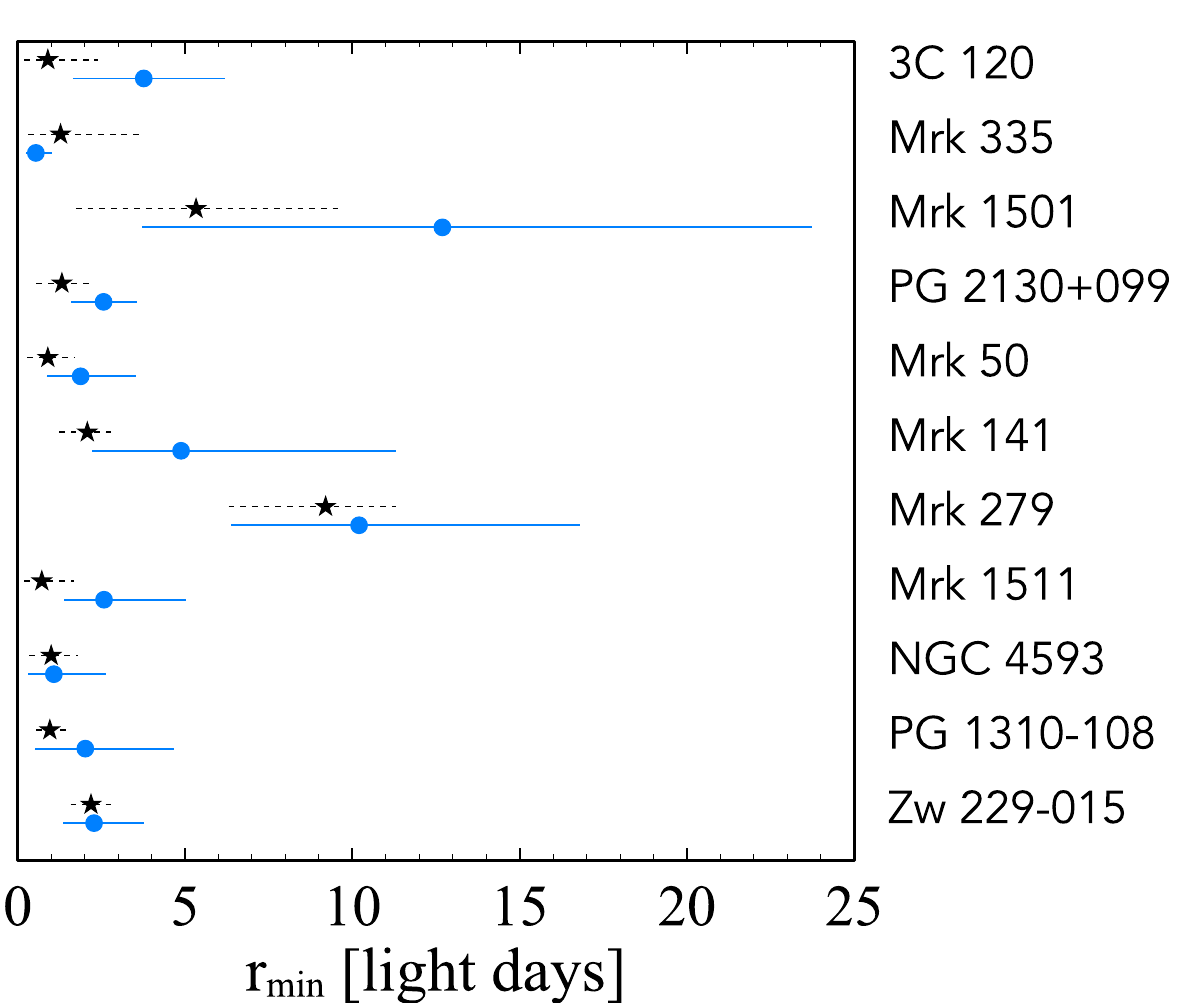}
\includegraphics[width=0.33\textwidth]{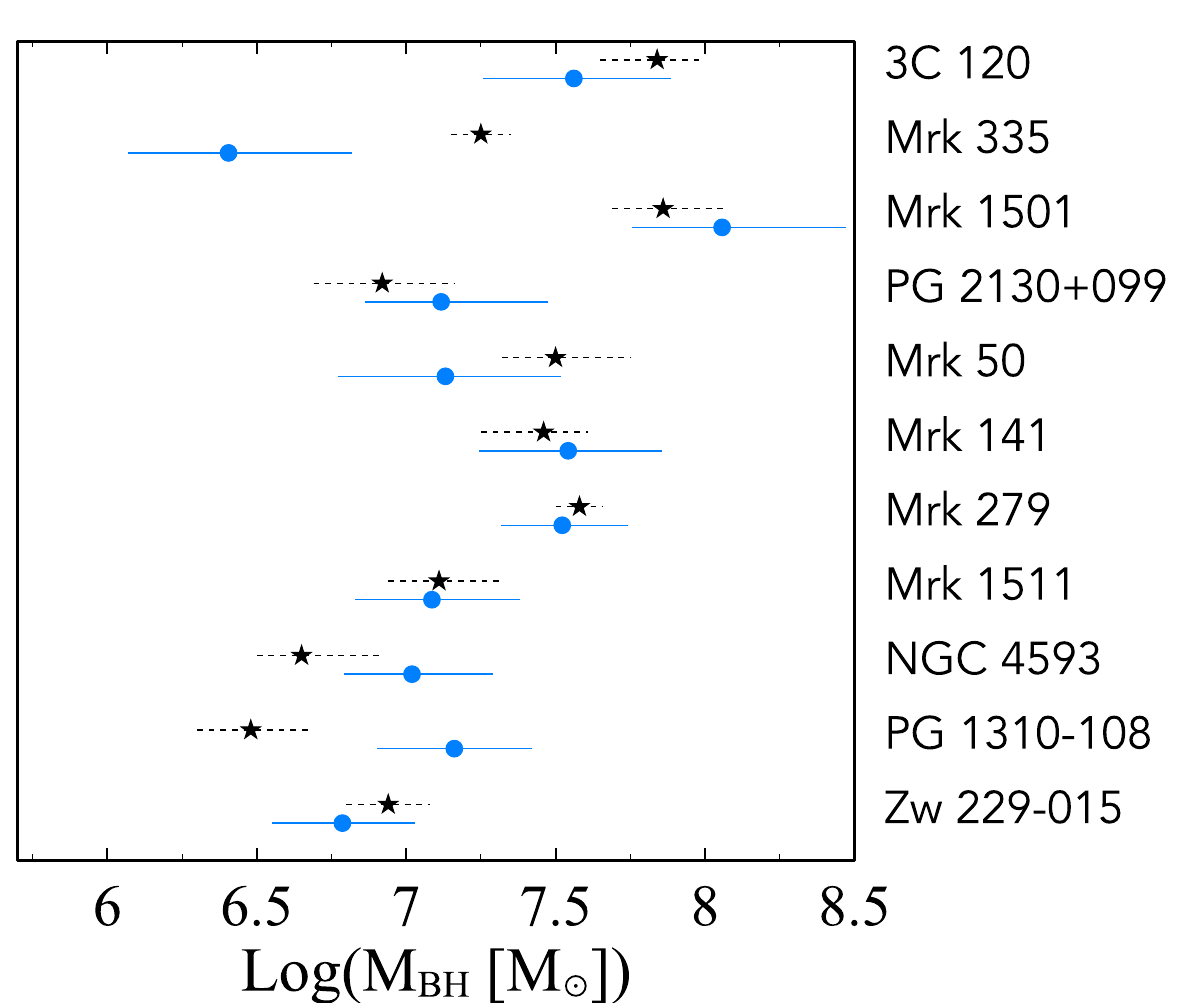}
\includegraphics[width=0.33\textwidth]{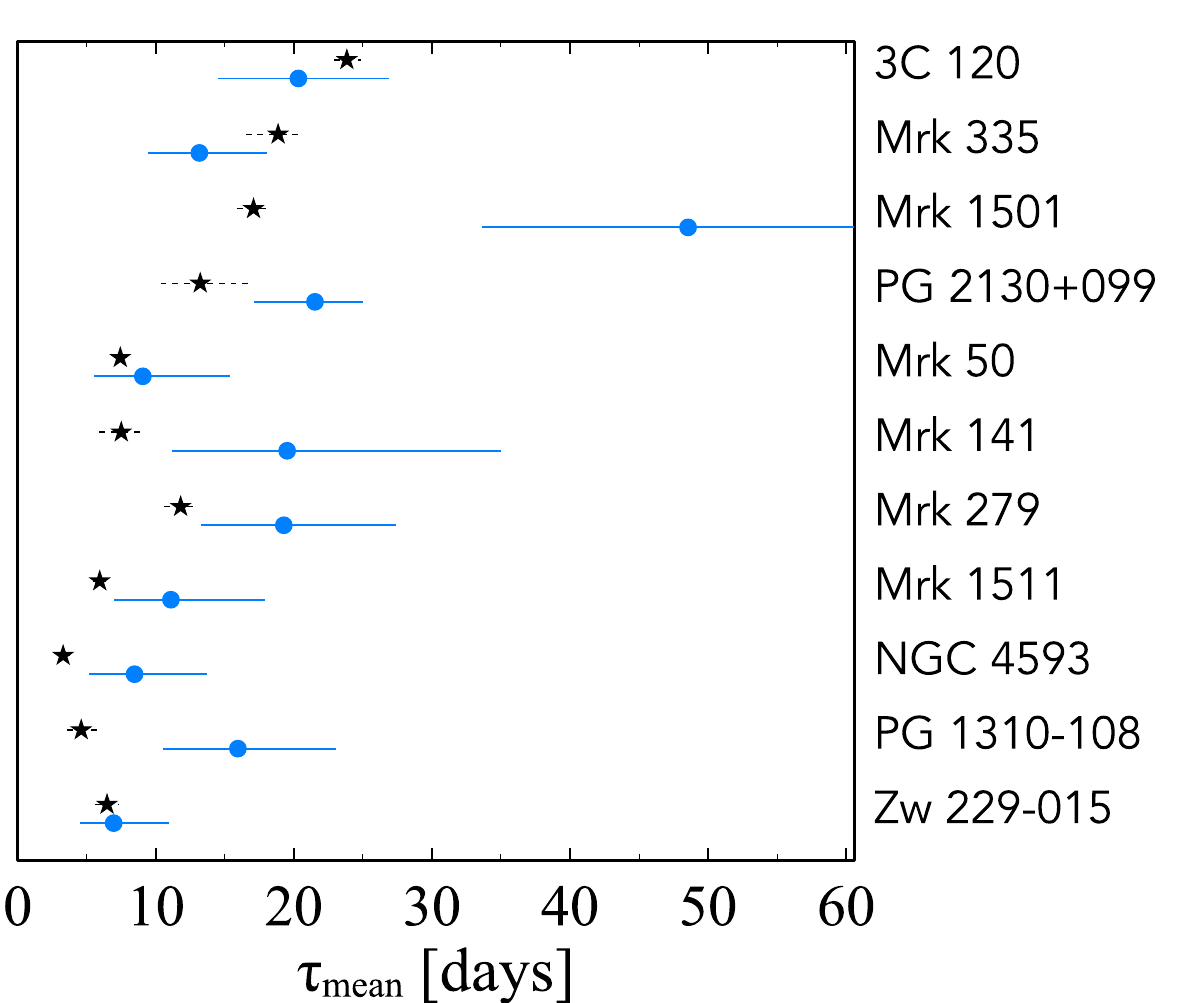}
\caption{Inferred values for the parameters and their respective 68\% confidence regions for each of the objects analysed. The first 4 objects from the top are part of the AGN10 monitoring campaign while the remaining 7 objects are from LAMP 2011. The filled blue circles are the inferred values using the single-epoch analysis while the black star symbols are the inferred values found by previous work using the full light-curve for the modelling of the BLR (\citealt{grier17} for AGN10 and \citealt{williams18} for LAMP 2011). The values inferred by \citealt{williams18} refer to their combined posterior probability distribution. We consider that a parameter cannot be constrained when the 68\% confidence range covers more than 50\% of the parameter space. Scale in the y-axis is arbitrary for visualisation purposes.}
\label{inferred_parameters}
\end{figure*}

\subsection{Notes on individual BLR parameters}
\label{sec:individual_parameters}
For future applications of our model, it is important to know which parameters can be reliably constrained using single-epoch spectra. In Section~\ref{sec:results} we presented the model results for each AGN, in this section we summarise and discuss our findings for each of the BLR parameters. Hereafter, when we write that the inferred value agrees with the full-lightcurve result, it means that the inferred parameter value from the single-epoch modelling agree within the uncertainties with the inferred parameter value from the full light-curve modelling. We note that we define agreement within the uncertainties if there is an overlap of the 68\% confidence region of our results and the 68\% confidence region of the full light-curve result.
\paragraph*{$\gamma$} $-$ The parameters that have very broad posterior probability distributions are considered to be unconstrained, because the model does not find a range in the parameter space that clearly has higher probability. In that case the entire parameter range explored shows comparable probability of occurring. This is notable for the parameter $\gamma$, which describes the angular distribution of particles. The parameter $\gamma$ is not constrained by the single-epoch spectra for any of the AGN analysed. We note that even the full light-curve modelling only marginally constrains this parameter, as can be seen from the wide 68\% uncertainty ranges shown as the dashed lines in Fig.~\ref{inferred_parameters}. It appears that both the single-epoch spectrum and full light-curve data do not provide enough unique information to constrain $\gamma$. We conclude that the parameter $\gamma$ cannot be constrained using single-epoch spectra, based on our results for the 11 AGN in this work and the three epochs of Arp 151 \citepalias{raimundo19}.

\paragraph*{\inc} $-$ The inclination angle (\inc) is constrained for most objects. We find that 8 AGN (3C120, Mrk 1501, PG 2130+099, Mrk 50, Mrk 141, Mrk 279, NGC 4593 and Zw 229-015) have well constrained inclination angles that agree with the full light-curve result. For 3 AGN (Mrk 335, Mrk 1511 and PG 1310-108) we consider that \inc\ cannot be constrained as the 68\% confidence range covers more than 50\% of the parameter space. There are no AGN with inferred \inc\ that are \emph{inconsistent} with the full light-curve result within the 68\% confidence range.

\paragraph*{\opening} $-$ Similar to the inclination angle, the BLR opening angle (\opening\,) is also constrained for most objects. We find that 6 AGN (3C120, Mrk 1501, PG 2130+099, Mrk 141, NGC 4593 and Zw 229-015) have well constrained opening angles that agree with the full light-curve result. For 3 AGN (Mrk 279, Mrk 1511 and PG 1310-108) the prior probability distribution is broad and \opening\ can only be marginally constrained. Still, the inferred \opening\ agrees with the full light-curve result. For 2 AGN (Mrk 335 and Mrk 50) we consider that \opening\ cannot be constrained as the 68\% confidence range covers more than 50\% of the parameter space. Mrk 1511 is a puzzling case as it shows a marginally constrained \opening\ that is not consistent within the 68\% confidence range with the full light-curve result. We note that, as for the full light-curve result, we see a correlation between \inc\ and \opening. A discussion on this correlation is presented by \citetalias{raimundo19}.

\paragraph*{$\kappa$} $-$ The parameter $\kappa$ is constrained for 4 AGN (Mrk 1501, Mrk 50, Mrk 141 and Zw 229-015) and the inferred values agree with the full light-curve result. For one AGN (NGC 4593) $\kappa$ is only marginally constrained but the 68\% confidence range from the full light-curve modelling is as broad as for the single-epoch modelling, which indicates that the full light-curve modelling is also not able to strongly constrain $\kappa$ for this AGN. For the remaining 6 AGN (3C120, Mrk 335, PG 2130+099, Mrk 279, Mrk 1511 and PG 1310-108), we consider that the single-epoch modelling is not able to constrain $\kappa$ as the 68\% confidence region covers more than 50\% of the parameter space.

\paragraph*{$\xi$} $-$ The mid-plane transparency ($\xi$) is only constrained for 2 of our AGN (Mrk 141 and Zw 229-015) with inferred values that agree with the full light-curve result. We note that these two AGN have the most asymmetric line shapes which may be one of the reasons why $\xi$ (and $\kappa$) are constrained. For Mrk 1501, Mrk 50, Mrk 279 and NGC 4593, only marginal constraints can be obtained, still, the inferred values agree with those found from the full light-curve result. We note that for NGC 4593 the 68\% confidence range from the full light-curve result is also wide and comparable with the uncertainties found in the single-epoch modelling. For the remaining 5 AGN (3C120, Mrk 335, PG 2130+099, Mrk 1511 and PG 1310-108) $\xi$ cannot be constrained based on the single-epoch spectrum.

\paragraph*{\fellip} $-$ The fraction of near-circular elliptical orbits can be constrained for two of the AGN in our sample (Mrk 141 and Zw 229-015), with inferred values that agree with the full light-curve result. These two AGN are those that show the strongest asymmetry and structure in the line profile. For Mrk 50 \fellip\ can only be marginally constrained but the value agrees with that of the full light-curve modelling. Due to their broad posterior probability distributions we consider that \fellip\ cannot be constrained for 8 AGN: 3C120, Mrk 335, Mrk 1501, PG 2130+099, Mrk 279, Mrk 1511, NGC 4593 and PG 1310-108. 

\paragraph*{\fflow} $-$ The parameter \fflow\ is binary and we consider that it is constrained for an AGN if the inferred parameter and 68\% confidence range indicate \fflow\ values that are exclusively $<0.5$ or $>0.5$. For 6 AGN \fflow\ can be constrained: Mrk 1501, Mrk 50, Mrk 141, Mrk 279, NGC 4593 and Zw 229-015. For the remaining 5 AGN \fflow\ cannot be constrained: 3C120, Mrk 335, PG 2130+099, Mrk 1511 and PG 1310-108. We can see that \fflow\ is constrained in some AGN for which \fellip\ is only marginally constrained or not constrained at all. This indicates that the model is able to detect if there are signatures of particles with inflowing or outflowing trajectories, even though the exact fraction of these particles (1$-$\fellip) is difficult to constrain accurately for most AGN in the sample.

\paragraph*{\thetae} $-$ The parameter \thetae\ is constrained for 5 AGN: Mrk 50, Mrk 141, Mrk 279, NGC 4395 and Zw 229-015 with values that agree with the full light-curve result, and not constrained for 6 AGN: 3C120, Mrk 335, Mrk 1501, PG 2130+099, Mrk 1511 and PG 1310-108. The parameter \thetae\ is only constrained for AGN where \fflow\ is also constrained. This is not surprising since these two parameters characterise the same type of gas orbits.

\paragraph*{\rmin} $-$ We find that the minimum radius of the BLR is constrained for all AGN, and with values that agree with the full light-curve result. Even though for one AGN (Mrk 1501) \rmin\ has a broad posterior distribution, we consider the parameter to be constrained since the prior for \rmin\ is relatively wide. The parameter \rmin\ is defined from $F$ and $\mu$ in the Gamma radial distribution of particles, and typically has a prior from 10$^{-3} - 10^{2}$ light-days.

\paragraph*{\Mbh} $-$ The black hole mass is the parameter that is most correlated with \taumean. Therefore, the approach of the single-epoch modelling to define a prior probability distribution on \taumean\ will necessarily influence \Mbh. This has also been observed by \citetalias{raimundo19}. There is some information in the emission line to constrain \Mbh, such as the relativistic effects caused by the black hole. However, since the \taumean\ prior is defined a priori we do not consider the inferred \Mbh\ to be independently reliable as \taumean\ necessarily favours a specific absolute size of the BLR and therefore a specific \Mbh\ range. As shown by \citetalias{raimundo19}, for the three epochs of Arp 151 analysed, the model infers \Mbh\ values that agree with the full light-curve result. The only exception is for the case where the \taumean\ inferred from the R$-$L relation is significantly different from the intrinsic \taumean\ and the assumed \taumean\ prior width is narrow, which causes the \Mbh\ value to differ from the intrinsic value. Here in this work we assume a \taumean\ prior width of 0.2 dex to mimic the scatter in the R$-$L relation. Since this width is relatively narrow, we expect situations where the inferred \Mbh\ value differs from the intrinsic value if the R$-$L estimate is significantly under- or overestimated. We find that for 9 out of 11 AGN the model finds \Mbh\ inferred values that agree within the uncertainties with the full light-curve result. For 2 AGN, Mrk 335 and PG 1310-108, the inferred \Mbh\ value and its 68\% confidence range are not consistent with the full light-curve result. For PG 1310-108 this could be due to the prior central value of \taumean\ being overestimated with respect to the full light-curve result and the line profile not having enough information to infer \Mbh\ correctly. For Mrk 335 the case appears to be different: the prior value for \taumean\ is similar to what was found from the full light-curve analysis, however, \taumedian\ and \rmedian\ are significantly different from the full light-curve result (Table~\ref{table_results}). This indicates that the time delay distribution that the model is finding for the BLR is complex, such that \taumean\ and \taumedian\ differ significantly. The parameter \taumean\ alone may not be a good probe of the size of the BLR for Mrk 335.

\begin{figure*}
\centering
\includegraphics[width=0.8\textwidth]{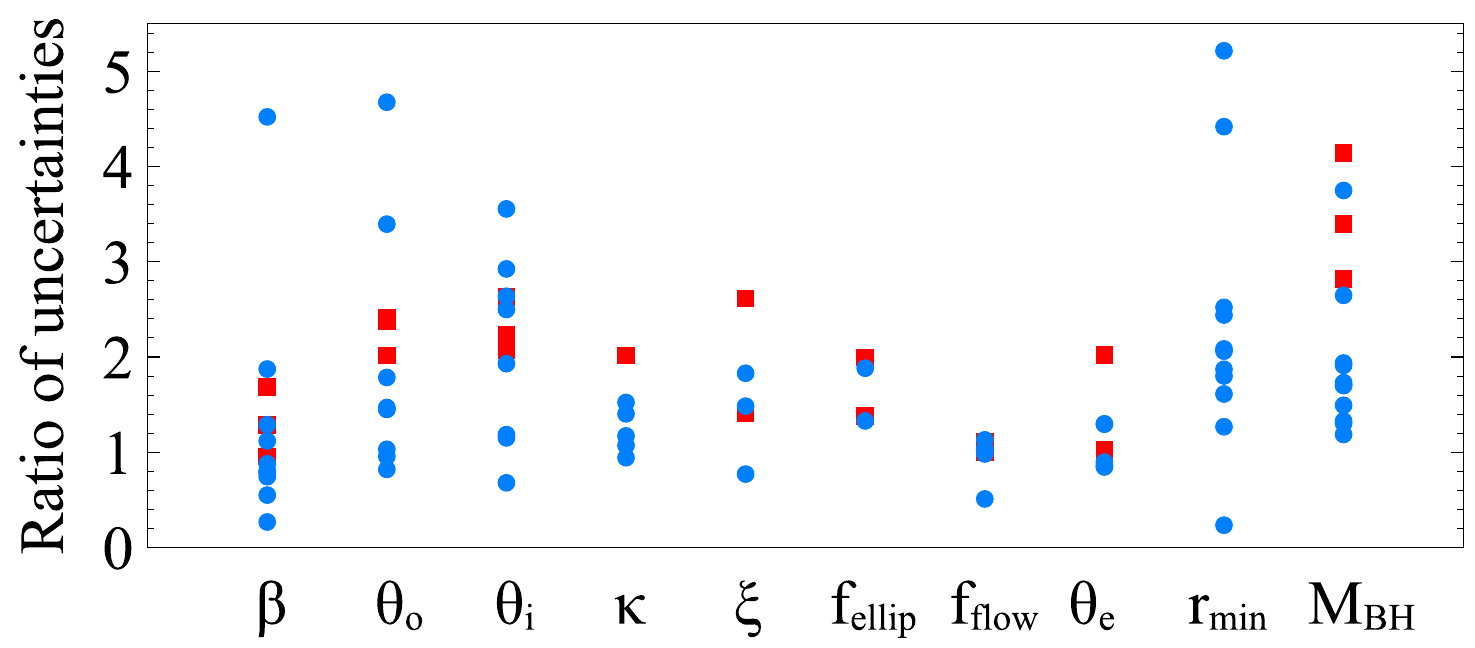}
\caption{Ratio of uncertainties between the single-epoch modelling method and the monitoring data modelling for each parameter. The uncertainties are defined as the 68\% confidence regions and the ratio calculated as the uncertainties of the single-epoch modelling divided by the uncertainties of the monitoring data modelling. The blue circles represent the results for the AGN in the AGN10 and LAMP2011 samples while the red squares represent the results for the three epochs of Arp 151 analysed by \citetalias{raimundo19}. The results are shown for the cases where the parameters are constrained or marginally constrained. Unconstrained parameters are not shown.}
\label{uncertainty_ratio}
\end{figure*}

\paragraph*{$\beta$} $-$ This parameter determines the shape of the radial distribution of particles in the BLR. It can be understood as the standard deviation of the Gamma distribution at the origin ($r = 0$). \citetalias{raimundo19} found that the inferred $\beta$ value depends on the epoch. Here we find something similar: for 3 AGN (3C120, Mrk 335 and Mrk 1511) $\beta$ is constrained but does not agree within the uncertainties with the full light-curve result. This indicates that either $\beta$ cannot be accurately constrained from the model possibly due to the lack of timing information, or that $\beta$ depends on the epoch analysed. \citetalias{raimundo19} suggested that this could be due to each epoch corresponding to a specific section of the BLR that is illuminated (or emitting). Even though for 8 of the AGN in our sample, the inferred $\beta$ value agrees with the full light-curve result, we cannot rely on the $\beta$ value found from the model, as it may not reflect the average radial distribution of particles in the BLR. 
Therefore, the $\beta$ value obtained is the `instantaneous' value for the particular observed epoch and is not a general property of the BLR. This is particularly important to keep in mind when modelling AGN without monitoring data. We note that in the full light-curve modelling of three distinct monitoring campaigns of Arp 151, \cite{pancoast18} found a difference in the values of $\beta$ determined for each campaign. This is in line with the hypothesis above, that $\beta$ may not be constant for a BLR but an average value that depends on the duration and/or timing of the monitoring campaign.

\paragraph*{\taumean, \taumedian, \rmean, \rmedian, $\rm r_{\rm out}$} $-$ The four first parameters of this list are not free parameters in the model, but calculated from the BLR structure. The parameter \taumean\ is defined with a prior probability distribution and \taumedian, \rmean\ and \rmedian\ will naturally be correlated with \taumean, as they all probe the size of the BLR. The mean and median estimates tend to either both agree or both disagree with the full light-curve result for each AGN. The only clear exception is Mrk 335 (see discussion on \Mbh\ above). The parameter $\rm r_{\rm out}$ is defined a priori for each AGN and is related to the duration of the simulated continuum light curve. We find that for all our sources $\rm r_{\rm out}$/\rmin\ is typically larger than 10, which according to models by \cite{robinson95} is necessary to reproduce most observed line profiles.

\paragraph*{Continuum light curves} $-$ We find a broad range of continuum light curves that can give rise to the observed line profiles. Among the solutions there are simulated light curves that resemble the real continuum light curves for the AGN, but in general they are different and span a broad range of shapes. This variety of light curves is due to the fact that we have two free parameters C$_{\rm add}$ and C$_{\rm mult}$ that normalise the absolute continuum and line fluxes and also because we only have a prior on the mean time delay. Since we do not have a strong constraint on all the time delays between the continuum and the line emission from each portion of the BLR, the model is free to select slightly different portions of the continuum light curve for each solution.

\subsection{Summary of results on the full sample}
\label{sec:full_sample}
Our work on the single-epoch spectral modelling of the BLR parameters now includes results on 12 AGN: the sample of 11 from this work and Arp 151 analysed by \citetalias{raimundo19}.

Based on the single-epoch analysis only, we find that the typical BLR for the sample is a low to intermediate inclination (\inc\ $<$ 50$-$60 degrees) and relatively thick (\opening $>$ 15 degrees) disc-like structure, with the majority of sources (7 out of 12) showing a significant fraction of orbits with velocities close to the inflowing or outflowing velocity distribution. These general properties agree with what was found from the modelling of monitoring data.

The inclination and opening angles are constrained for the majority of the AGN in our sample. The parameter \fflow\ is also constrained for the majority of the sample, and \thetae\ tends to only be constrained for AGN for which \fflow\ is determined. This is somewhat expected as \fflow\ identifies the presence of inflowing/outflowing type orbits, and \thetae\ determines the angle between this type of orbit and the circular orbit value, controlling the fraction of bound orbits. The parameter \fellip\ is constrained for a smaller number of AGN than \fflow. This means that the model is not always able to constrain the exact fraction of near-circular versus inflowing/outflowing orbital configurations, but is often able to identify \emph{the presence} of inflowing/outflowing orbits.

We find that in general, the AGN with the lower Eddington ratios (i.e. lower mass accretion rates scaled by their black hole mass): Mrk 50, Mrk 141, NGC 4593 and Zw 229-015 tend to have a larger number of BLR parameters constrained than the AGN with the highest Eddington ratios: 3C 120, Mrk 335, Mrk 1501, PG 2130+099, Mrk 279 and PG 1310-108. The only exception is Mrk 1511 for which even though the Eddington ratio is low, only one parameter is constrained. This could be due to other causes that make it difficult for the model to constrain the BLR parameters, such as a residual Fe II contribution to the line profile, or to the overestimated R$_{\rm BLR}$ from the R$-$L relation, as outlined in Section~\ref{sec:mrk1511}.
The overall trend with Eddington ratio may be related to the different line profiles that high and low Eddington ratio AGN have. From Fig.~\ref{epoch_spec} we can see that the high Eddington ratio AGN tend to have smoother and more symmetric line profiles. Symmetric line profiles may not contain enough information to independently constrain all the parameters that in some BLR configurations can cause asymmetries, such as $\kappa$, $\xi$ and \fflow\ and its associated \fellip\ and \thetae.

The uncertainties of the single-epoch model constrained parameters are typically higher than those found from the full light-curve modelling. This can be seen from the wider 68\% confidence ranges associated with the parameters in the single-epoch modelling. Fig.~\ref{uncertainty_ratio} shows a graphical comparison of the single-epoch versus full light curve uncertainties for each of the constrained or marginally constrained parameters. The y-axis is the ratio between the single-epoch 68\% confidence range and the full light-curve 68\% confidence range. The blue circles are the results for the AGN in this paper while the red squares are the results for the three epochs of Arp 151 from \citetalias{raimundo19}. The ratio of uncertainties goes up to $\sim$5 but in general is lower than $\sim$3. The extra information that we add, on the form of the prior probability distribution on \taumean\ also contributes to lower the ratio of uncertainties. We expect in general to observe uncertainty ratios higher than 1 when comparing the single epoch with the full light curve results. For a few of the AGN and some of the parameters we see an uncertainty ratio lower than 1, which could indicate that the single-epoch model is inferring parameter values with lower uncertainties than the modelling of monitoring data. This is due to the particular comparison that we are making. The AGN that have uncertainty ratios lower than 1 are typically from the LAMP 2011 sample (and mostly NGC 4593). The uncertainties for the full light-curve modelling are in this case those of the combined posterior in \cite{williams18}, which includes the results using three different spectral decompositions. For some AGN this results in wider uncertainties than when using a single spectral decomposition as we do in our single-epoch study. The object NGC 4593 in particular shows differences in the posterior probability distributions (red and blue histograms of Fig.~\ref{posterior_ngc4593}) for the parameters \inc, $\xi$ and \fflow, for which we measure uncertainty ratios lower than one in Fig.~\ref{uncertainty_ratio}. A low uncertainty ratio is also observed for $\beta$, likely due to the fact that the $\beta$ we measure corresponds to a single-epoch and not to the average $\beta$ of the monitoring campaign.

\begin{figure*}
\centering
\includegraphics[width=1.0\textwidth]{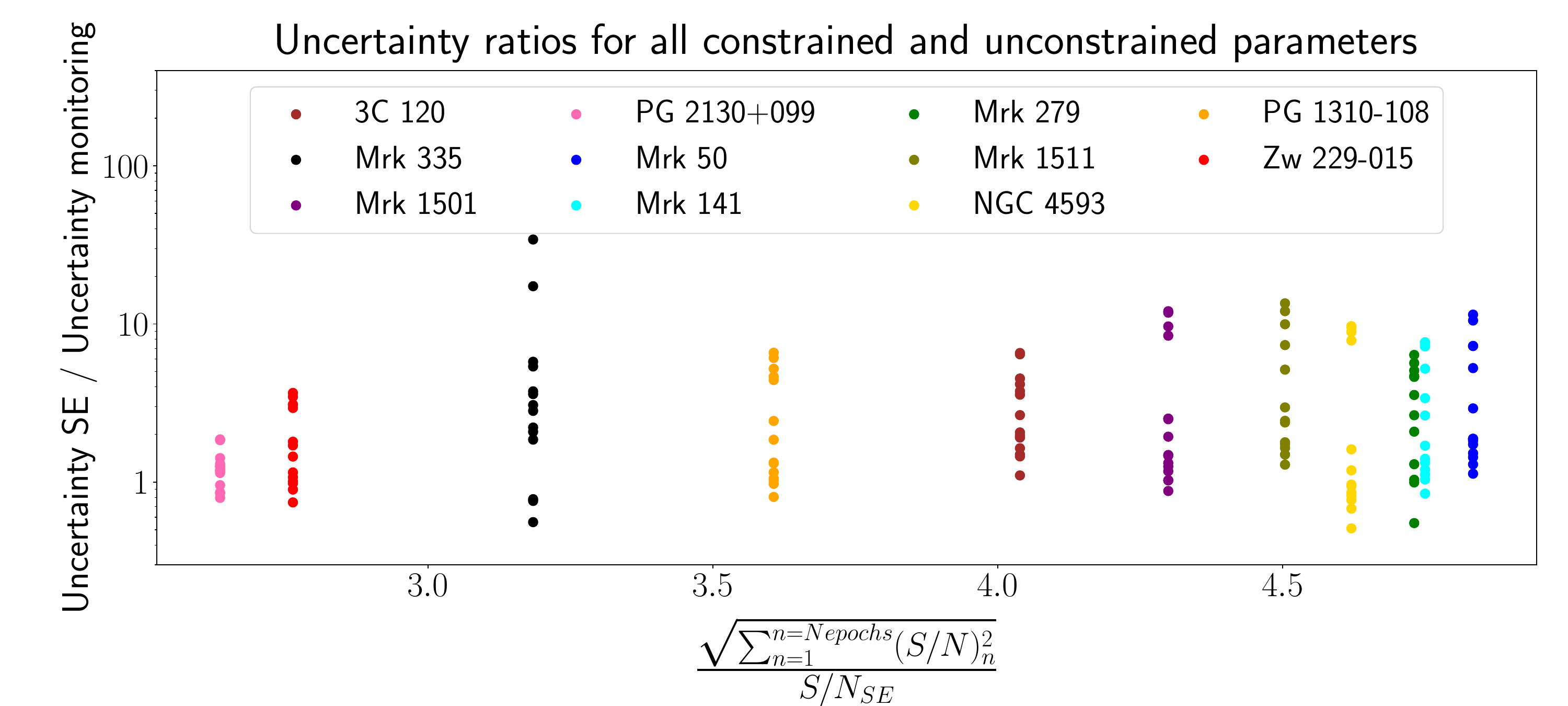}\\
\caption{Ratio of uncertainties between the single-epoch (SE) and the monitoring inferred parameters as a function of the statistical S/N for all epochs divided by the S/N of the single-epoch spectrum. The filled circles represent the uncertainty ratios for the inferred parameters colour coded as a function of the AGN. Each vertical line of filled circles of the same colour corresponds to the inferred parameters for a single AGN in the sample. All constrained and unconstrained parameters are shown.}
\label{uncertainties_nepochs_SN_scaled}
\end{figure*}

The ratio of uncertainties between the single-epoch and the full light-curve results is not uniquely determined by the relative loss of S/N that results from using one single spectrum instead of a set of spectra from a monitoring campaign. To illustrate this, we show the ratio of uncertainties as a function of the statistical S/N for the monitoring campaign, calculated as the square root of the square sum of the mean S/N values for each epoch, and normalised to the S/N of the single-epoch spectrum (Fig.~\ref{uncertainties_nepochs_SN_scaled}). There is no strong trend between the variables shown in Fig.~\ref{uncertainties_nepochs_SN_scaled}. We find that the constraints we obtain for the parameters are determined by the information contained in the line profile, and therefore depends on the BLR characteristics as seen at that particular epoch. Using the spectral monitoring data adds information by modelling line profiles of different shapes and probing the BLR at different epochs. The tighter parameter constraints obtained by the full light-curve modelling are due to this effect and not simply due to an increased S/N obtained after combining multiple but similar line profiles.

\begin{figure*}
\centering
\includegraphics[width=0.32\textwidth]{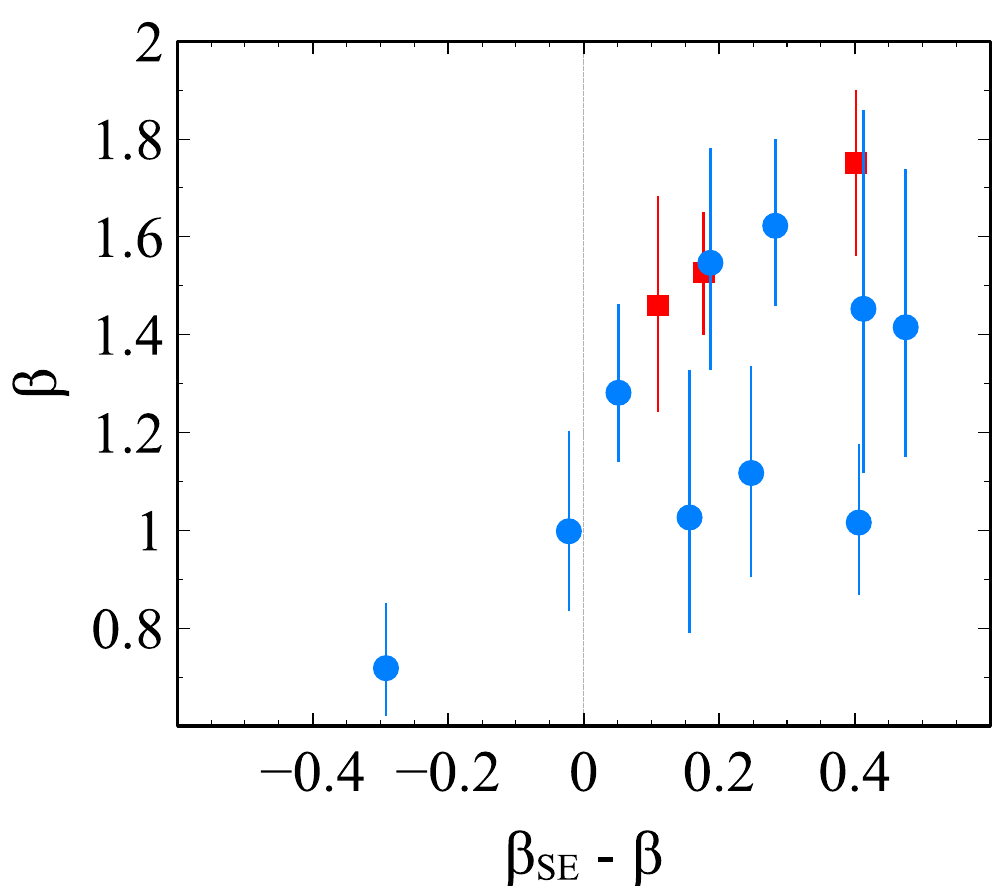}\hspace{0.25cm}
\includegraphics[width=0.32\textwidth]{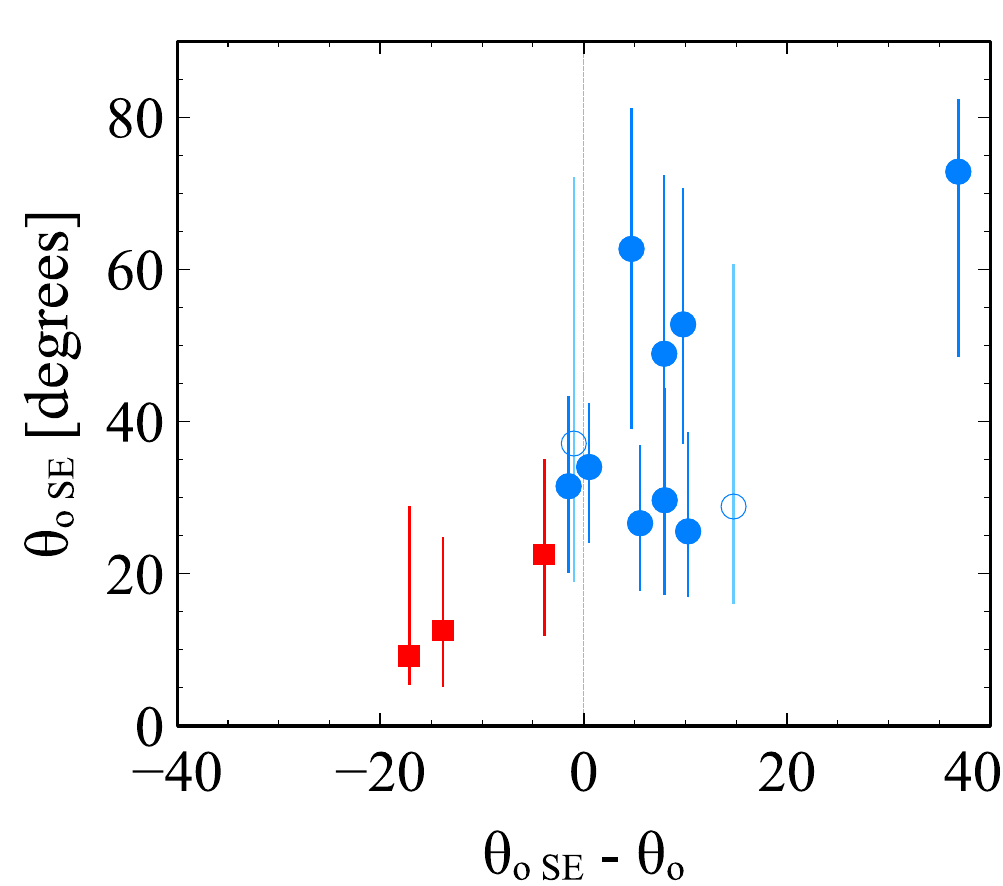}\hspace{0.25cm}
\includegraphics[width=0.32\textwidth]{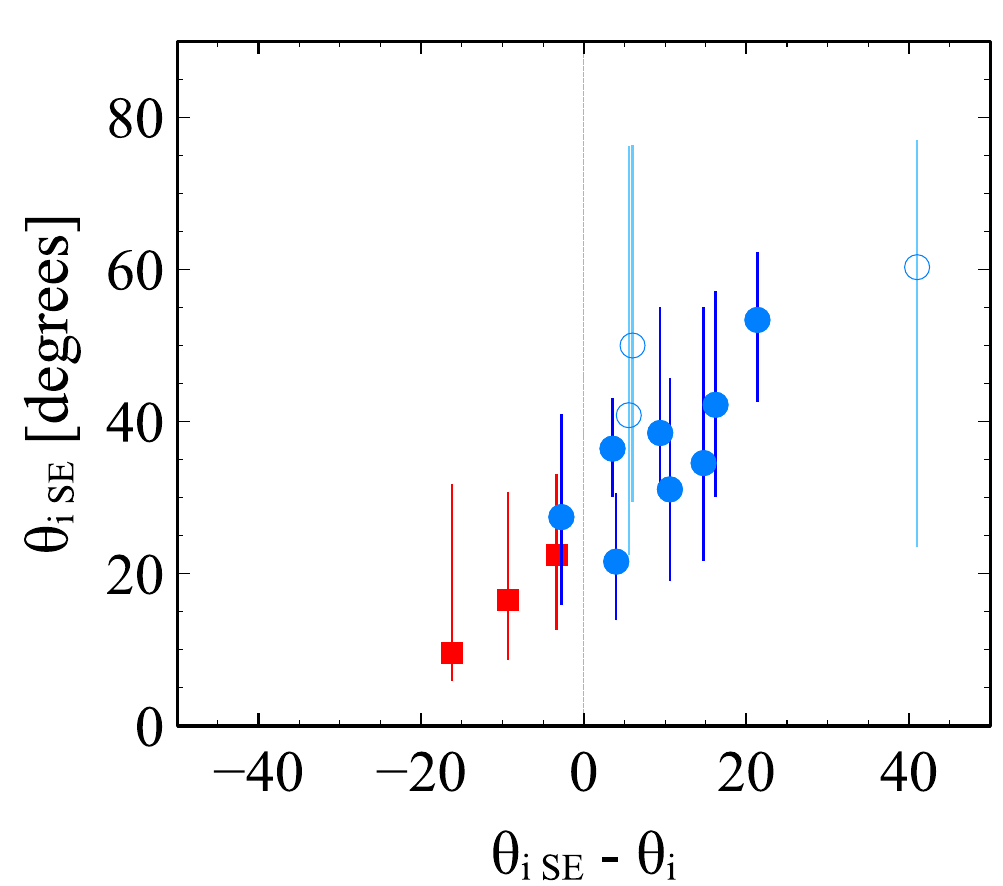}\\[0.1cm]
\includegraphics[width=0.32\textwidth]{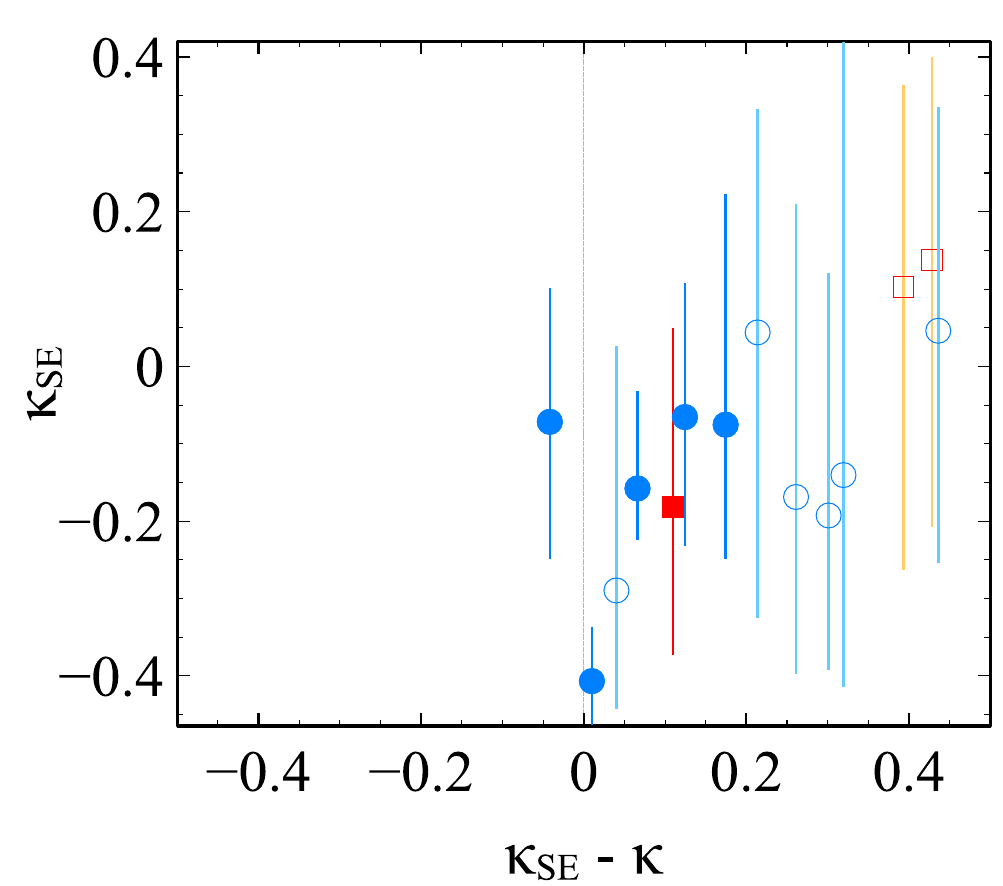}\hspace{0.25cm}
\includegraphics[width=0.32\textwidth]{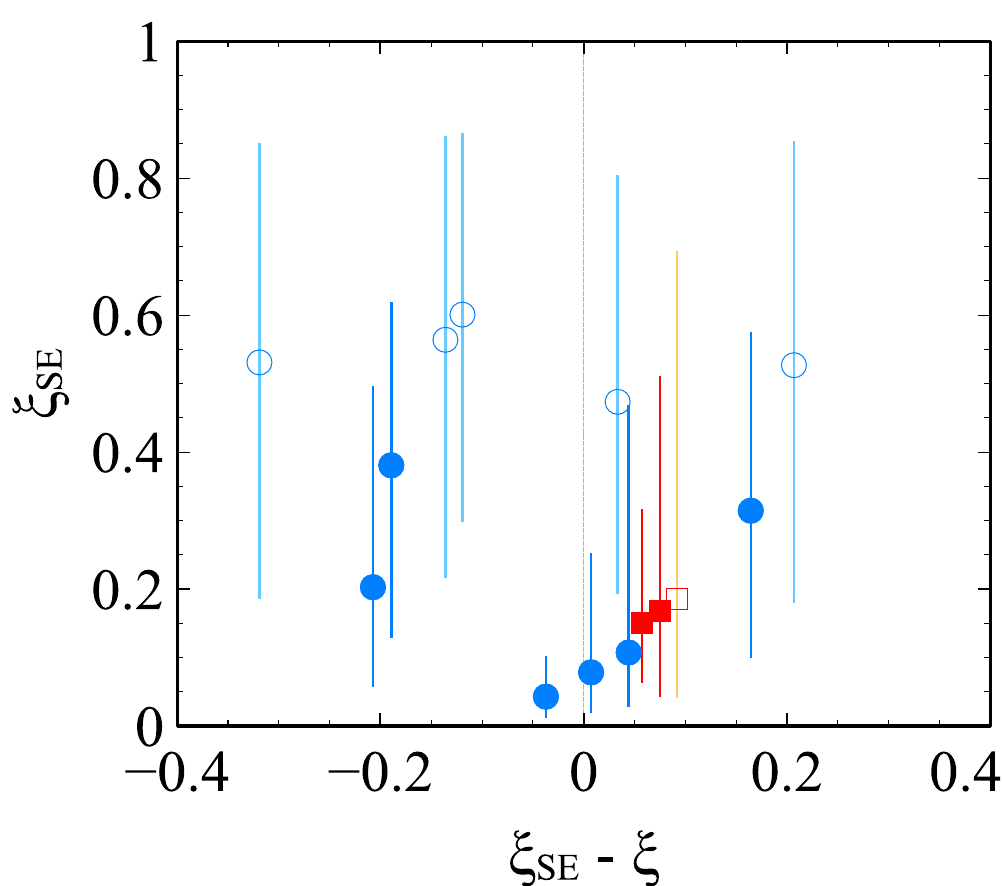}\hspace{0.25cm}
\includegraphics[width=0.32\textwidth]{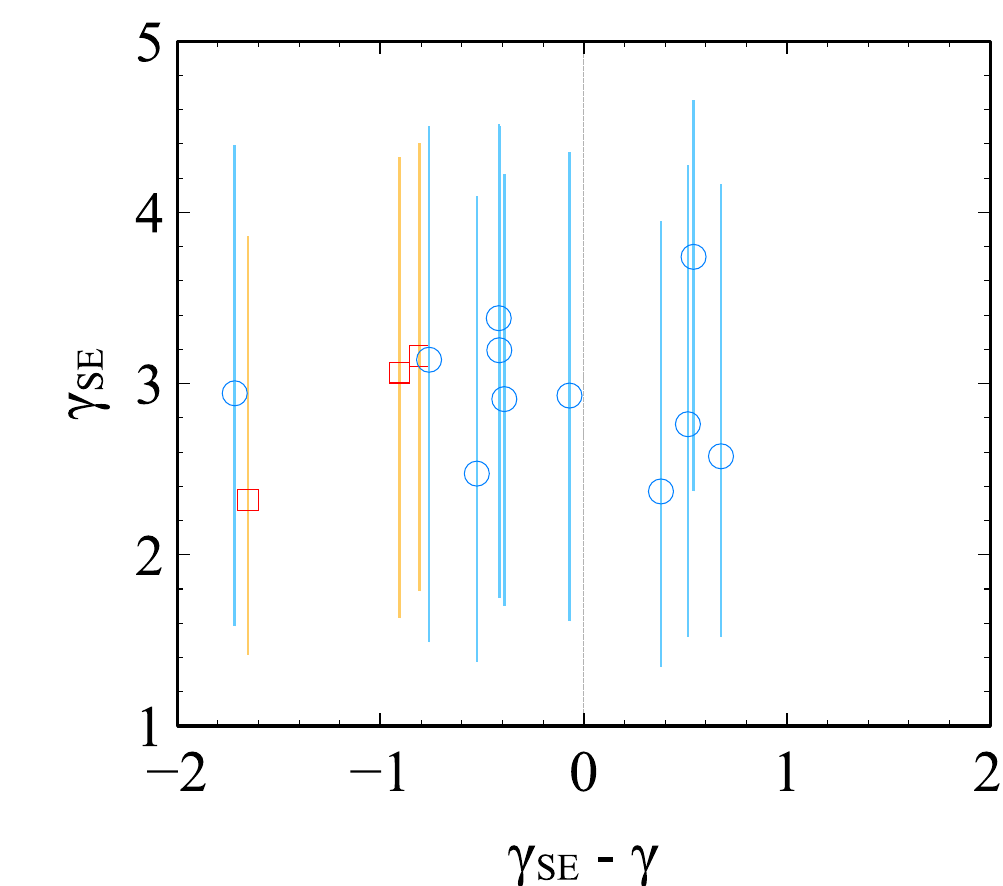}\\[0.1cm]
\includegraphics[width=0.32\textwidth]{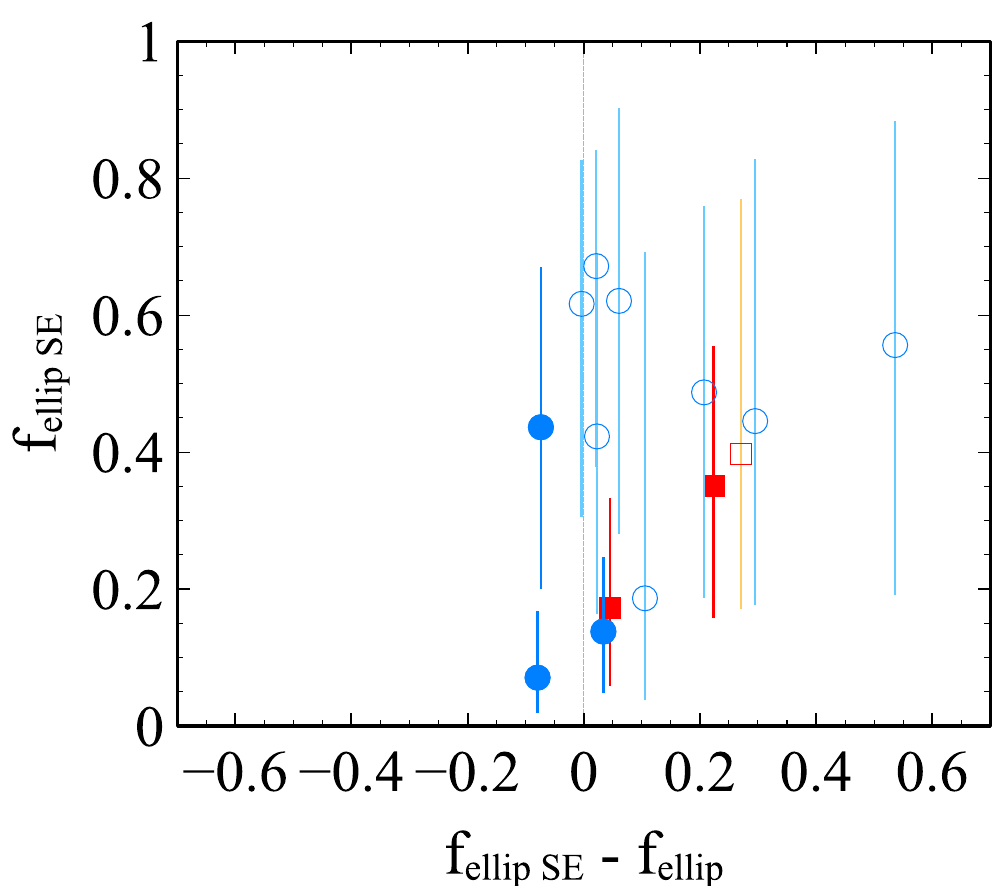}\hspace{0.25cm}
\includegraphics[width=0.32\textwidth]{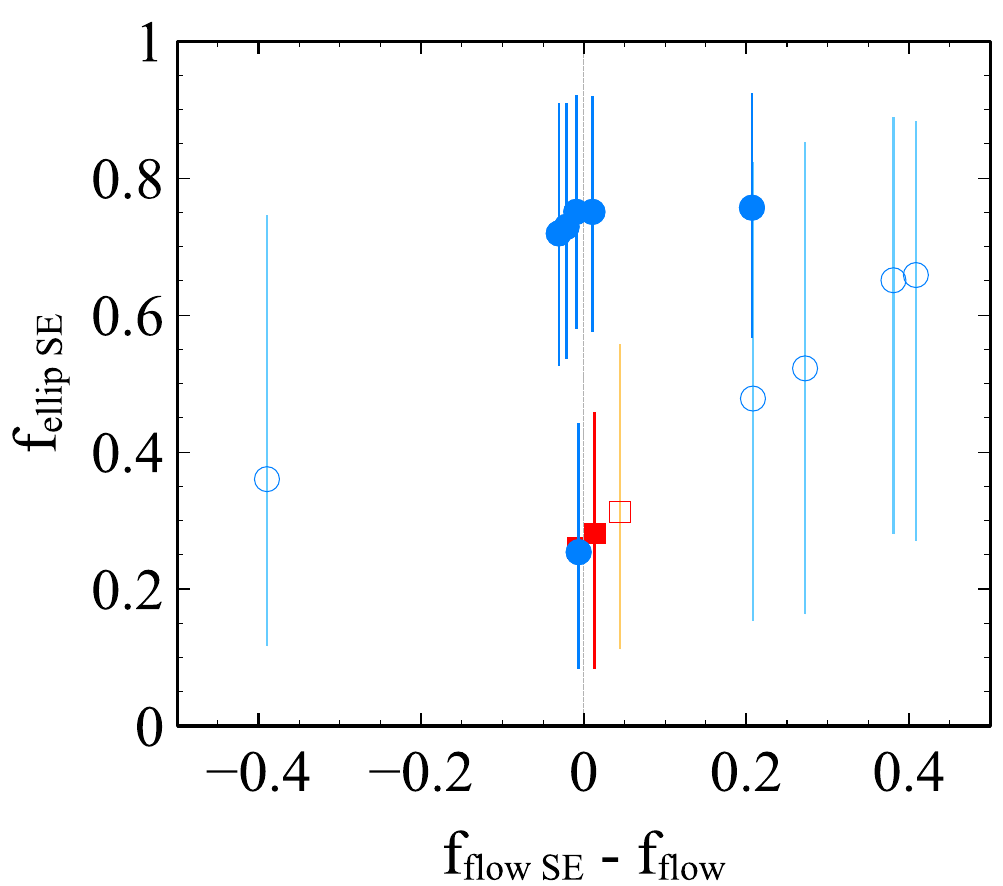}\hspace{0.25cm}
\includegraphics[width=0.32\textwidth]{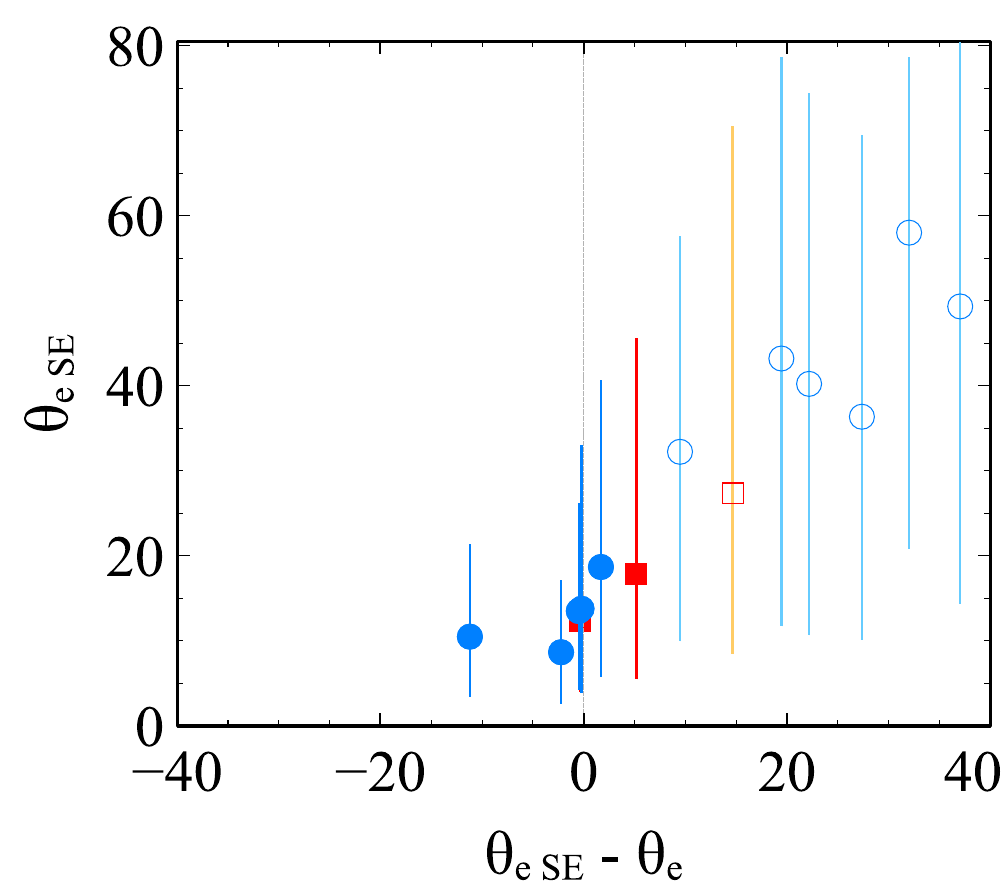}\\[0.1cm]
\includegraphics[width=0.32\textwidth]{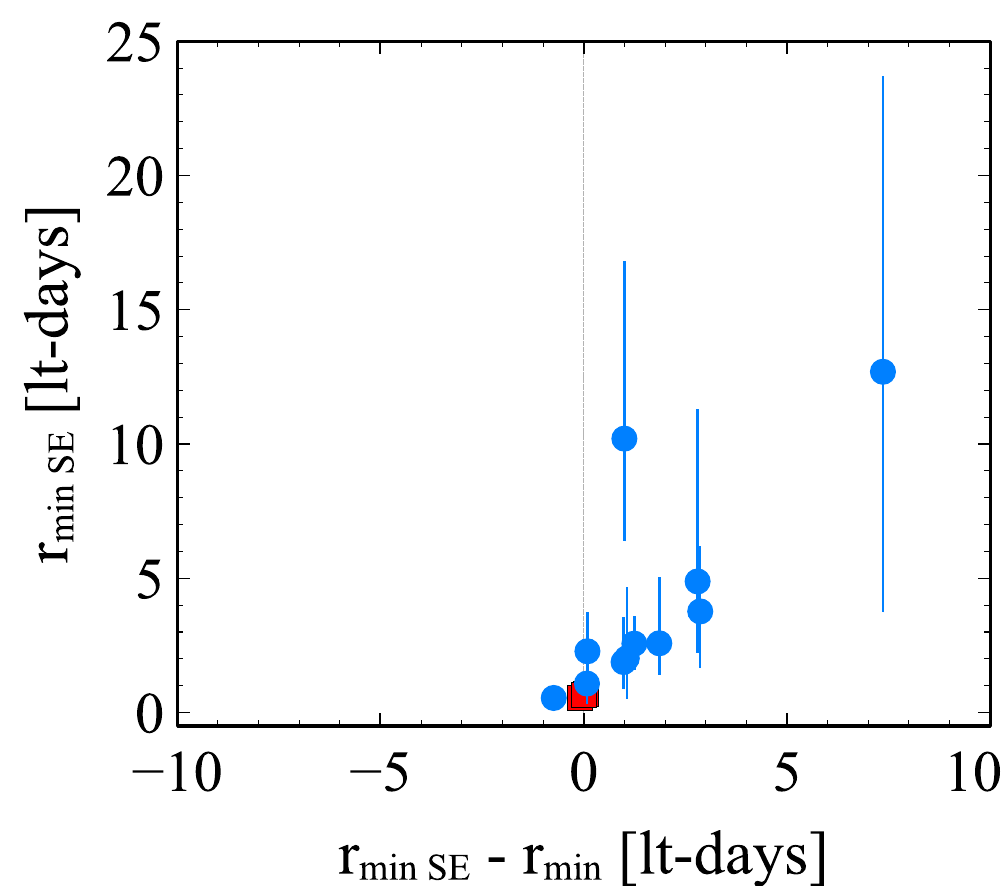}\hspace{0.25cm}
\includegraphics[width=0.32\textwidth]{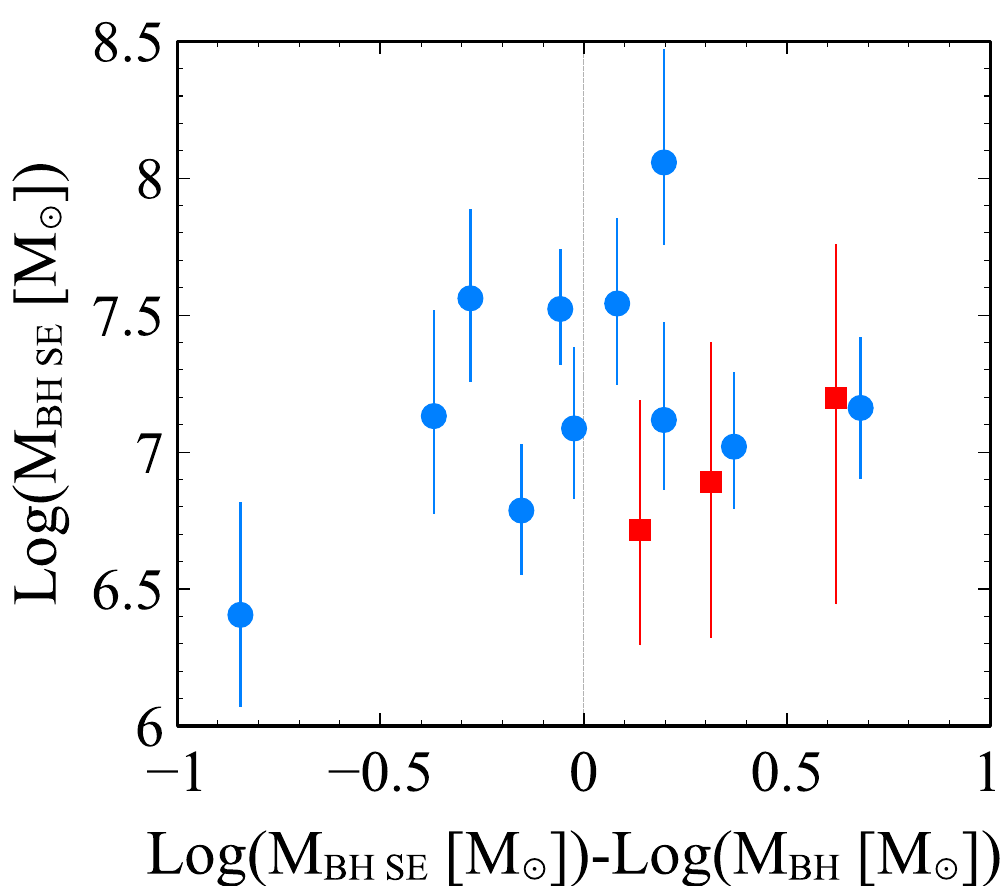}\hspace{0.25cm}
\includegraphics[width=0.32\textwidth]{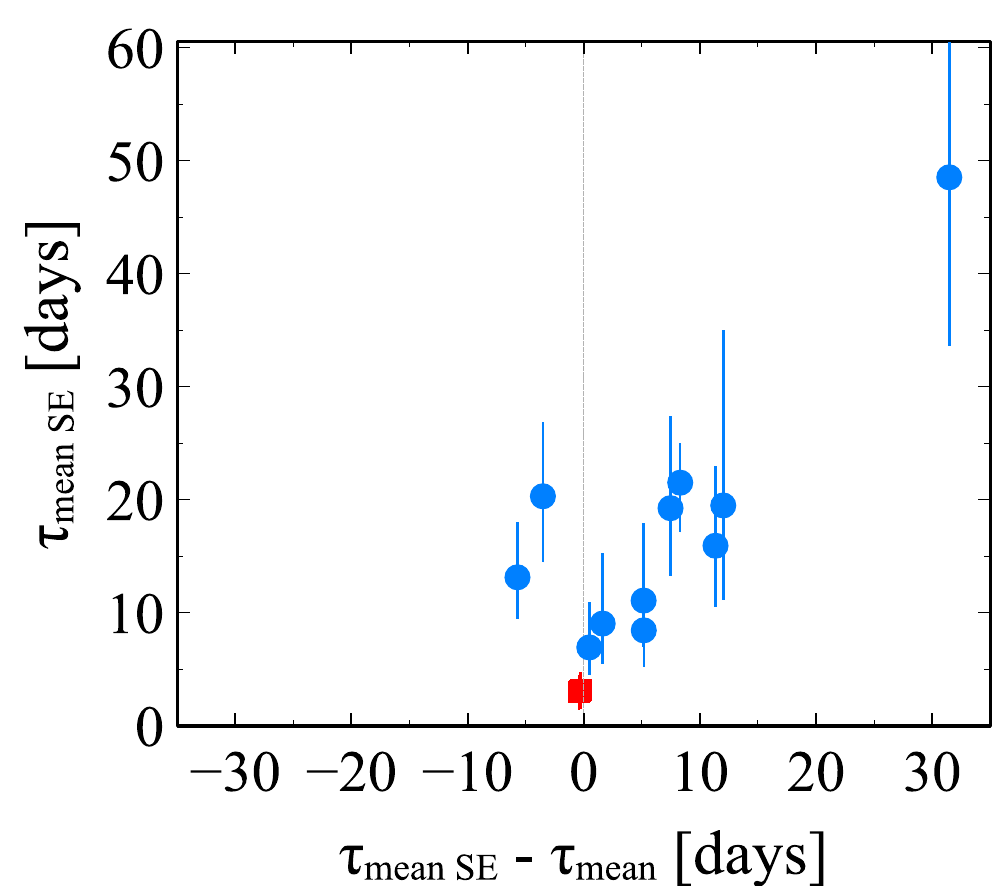}\\[0.1cm]
\caption{Difference between the single epoch (SE) and the full light-curve inferred parameter values. The y-axis indicates the since epoch modelling inferred parameter value and its respective 68\% confidence range. The x-axis indicates the difference between the single-epoch inferred value and the full lightcurve result. The filled symbols are the constrained and marginally constrained parameters while the open symbols are the unconstrained parameters. The blue circles represent the AGN10 and LAMP2011 samples while the red squares are the 3 epochs of Arp 151 from \citetalias{raimundo19}.}
\label{parameter_scatter}
\end{figure*}

With the sample of 12 AGN we investigate the scatter in the values of the parameters constrained by the single-epoch model. Fig.~\ref{parameter_scatter} shows the parameter value determined from the single-epoch (SE) model, as a function of the difference between the single-epoch inferred value and the full-lightcurve inferred value. The filled symbols are the constrained and marginally constrained parameters while the open symbols are the unconstrained parameters. The blue circles represent the AGN10 and LAMP2011 samples while the red squares are the 3 epochs of Arp 151 from the work of \citetalias{raimundo19}. For the constrained parameters there does not seem to be a significant bias in terms of under- or overestimated values, with the exception of $\beta$. We note that in the x-axis in Fig.~\ref{parameter_scatter} we are showing the difference in the median values of the posterior probability distribution. The uncertainties in the median values (as shown in the y-axis) should be taken into account in the analysis of the scatter diagrams. The parameter $\beta$ tends to be overestimated with respect to the full light-curve result. This indicates that the single-epoch model in general prefers radial particle distributions that decay more steeply with radius than the full light-curve modelling, likely due to $\beta$ being dependent on the epoch. One of the possible explanations is a `breathing BLR' (\citealt{netzer&maoz90}, \citealt{korista&goad04}, \citealt{cackett&horne06}), where the responsivity of the emitted line varies as a function of the radius in the BLR and with the continuum flux history. The parameter $\beta$ could be tracing the BLR zone that is responding at a specific epoch (e.g. \citealt{baldwin95}). The parameter \rmin\ also tends to be overestimated. This can be understood based on the relation between \taumean, \rmin\ and $\beta$. An increase in the mean time delay of the BLR can be represented by a steeper radial distribution of particles (higher $\beta$) and a larger \rmin. The mean time delay will also be a natural upper limit for \rmin. When the single-epoch assumed prior for \taumean\ is close to the value inferred from the full light-curve analysis (as for the case of Arp 151 shown with the red squares), the model is able to find an \rmin\ which is similar to the full light-curve result and is tightly constrained. For the cases where \taumean\ is overestimated in comparison with the full light-curve result, \rmin\ tends to also be overestimated but the confidence regions associated with the inferred \rmin\ are also wider showing that the model's inferred value has larger associated uncertainties. The scatter in the \taumean\ parameter is due in part to our assumed prior value from the R$-$L relation and is shown here just for reference. 

\begin{figure*}
\centering
\includegraphics[width=0.33\textwidth]{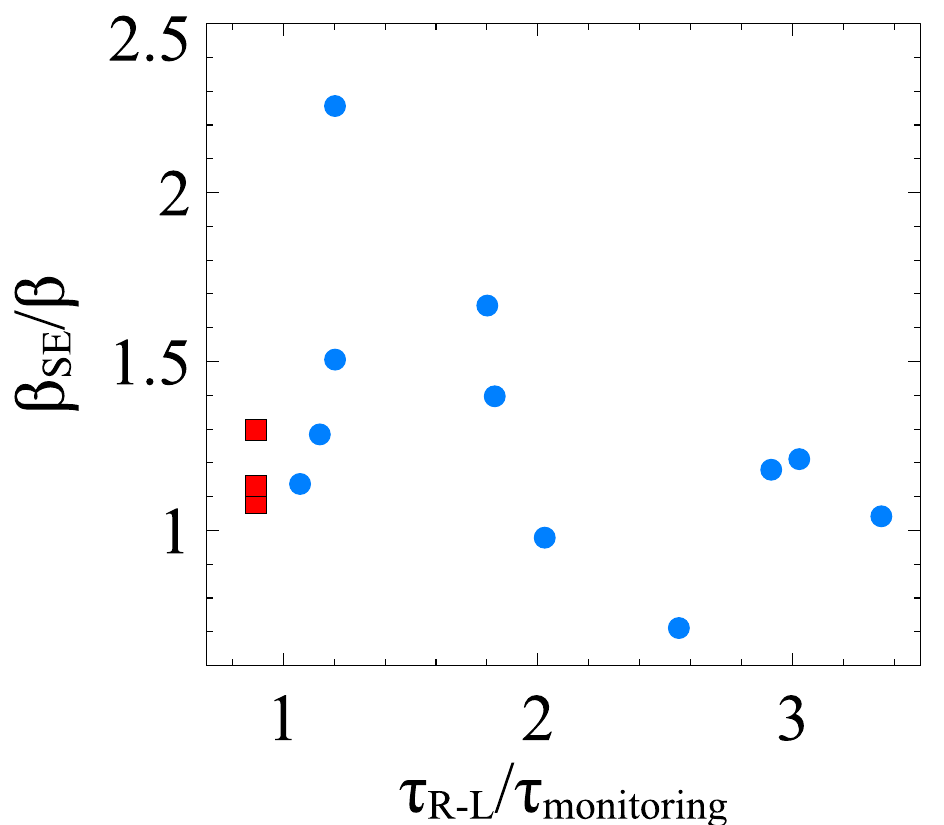}
\includegraphics[width=0.33\textwidth]{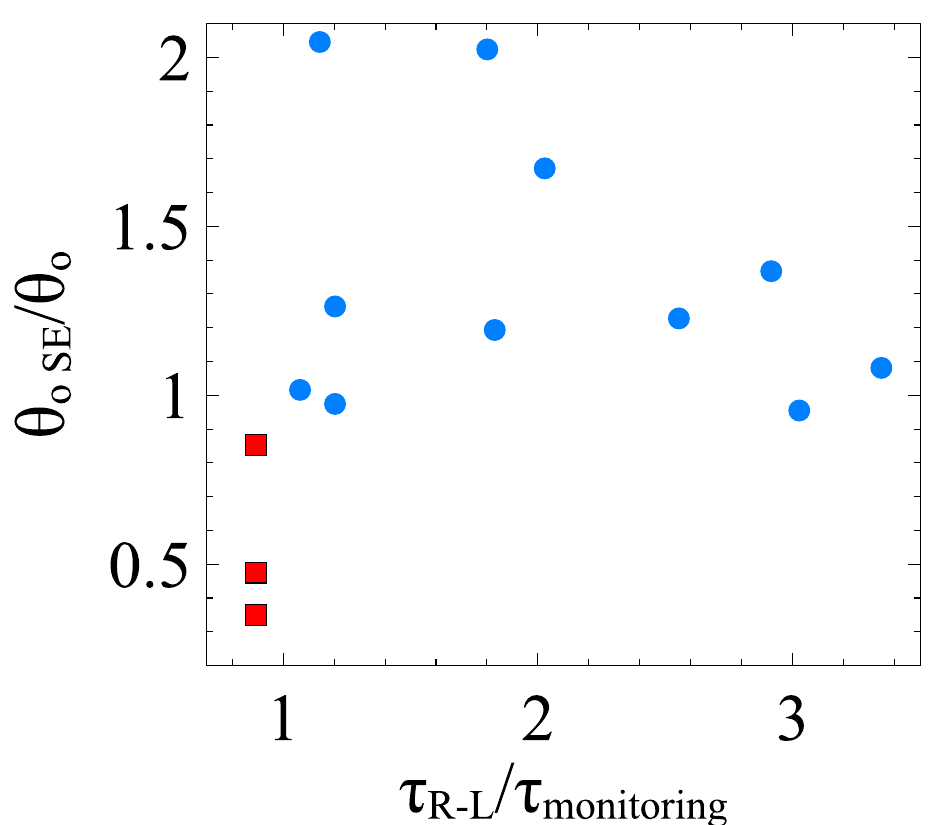}
\includegraphics[width=0.33\textwidth]{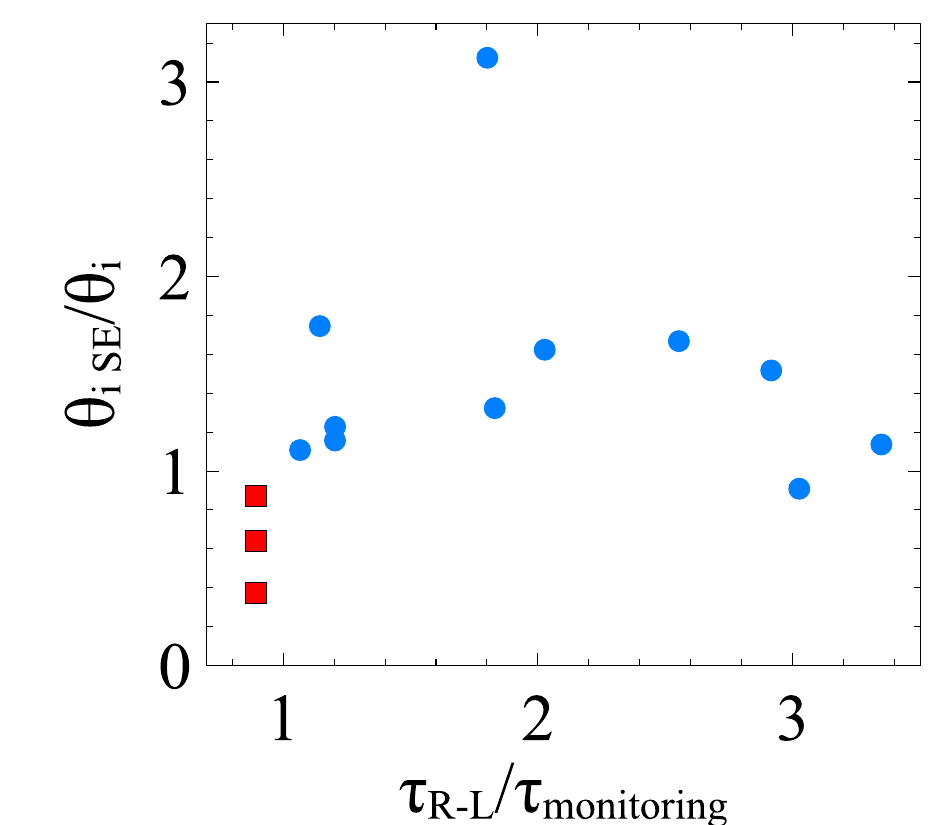}\\[0.2cm]
\includegraphics[width=0.33\textwidth]{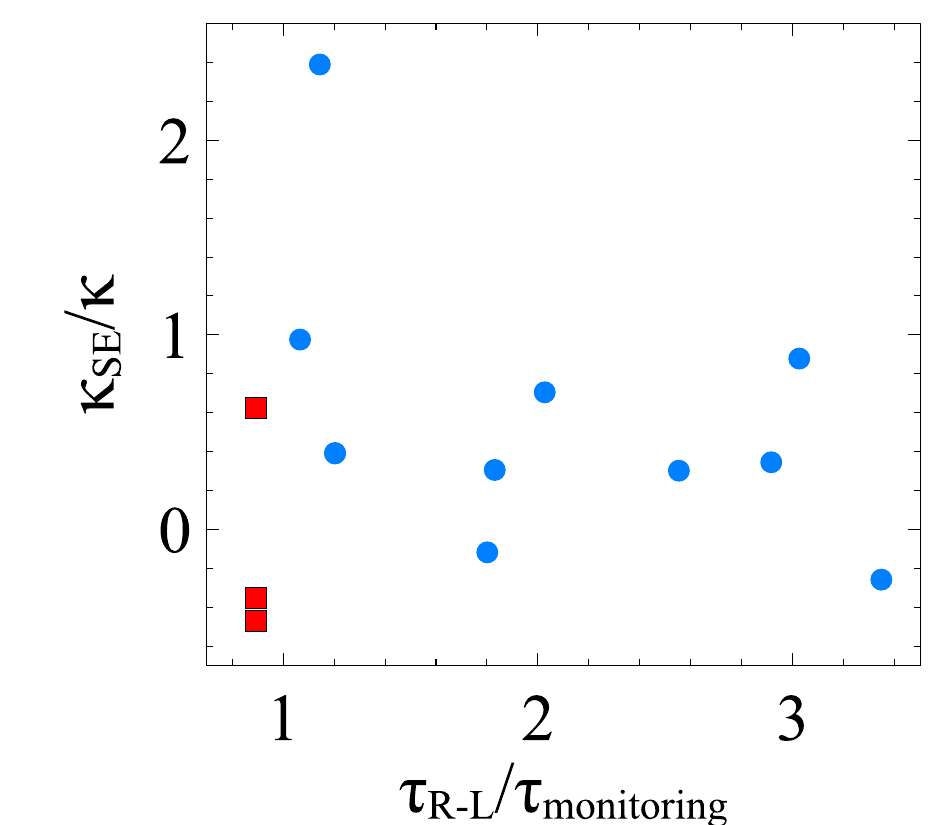}
\includegraphics[width=0.33\textwidth]{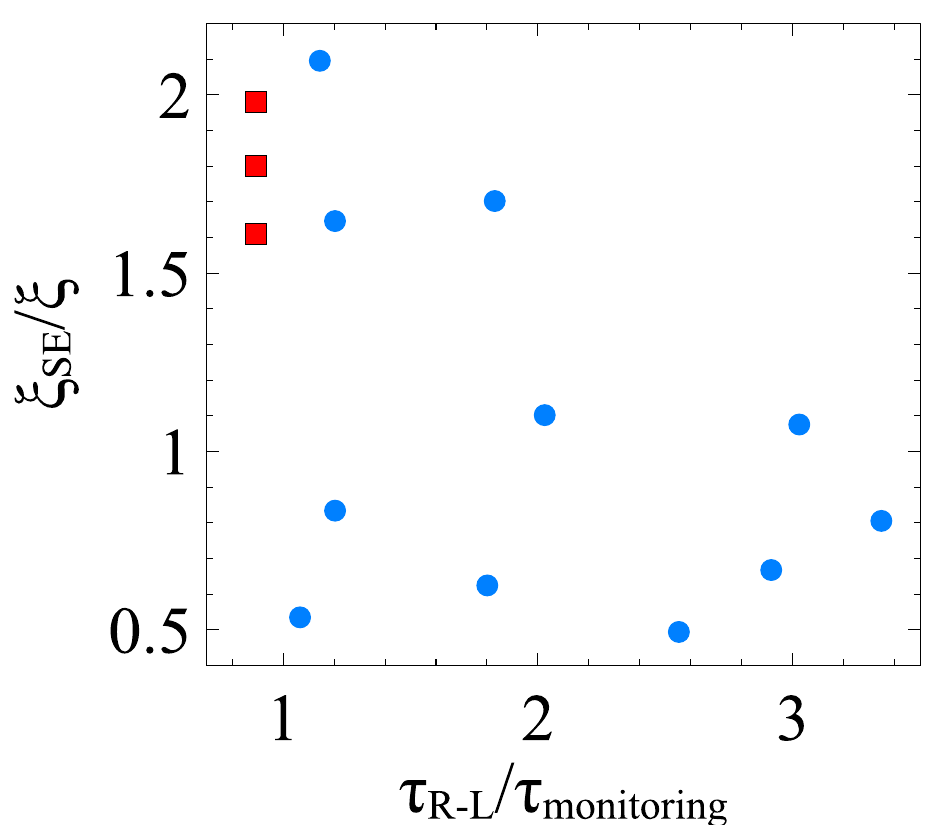}
\includegraphics[width=0.33\textwidth]{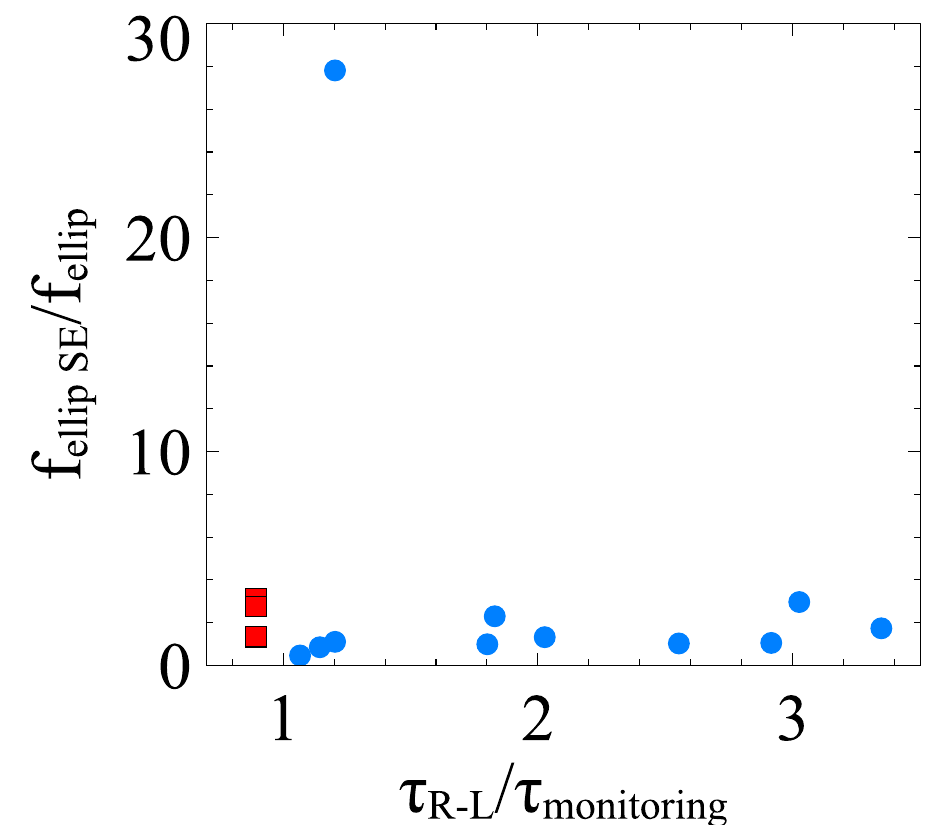}\\[0.2cm]
\includegraphics[width=0.33\textwidth]{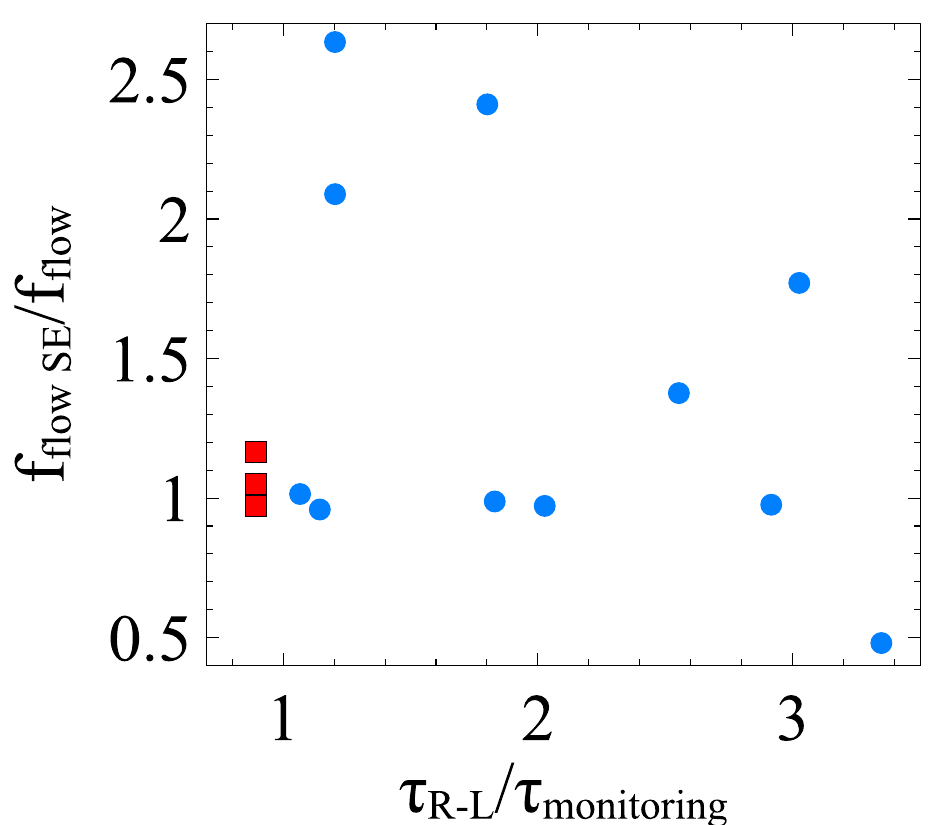}
\includegraphics[width=0.33\textwidth]{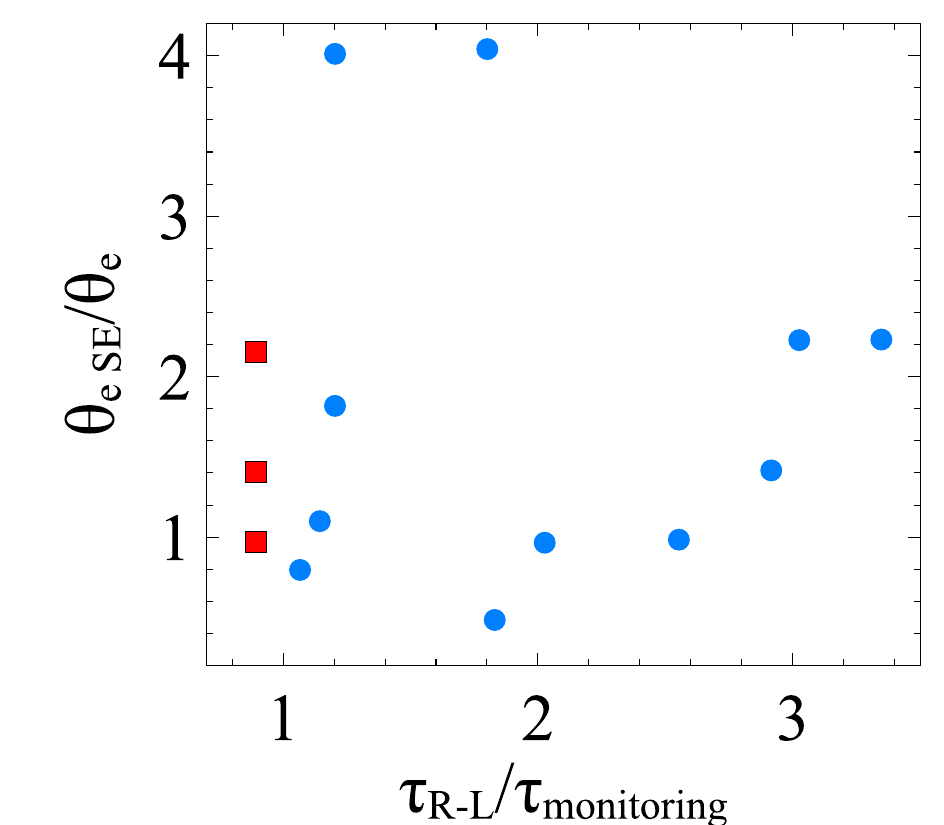}
\includegraphics[width=0.33\textwidth]{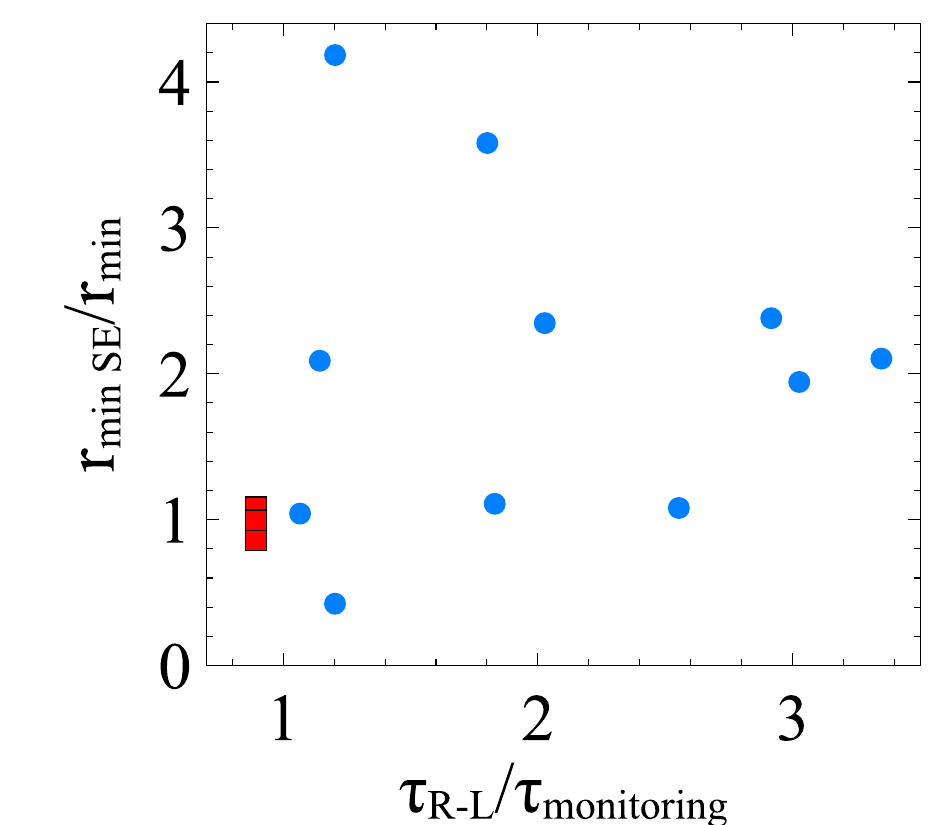}\\[0.2cm]
\includegraphics[width=0.33\textwidth]{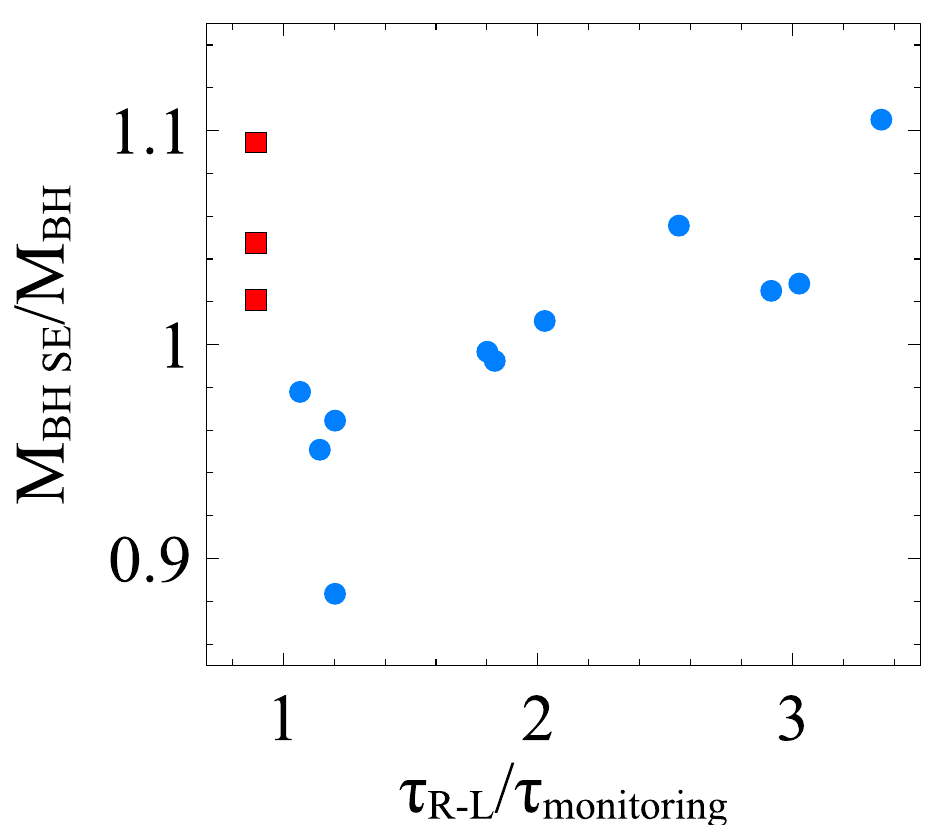}
\includegraphics[width=0.33\textwidth]{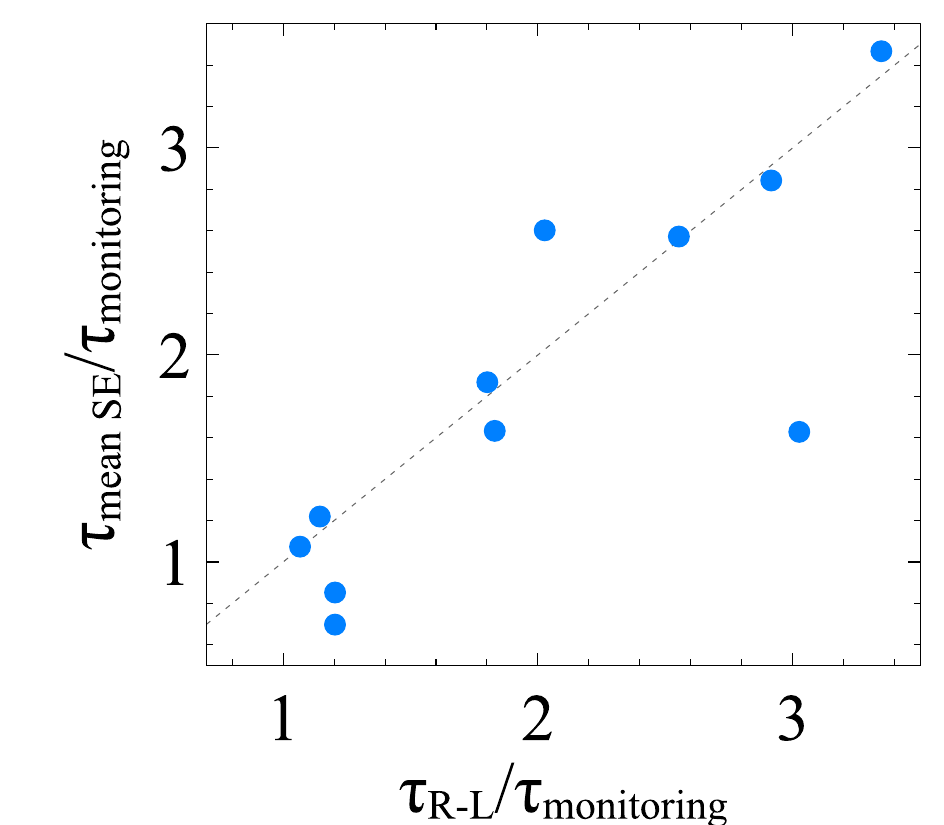}
\caption{Ratio of the mean value of the parameters as a function of the ratio between the assumed $\tau_{\rm mean}$ from the R$-$L relation and the $\tau_{\rm mean}$ from the modelling of the monitoring data. The \taumean\ from the R$-$L relation corresponds to the central value assumed for our prior probability distribution on \taumean. Each filled circle represents one AGN. The blue circles are the AGN from this work while the red squares are the results from the three epochs of Arp 151 \citepalias{raimundo19}. All constrained and unconstrained parameters are shown. The only parameters omitted are those for which the full light-curve modelling can only determine upper or lower limits for the parameters.}
\label{parameter_RL_offset}
\end{figure*}

We also investigated if the inferred parameter values were influenced by how well the R$-$L relation assumed mean time delay (i.e. the central value of our prior probability distribution) represents the full-lightcurve determined \taumean. In Fig~\ref{parameter_RL_offset} we show the ratio of the inferred parameter values as a function of the ratio between the R$-$L relation mean time delay ($\tau_{\rm R-L}$) and the full-lightcurve mean time delay determined from modelling of the monitoring data ($\tau_{\rm monitoring}$). All the inferred parameters are shown, both the constrained and unconstrained parameters. The only data points omitted in this figure are the cases where the full light-curve modelling only obtained lower or upper limits for the parameter values. There is no evidence from this figure that an under- or overestimated \taumean\ results in worse parameter estimates. In general the model's ability to infer parameter values that agree with the full light-curve result seems to be independent of the choice of \taumean. The expected exception is \Mbh\ which depends on \taumean.

As was done for Arp 151 \citepalias{raimundo19}, we model additional epochs for each of the AGN in our sample. In particular we select an additional epoch with lower average S/N (typically 9 or 20) and use the same modelling procedure (see Section~\ref{sec:epoch_selection}). We find that the results for the two epochs (the default highest S/N epoch and the lower S/N epoch) are fully consistent for each AGN. A comparison of the inferred parameters for each epoch and the respective 68\% confidence regions can be found in Fig.~\ref{inferred_parameters_SN20}. The inferred parameters from both epochs agree, with minor differences in terms of the confidence regions. The differences in the confidence regions are not systematic, in the sense that there is no one epoch that systematically shows smaller confidence ranges than the other. The relative width of the confidence regions vary from epoch to epoch and parameter to parameter, with some parameters where the maximum S/N epoch provides tighter constraints and other parameters where the lower S/N epoch provides the tighter constraints. This shows that the model obtains consistent results when independently modelling distinct spectral epochs of the same AGN. It also shows that the line flux S/N is not the dominant factor in determining which BLR parameters can be inferred and the accuracy level associated with that inference.

\subsection{Caveats}

\subsubsection{Underlying physical model}
In this work, we compare the performance of the single epoch model to the full light-curve model. One of the limitations of this approach is that both models rely on the same underlying physical prescription: a simplified model for the BLR. The advantage of using a simplified physical model is that the main BLR parameters have a physical meaning and are intuitive, and that the model is highly flexible. By using just a limited set of parameters, the model can reproduce a large variety of possible BLR geometric configurations and dynamical properties. The caveat is that we are limited to the physical reach of the model, i.e. physical processes that are not included in the model (see Section~\ref{sec:model_limitations}), will not be constrained. Nevertheless, the physical model we use and that was used in previous versions of the BLR modelling, has successfully constrained BLR parameters for 17 AGN based on monitoring data (\citealt{pancoast12}, \citealt{pancoast14b}, \citealt{grier17}, \citealt{pancoast18}, \citealt{williams18}). More importantly, the \Mbh, time delays, and dynamics found by these works are in agreement with reverberation mapping studies that used independent methods to determine the same properties, which indicate that the parameters determined by the model are representative of the intrinsic BLR parameters. The physical model has shown to be a useful basis to interpret BLR properties, however one should keep in mind that the results obtained are limited to the complexity level of the physical prescription as described in Section~\ref{sec:model_limitations}.

\subsubsection{Spectral decomposition}
\cite{williams18} show that for some AGN, the choice of templates in the spectral decomposition may affect the BLR modelling. The effect is more noticeable for AGN with prominent Fe II emission, since it will more strongly affect the resulting spectrally decomposed H$\beta$ line profile. In our work we find that the spectral decomposition may affect the BLR modelling, in agreement with \cite{williams18}. However the specific spectral decomposition adopted does not alone determine the BLR parameters. That is, selecting a specific spectral decomposition does not force the parameters to a specific solution. This can be seen in the results for the 7 AGN of the LAMP 2011 sample. We use as input a spectrum that has been decomposed using the \cite{kovacevic10} Fe II templates. Therefore our posterior probability distributions can be directly compared with the \cite{williams18} results using the same spectral decomposition (shown as the red histograms in Figs.~\ref{posterior_mrk141},  \ref{posterior_zw229} and Figs.~\ref{posterior_mrk335} to \ref{posterior_iisz10}). For most AGN and most parameters we see a general agreement between all three posterior distributions: the two sets of results from \cite{williams18} and our single-epoch results. However, if the spectral decomposition was driving the results, we would expect our inferred parameters to always match the \cite{williams18} results using the \cite{kovacevic10} templates. This is not the case as can be seen for example in Fig.~\ref{posterior_mrk141} for the $\kappa$ parameter for Mrk 141. There is a slight difference between the combined and \cite{kovacevic10} posterior distributions presented by \cite{williams18} and our result resembles the combined posterior more closely. The opposite also happens, where our distribution more closely resembles the \cite{kovacevic10} posterior distributions (for example for $\theta_{i}$ and $\kappa$ for NGC 4593 - Fig.~\ref{posterior_ngc4593}).
Additionally, there are instances where our single-epoch model is not able to constrain parameters, even though we are using the sample spectral decomposition of \cite{williams18}. Examples of this are $f_{\rm flow}$ for Mrk 1511 and PG 1310-108, or $\xi$ in Mrk 1511 (see Fig.~\ref{inferred_parameters}). This indicates that whatever information was present in the monitoring data that allowed the model to constrain the parameters, is not present or cannot be uniquely inferred from the single epoch spectrum. One possibility is that with relatively symmetric line profiles as those of the AGN mentioned above, it is inherently difficult to infer the direction of motion from a single spectral profile.

\section{Conclusions}
\label{sec:conclusions}
In this work we applied our single-epoch broad line region model \citepalias{raimundo19} to 11 nearby AGN. This sample of AGN have previously been modelled using reverberation mapping monitoring data (\citealt{grier17}, \citealt{williams18}), and therefore provide an opportunity to test our model's performance for a variety of broad line shapes, black hole masses and accretion rates. Our main conclusions are the following:\\[0.1cm]
$\bullet$ The single-epoch model is able to constrain some of the BLR physical parameters from single spectra assuming an underlying physical model, reinforcing our findings for the test case of Arp 151 (\citetalias{raimundo19}). The specific BLR parameters that are constrained vary for each AGN, and likely depend on the particular features of the broad line profile.\\[0.1cm]
$\bullet$ The model achieves reliable results in terms of the inferred values for the parameters. When there is not enough information in the line profile to constrain a parameter, the output from the model shows a broad posterior probability distribution which indicates that the parameter cannot be accurately constrained. When the model is able to constrain the BLR parameters, the inferred values of the parameters agree within the 68\% confidence range with the results from the modelling using monitoring data.\\[0.1cm]
$\bullet$ We find that two parameters cannot be independently constrained from our single-epoch modelling: $\beta$, which describes the radial distribution of particles in the BLR, and the black hole mass (\Mbh), in agreement with \citetalias{raimundo19}. The parameter $\beta$ may vary with selected epoch, possibly as a function of what portion of the BLR is illuminated or emitting at that epoch. The value of $\beta$ determined from single-epoch modelling reflects a temporary property of the BLR and does not reflect the \emph{average} radial distribution of particles of the BLR as probed by the full light-curve analysis of the original BLR modelling code. The black hole mass we infer is by default correlated with the mean time delay (\taumean). Since we need to define a specific prior probability distribution for \taumean\ to set the physical scale of the BLR, the inferred \Mbh\ will in part depend on the assumptions on \taumean. The constraints on \taumean\ in our model are obtained from the R$-$L relation. This is similar to what is done for single-epoch black hole mass measurements, that typically rely on the BLR radius from the R$-$L relation. Our uncertainties in the inferred \Mbh\ are similar to those associated with single-epoch mass measurements (e.g. \citealt{vestergaard&peterson06}, \citealt{vestergaard&osmer09}).\\[0.1cm]
$\bullet$ We find that for some AGN, when the single-epoch model can only marginally constrain or cannot constrain a BLR parameter at all, the modelling of the monitoring data also shows the same difficulty in constraining the parameter. This is notable in particular for the $\gamma$ parameter that describes the angular concentration of particles in the BLR. This shows that for a subsample of AGN and some of the physical parameters, having monitoring data does not significantly improve our ability to quantitatively infer a BLR parameter.\\[0.1cm]
$\bullet$ In general, the monitoring data have more information than that contained in a single spectrum, and if available should be the preferred dataset to use as input to the model to constrain the BLR parameters. However, considering the performance of the single-epoch modelling we show here, and its potential to be extended to a significantly larger number of AGN at low and high redshift, single-epoch modelling is an alternative and practical tool to constrain some of the BLR parameters.\\[0.1cm]
$\bullet$ We identify a qualitative trend in terms of the broad line profiles and the ability of the single-epoch model to constrain the BLR parameters. For the AGN with the more symmetrical and featureless line profiles (e.g. 3C120, Mrk 335 and PG 2130+099) the single-epoch model is only able to constrain a small number of BLR parameters. The better performance of the model, in terms of number of parameters constrained occurs for the AGN with most substructure and asymmetry in the broad line profiles (e.g. Mrk 141 and Zw 229+015). This qualitatively suggests that the more complex line profiles provide an advantage in terms of the amount of information provided to the model.\\[0.1cm]
$\bullet$ We find that in general, the AGN with the lower Eddington ratios tend to have a larger number of BLR parameters constrained than the AGN with the highest Eddington ratios. This is likely due to the line profiles of the highest Eddington ratio AGN, which typically show smoother and less asymmetric profiles.\\[0.1cm]

We have shown in this work and in our first paper \citepalias{raimundo19}, that single-epoch modelling of AGN broad line profiles can provide constraints on BLR parameters, assuming an underlying physical model for the BLR. The inferred values of the parameters agree with those determined from models of the monitoring data, which makes the single-epoch model reliable for the study of low and high redshift AGN for which no monitoring data are available. 
In follow-up work we will further investigate the possible limitations of this work and define strategies to improve the method, with the ultimate goal of constraining the typical BLR parameters for the AGN population at low and high redshift.
\section*{Acknowledgements}
The authors thank Anna Pancoast and Aaron Barth for helpful discussions and ideas that contributed to this work and for their significant contributions to LAMP2011. S.~I.~R. and M.V. gratefully acknowledge support from the Independent Research Fund Denmark via grant numbers DFF 4002-00275 and 8021-00130.

\bibliographystyle{mnras}
\bibliography{AGN}

\appendix

\section{Additional figures and table of results}

\begin{figure*}
\centering
\includegraphics[width=1.0\textwidth]{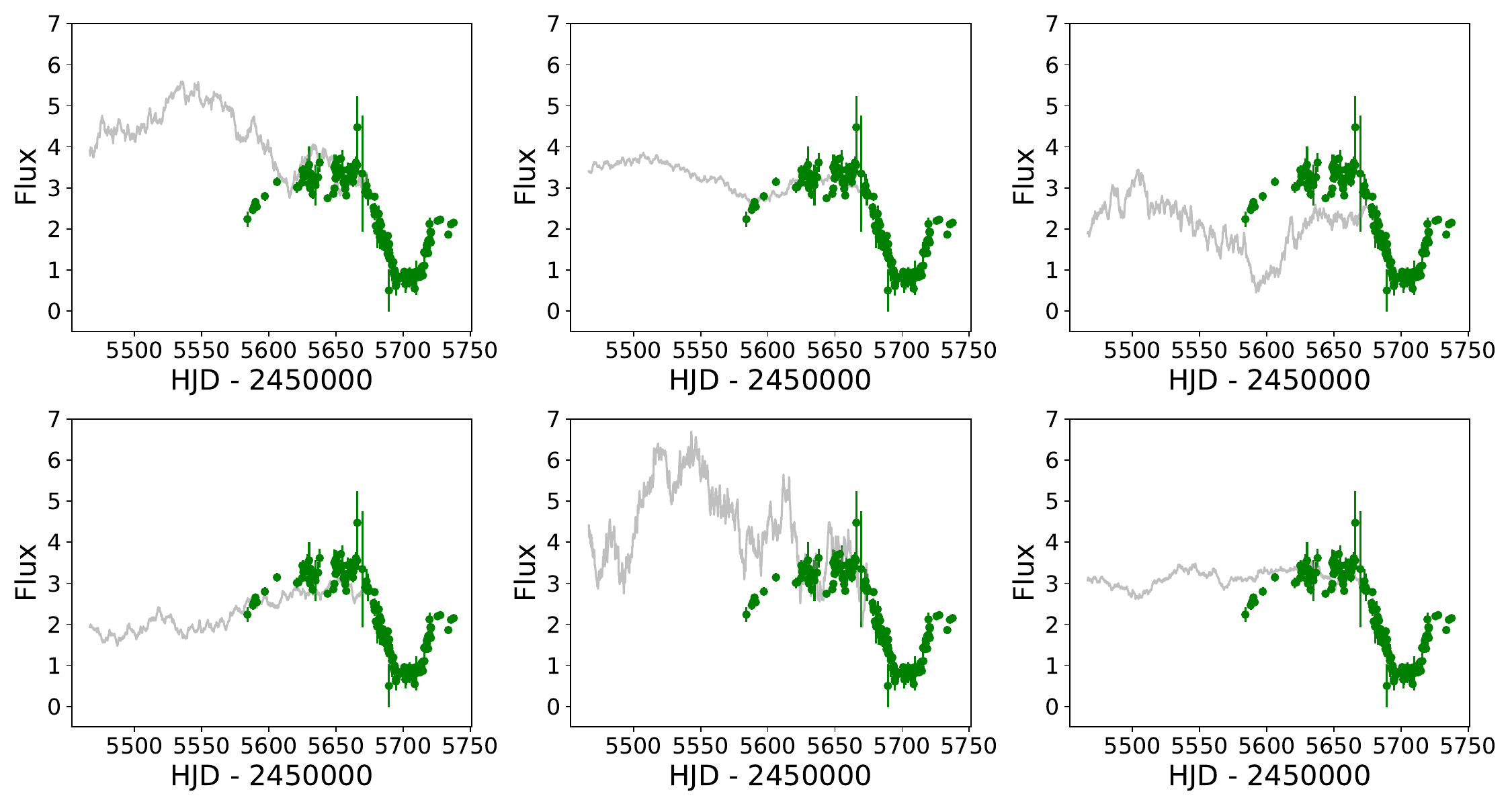}
\caption{Example of simulated continuum lightcurves for Mrk 50. This sub-sample of lightcurves were extracted from the final posterior distribution for Mrk 50. The simulated lightcurves are shown in grey while the observed light curve (from \citealt{williams18}) is shown in green. The continuum light curves are simulated backwards in time starting from the time instant corresponding to the single-epoch selected for the spectral modelling.}
\label{simulated_lightcurves_mrk50}
\end{figure*}

\begin{figure*}
\centering
\includegraphics[width=0.95\textwidth]{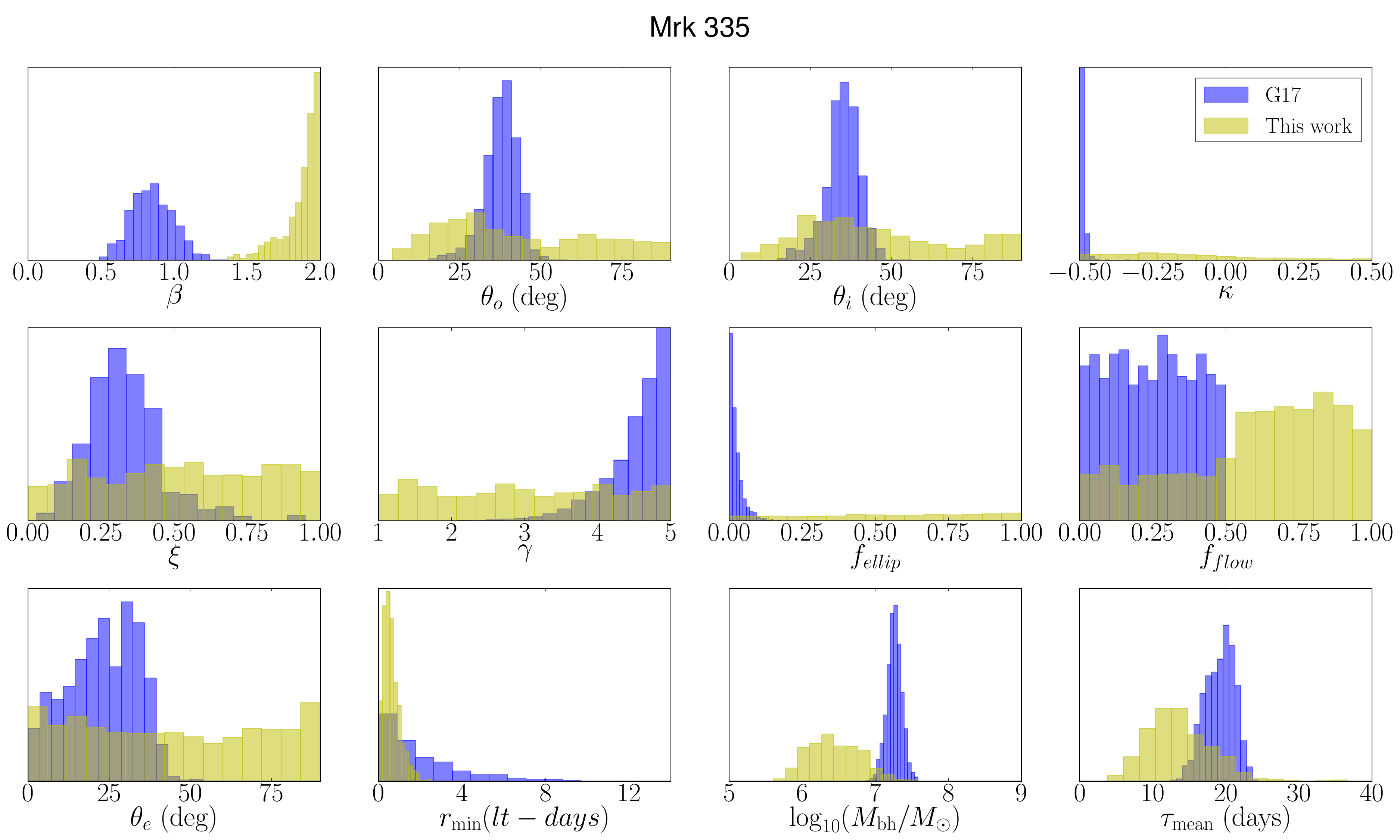}\\
\caption{Same as Fig.~\ref{posterior_3c120} but for Mrk 335.}
\vspace{-0.3cm}
\label{posterior_mrk335}
\end{figure*}

\begin{figure*}
\centering
\includegraphics[width=0.95\textwidth]{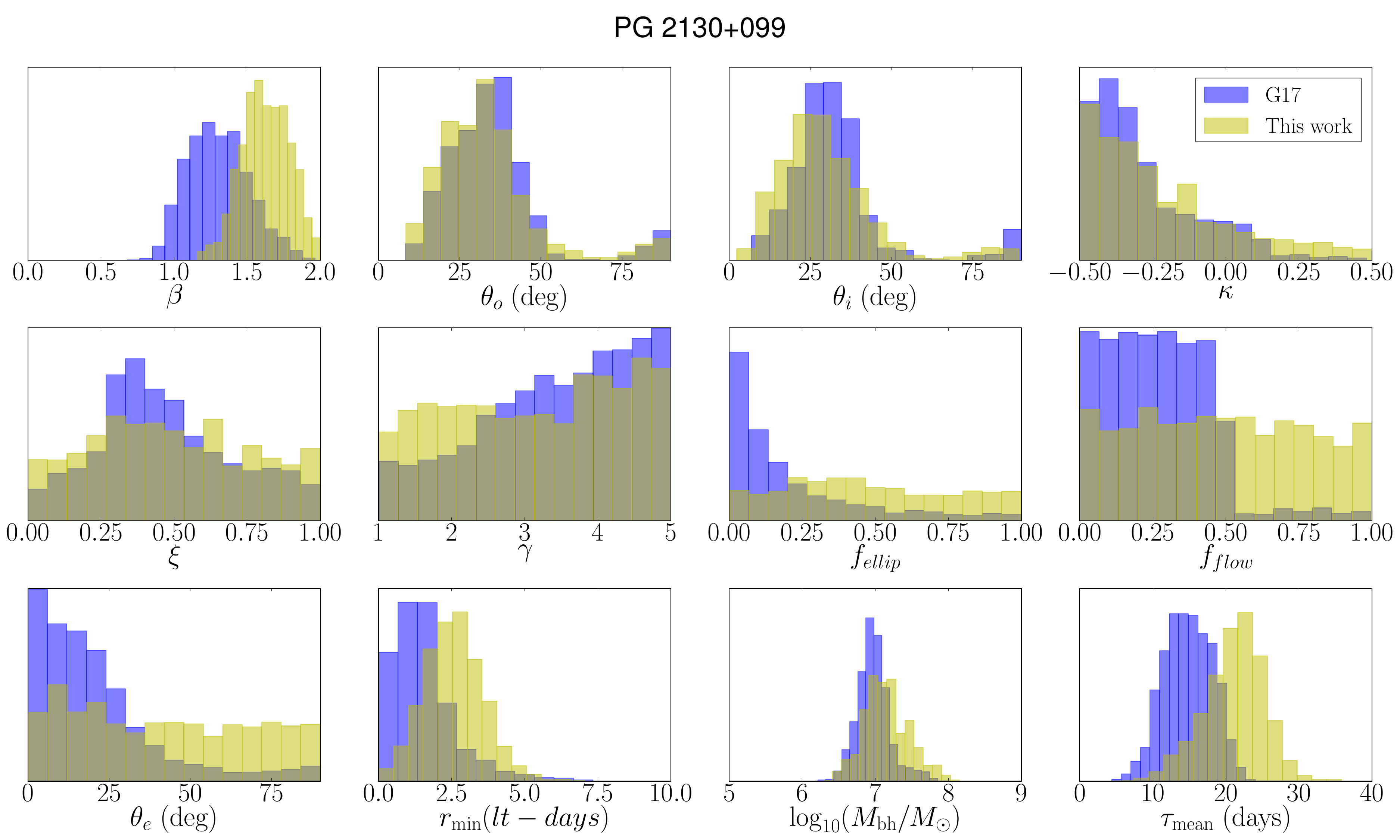}
\caption{Same as Fig.~\ref{posterior_3c120} but for PG 2130+099.}
\label{posterior_pg2130}
\end{figure*}

\begin{figure*}
\centering
\includegraphics[width=0.95\textwidth]{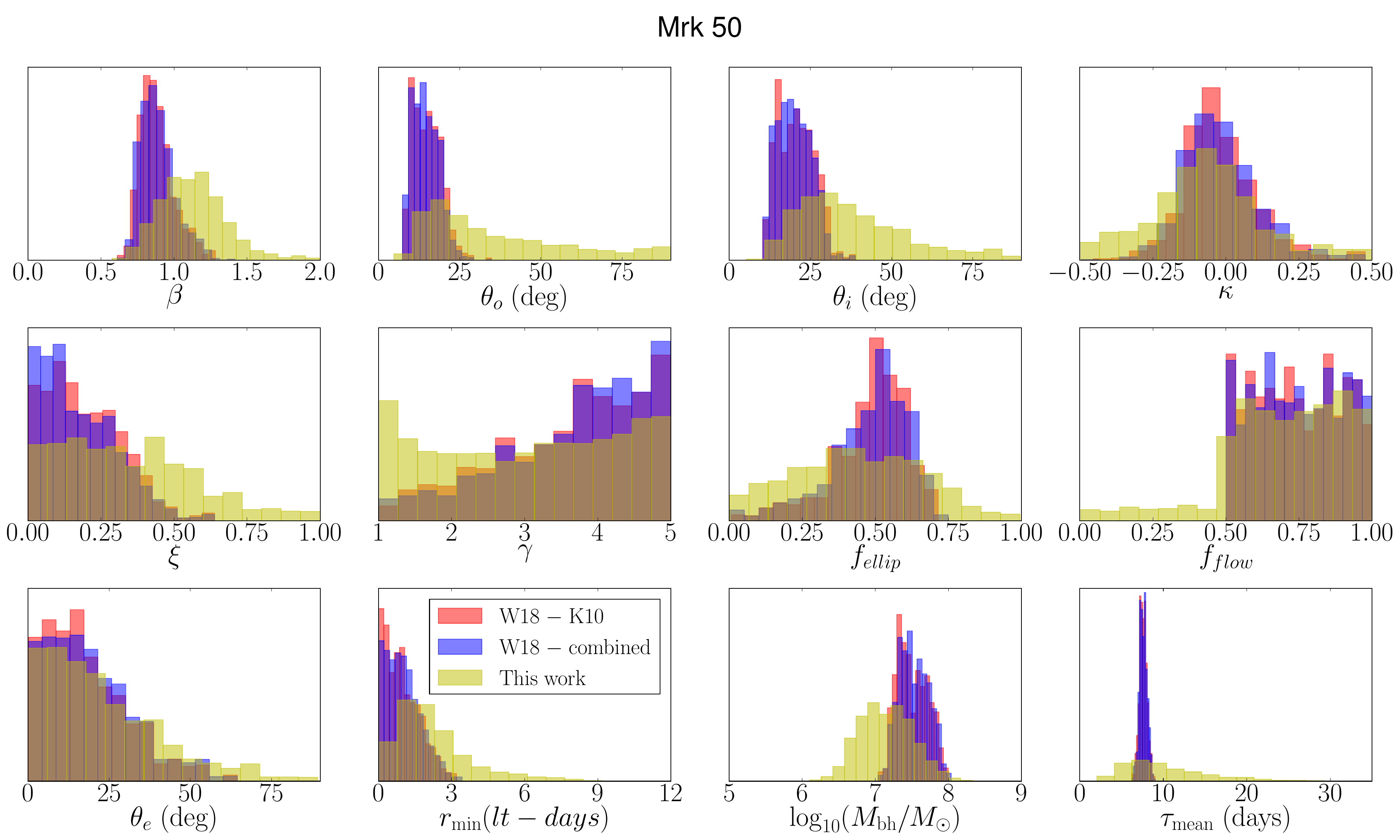}
\caption{Same as Fig.~\ref{posterior_3c120} but for Mrk 50.}
\label{posterior_mrk50}
\end{figure*}

\begin{figure*}
\centering
\includegraphics[width=0.95\textwidth]{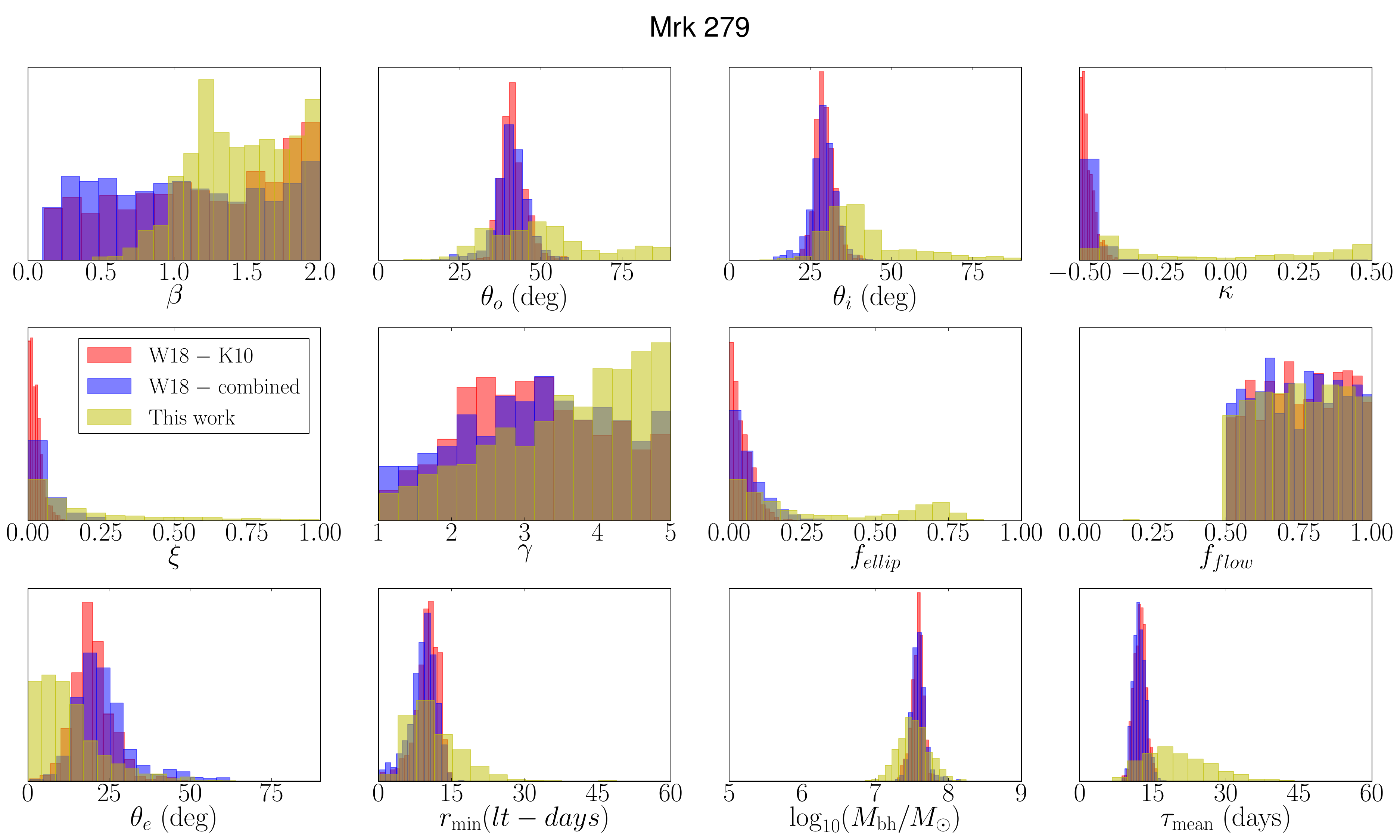}
\caption{Same as Fig.~\ref{posterior_3c120} but for Mrk 279.}
\label{posterior_mrk279}
\end{figure*}

\begin{figure*}
\centering
\includegraphics[width=0.95\textwidth]{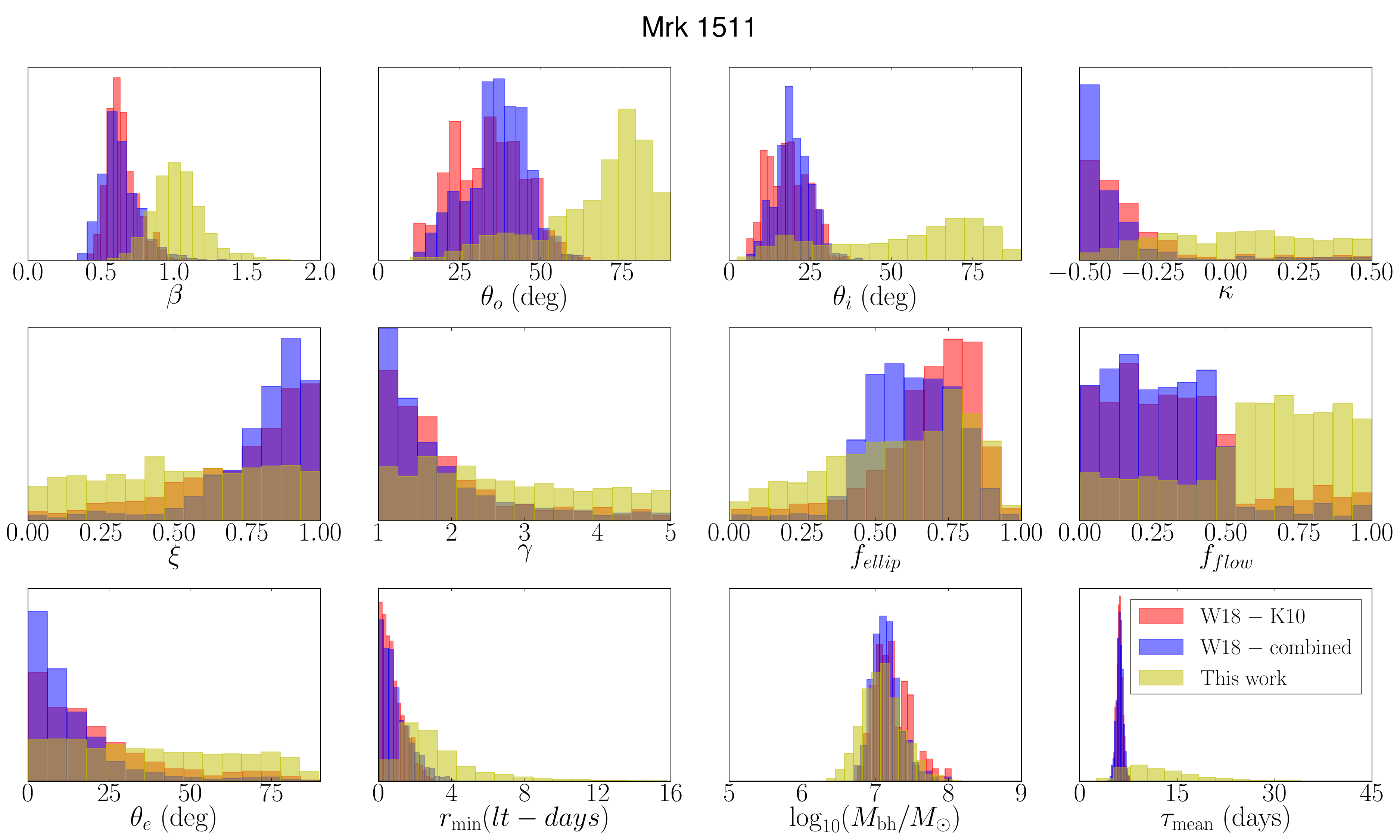}
\caption{Same as Fig.~\ref{posterior_3c120} but for Mrk 1511.}
\label{posterior_mrk1511}
\end{figure*}

\begin{figure*}
\centering
\includegraphics[width=0.95\textwidth]{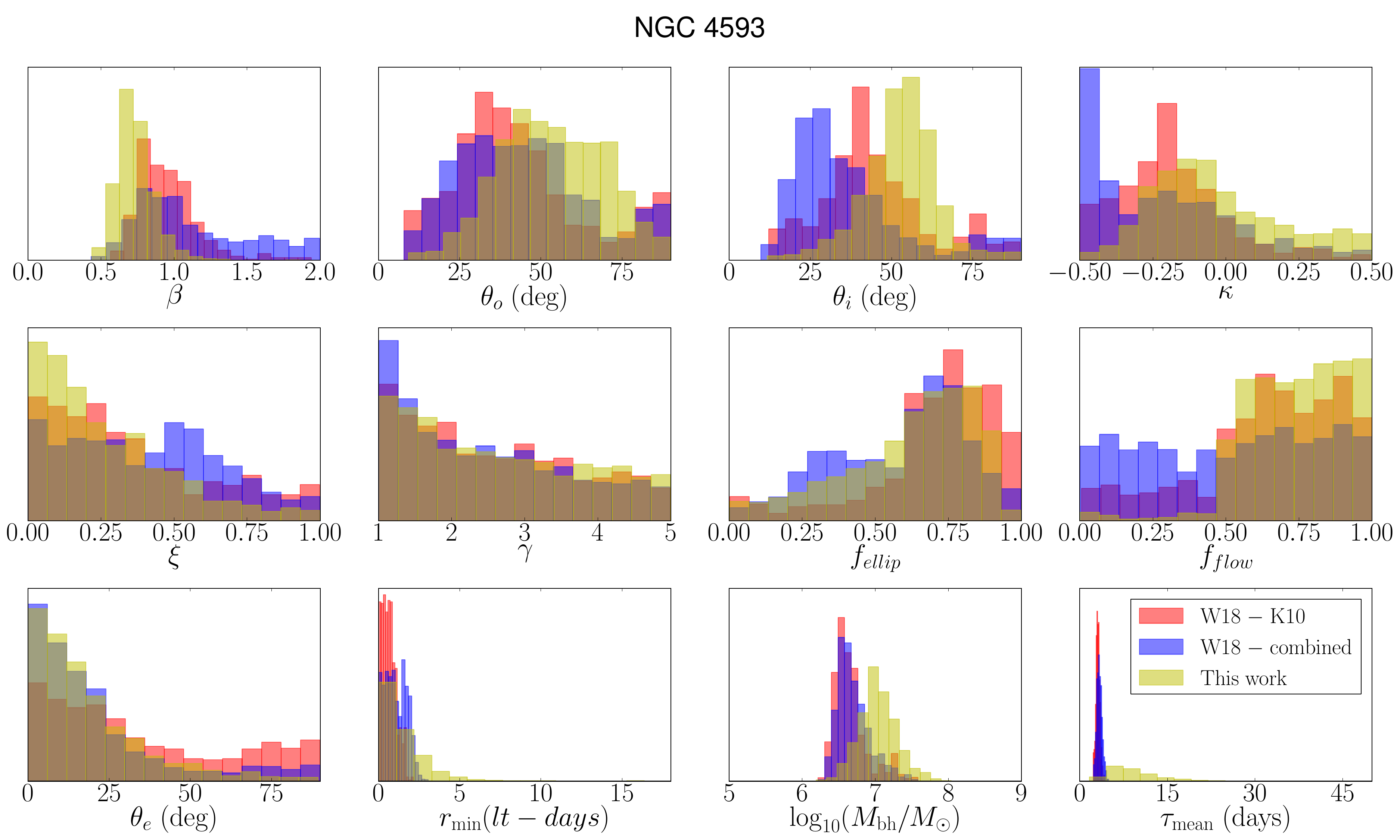}
\caption{Same as Fig.~\ref{posterior_3c120} but for NGC 4593.}
\label{posterior_ngc4593}
\end{figure*}

\begin{figure*}
\centering
\includegraphics[width=0.95\textwidth]{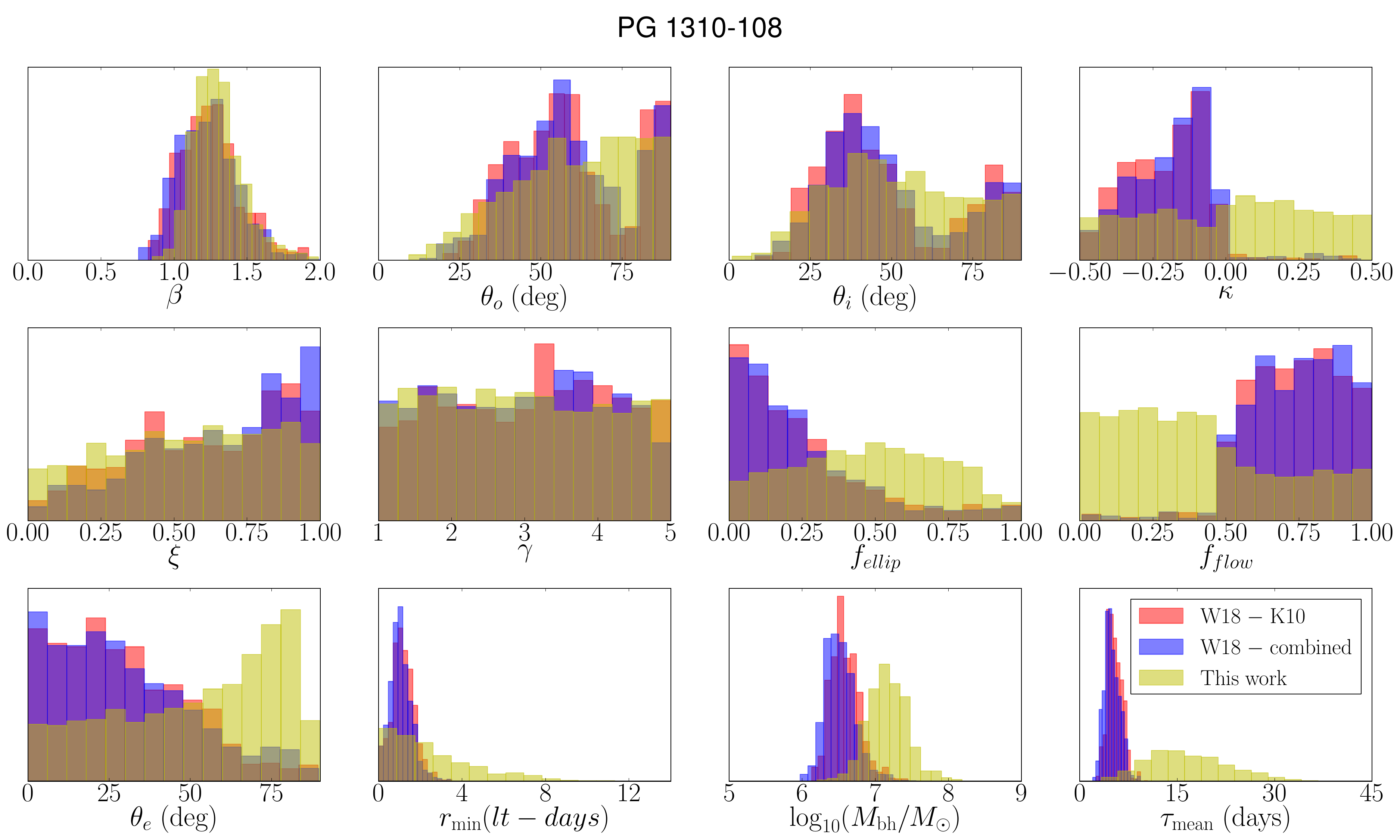}
\caption{Same as Fig.~\ref{posterior_3c120} but for PG 1310-108.}
\label{posterior_iisz10}
\end{figure*}

\begin{figure*}
\centering
\includegraphics[width=0.33\textwidth]{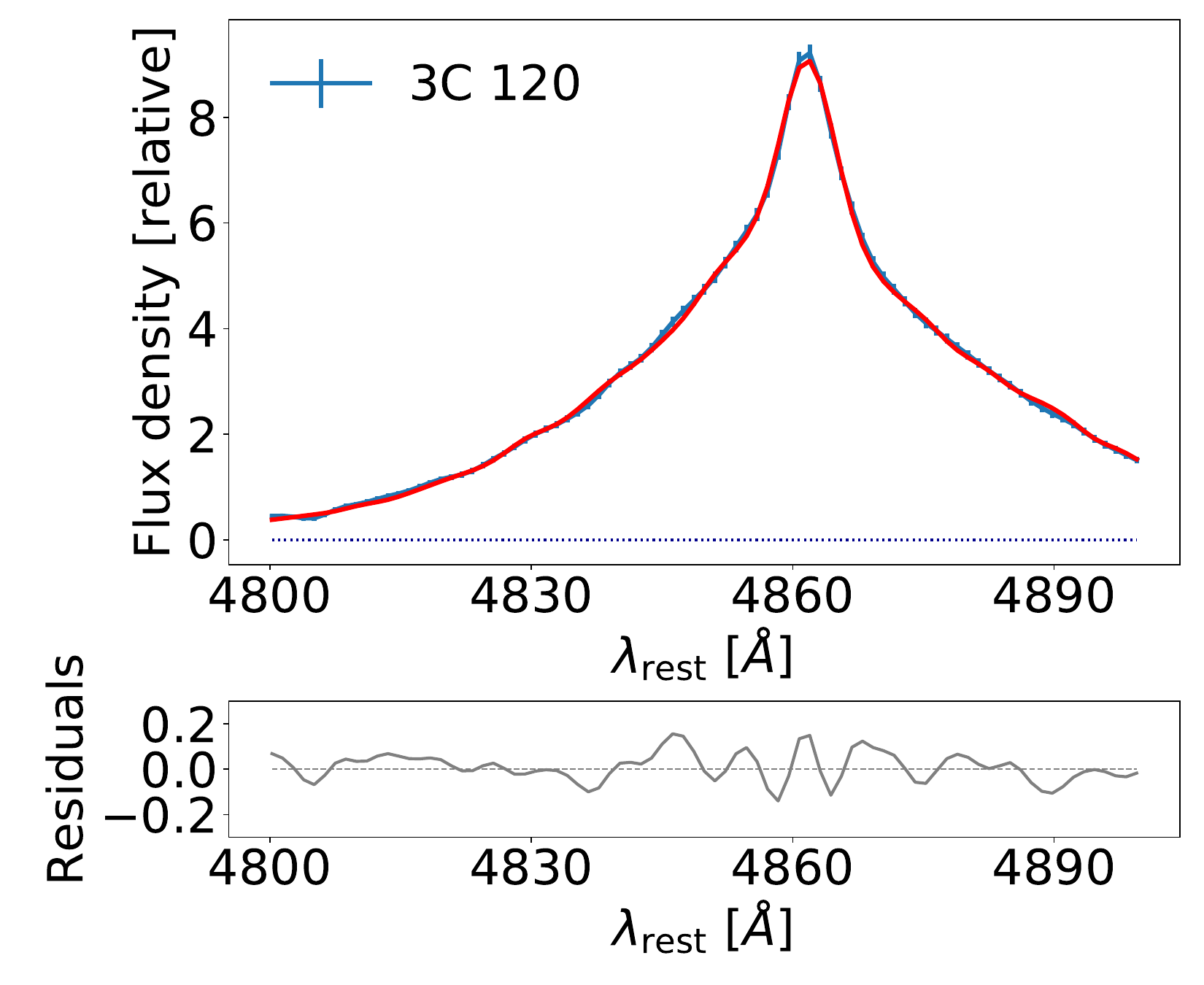}
\includegraphics[width=0.33\textwidth]{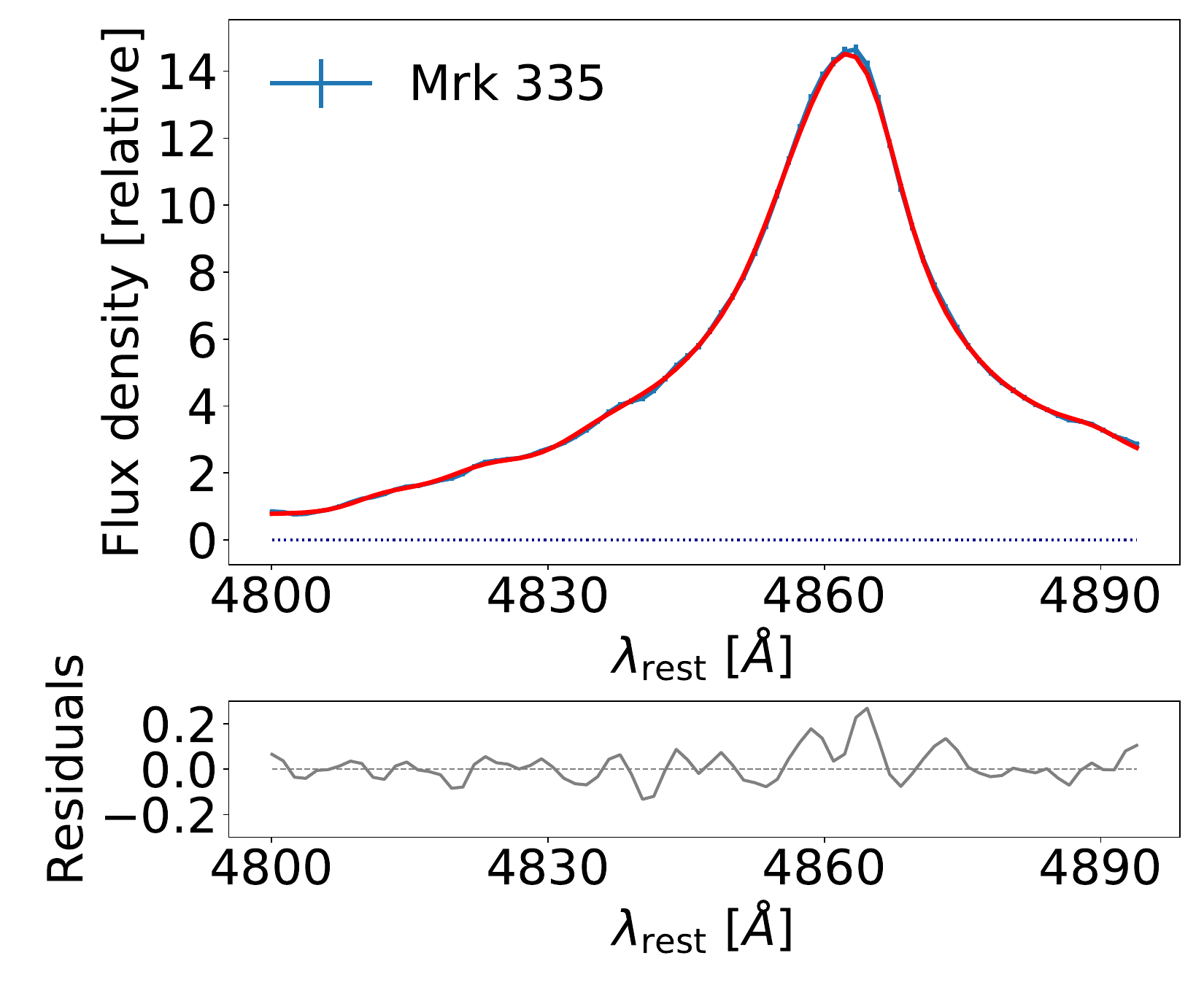}
\includegraphics[width=0.33\textwidth]{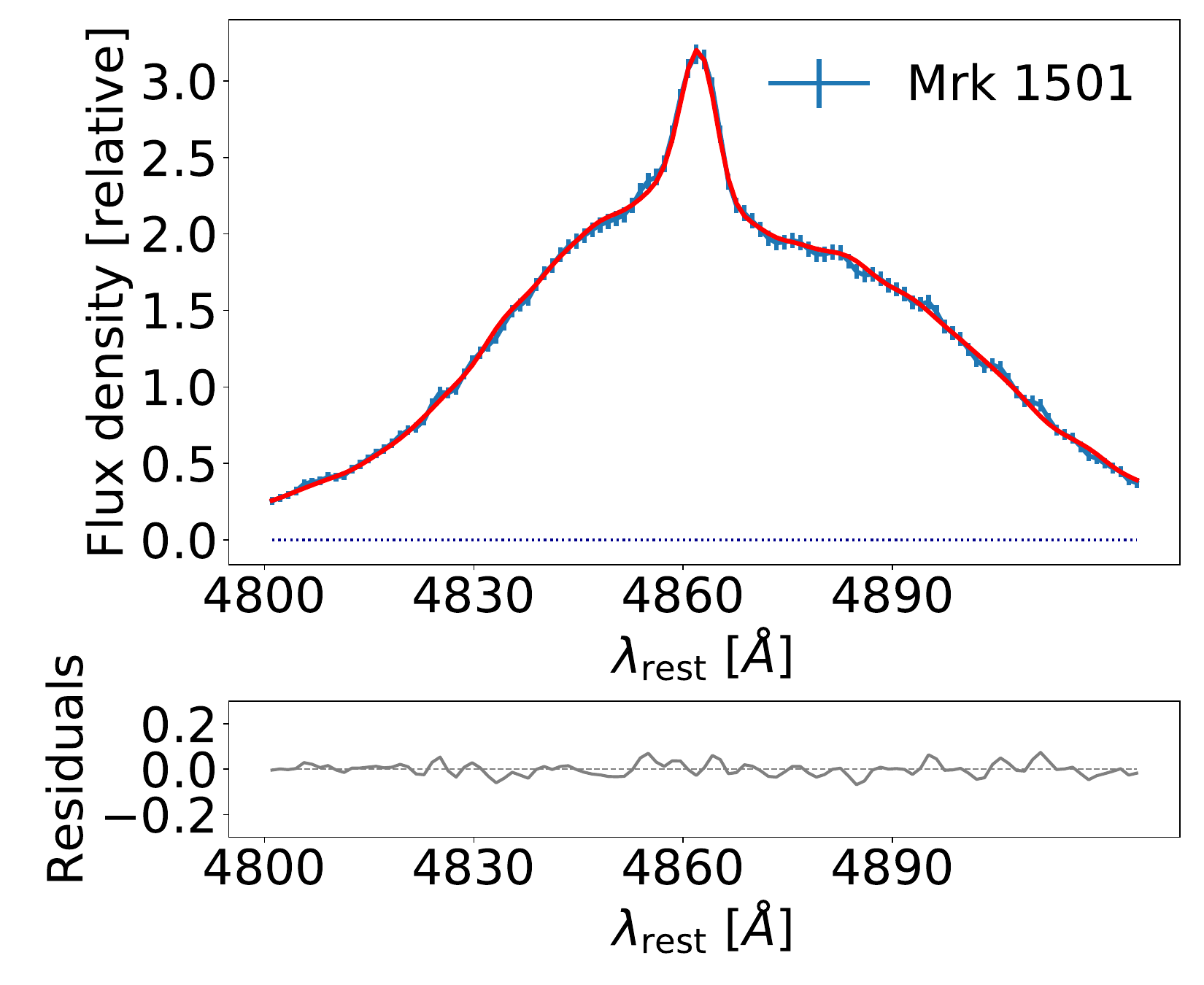}\\[0.2cm]
\includegraphics[width=0.33\textwidth]{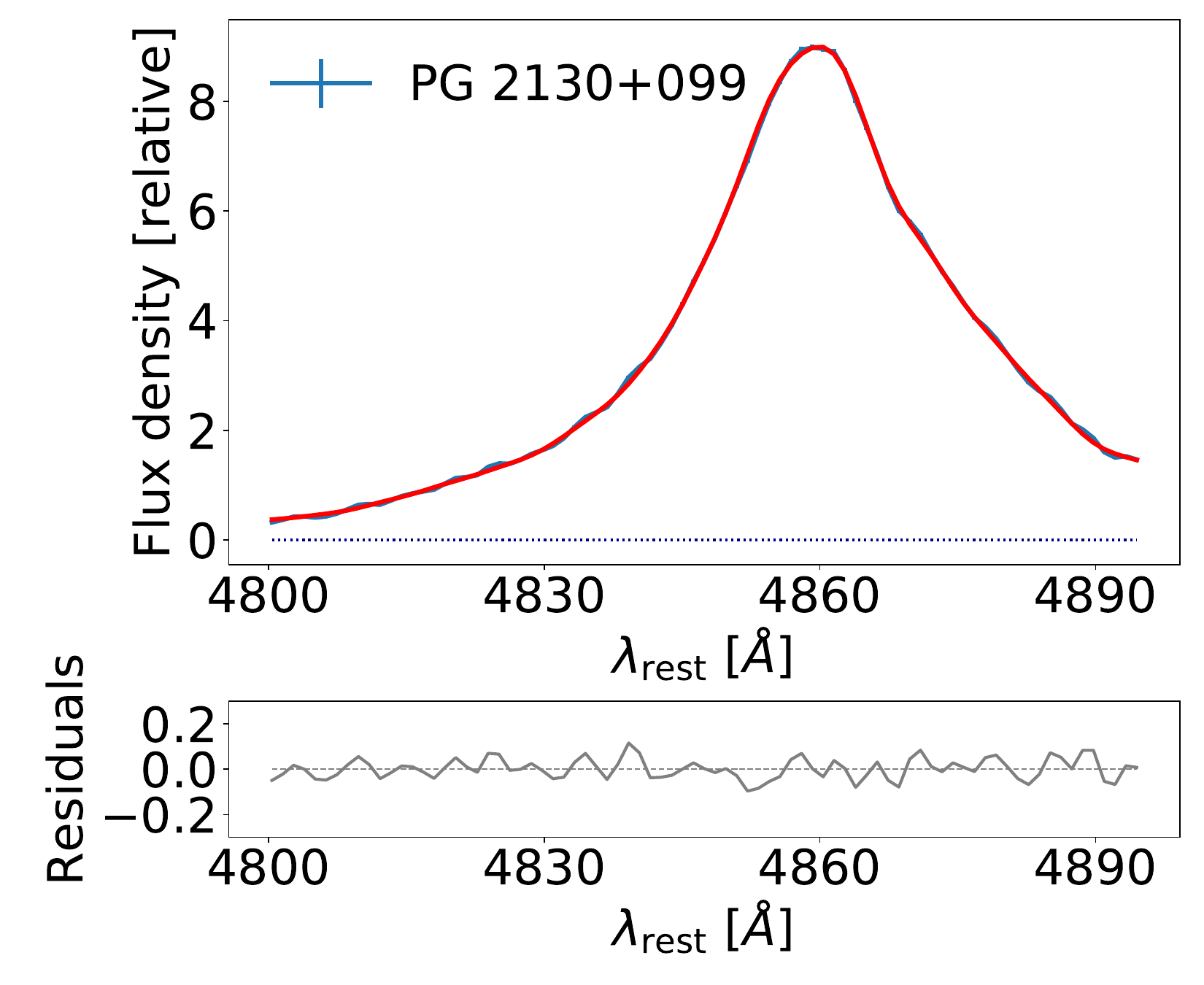}
\includegraphics[width=0.33\textwidth]{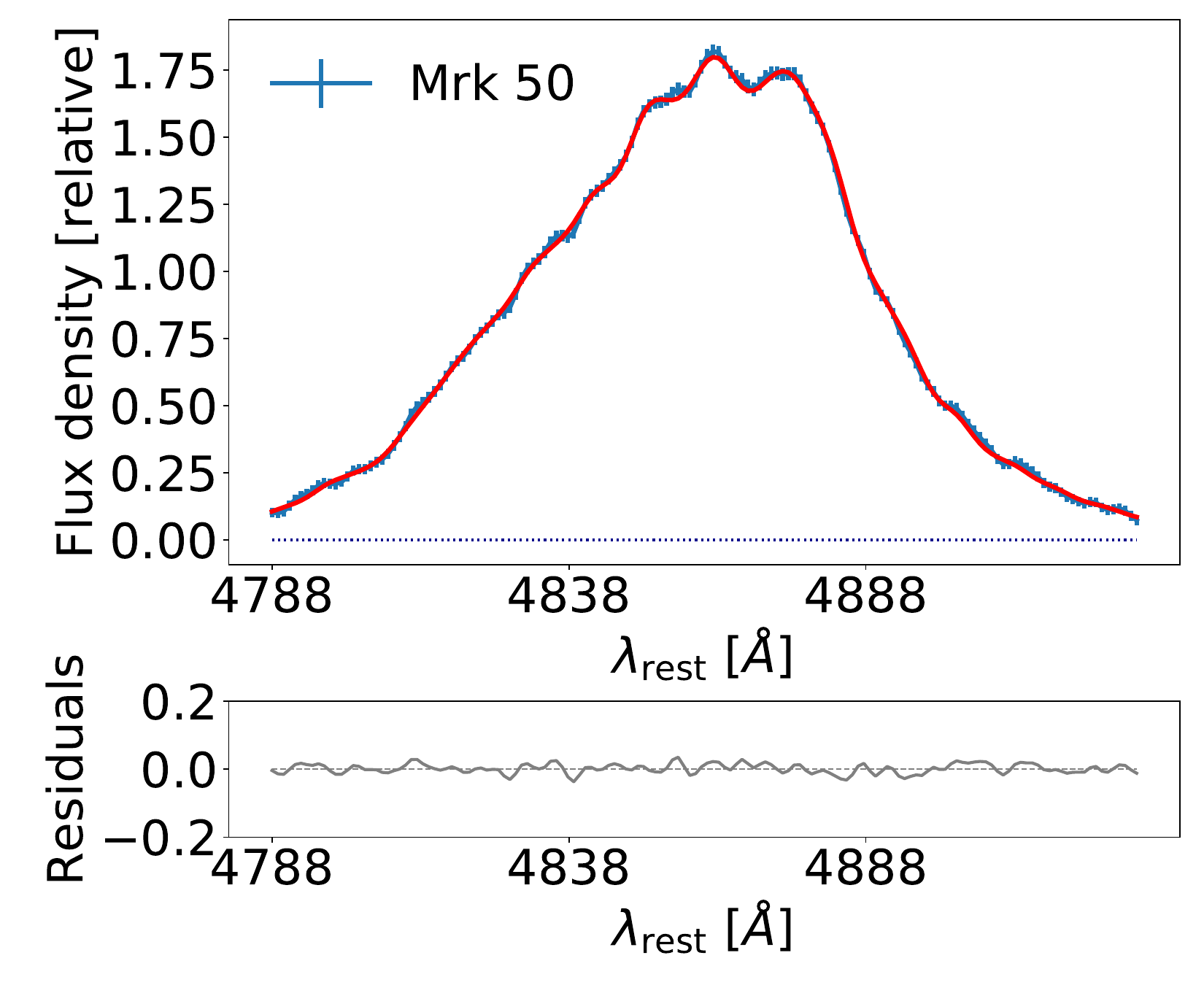}
\includegraphics[width=0.33\textwidth]{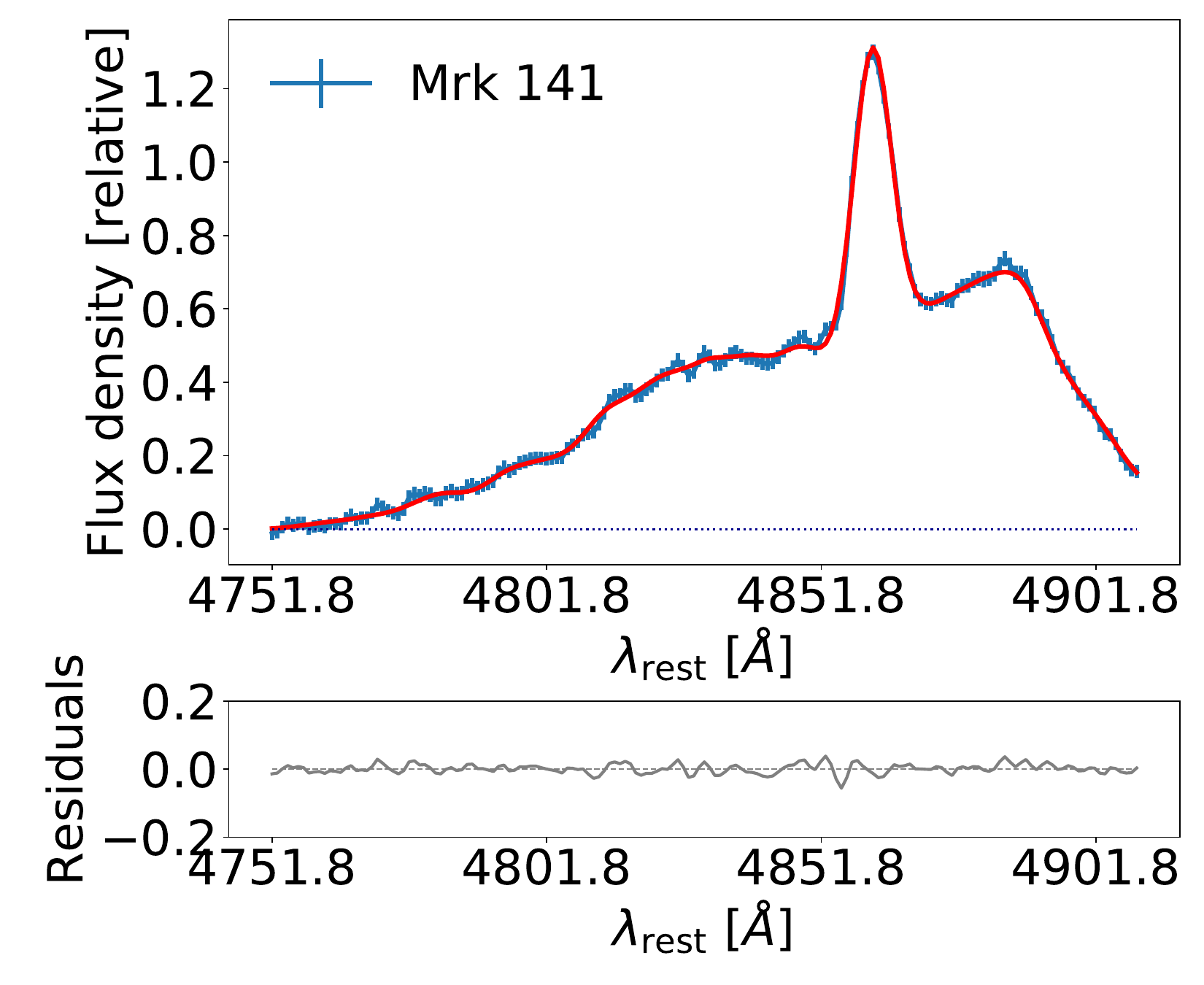}\\[0.2cm]
\includegraphics[width=0.33\textwidth]{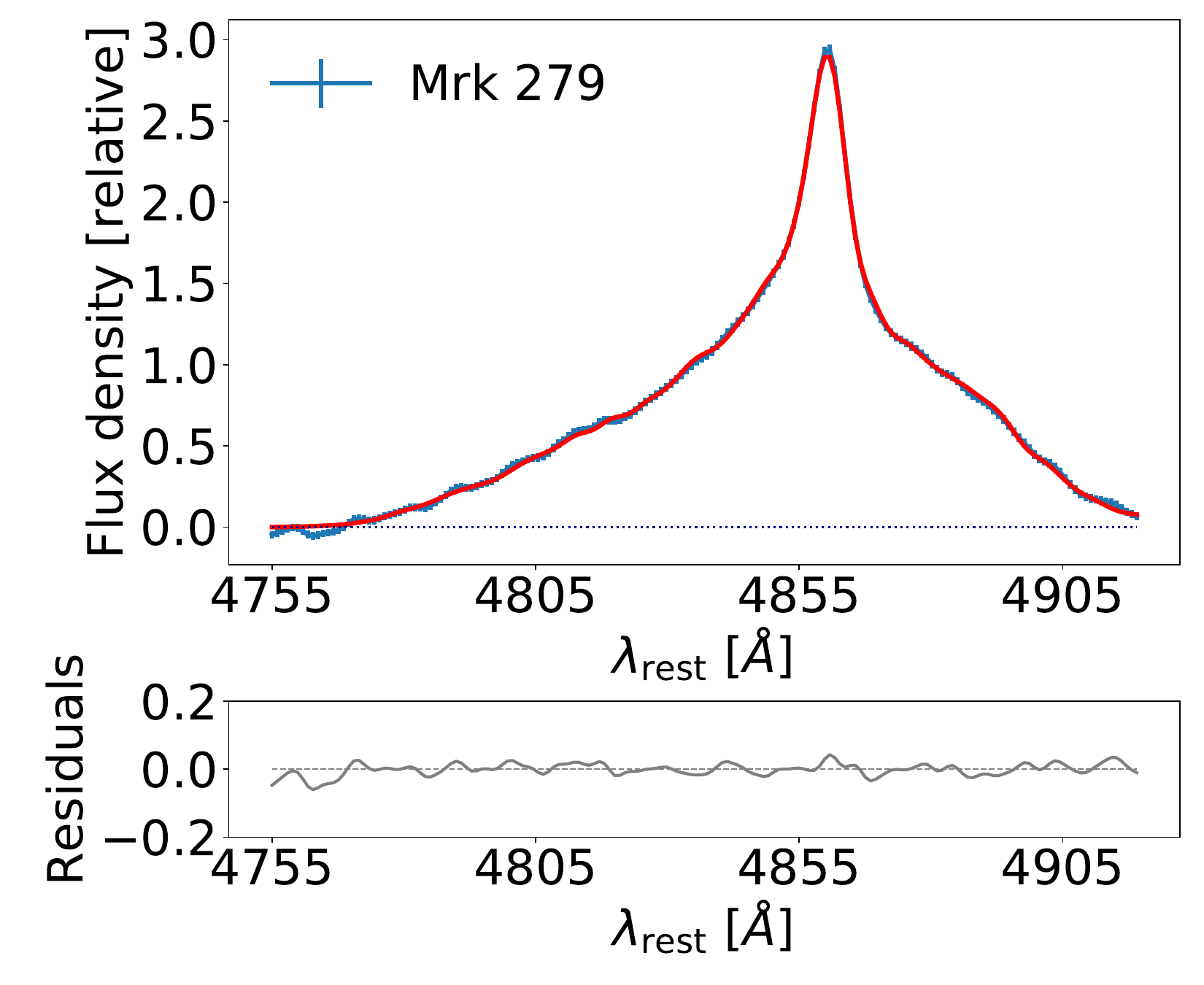}
\includegraphics[width=0.33\textwidth]{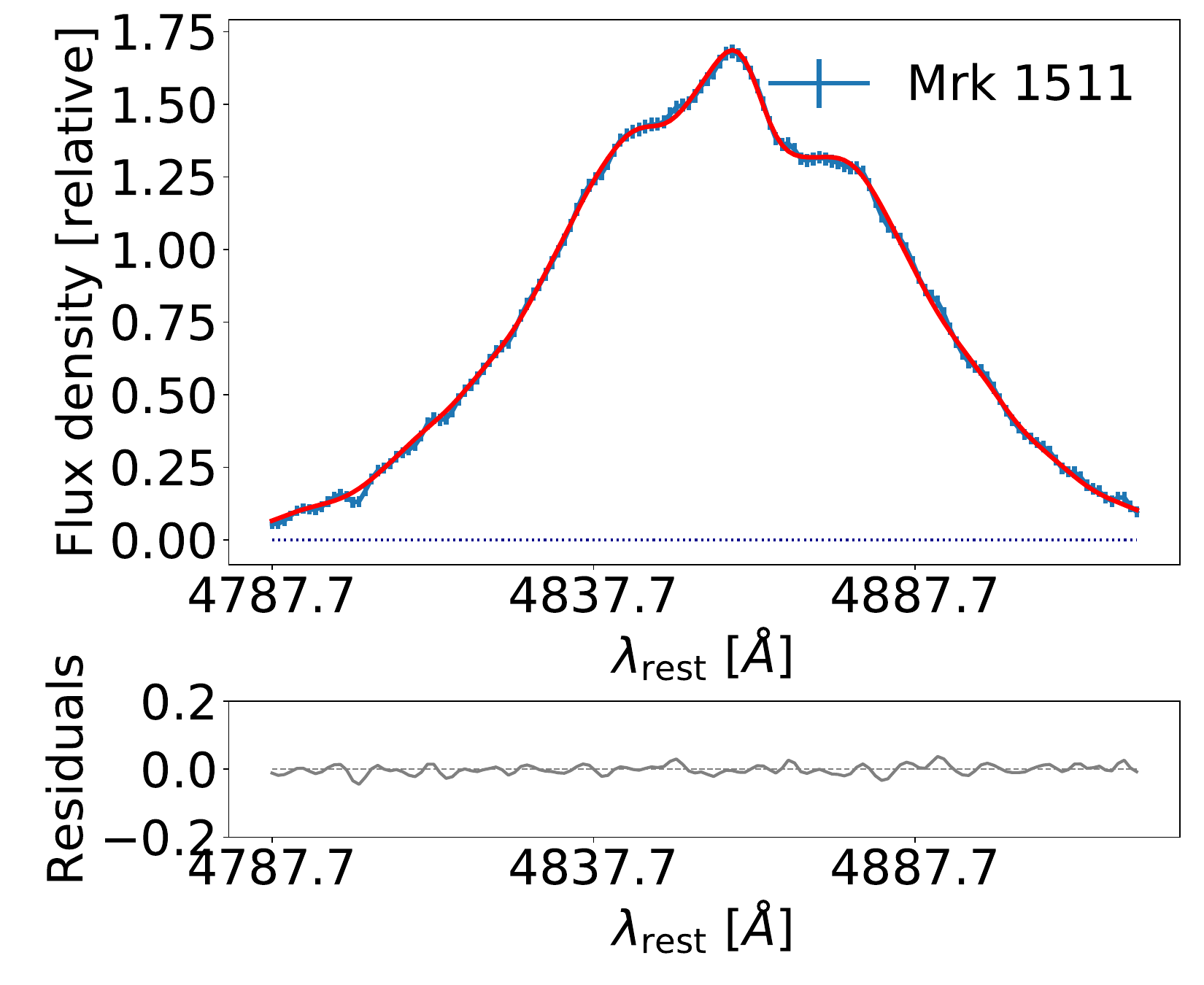}
\includegraphics[width=0.33\textwidth]{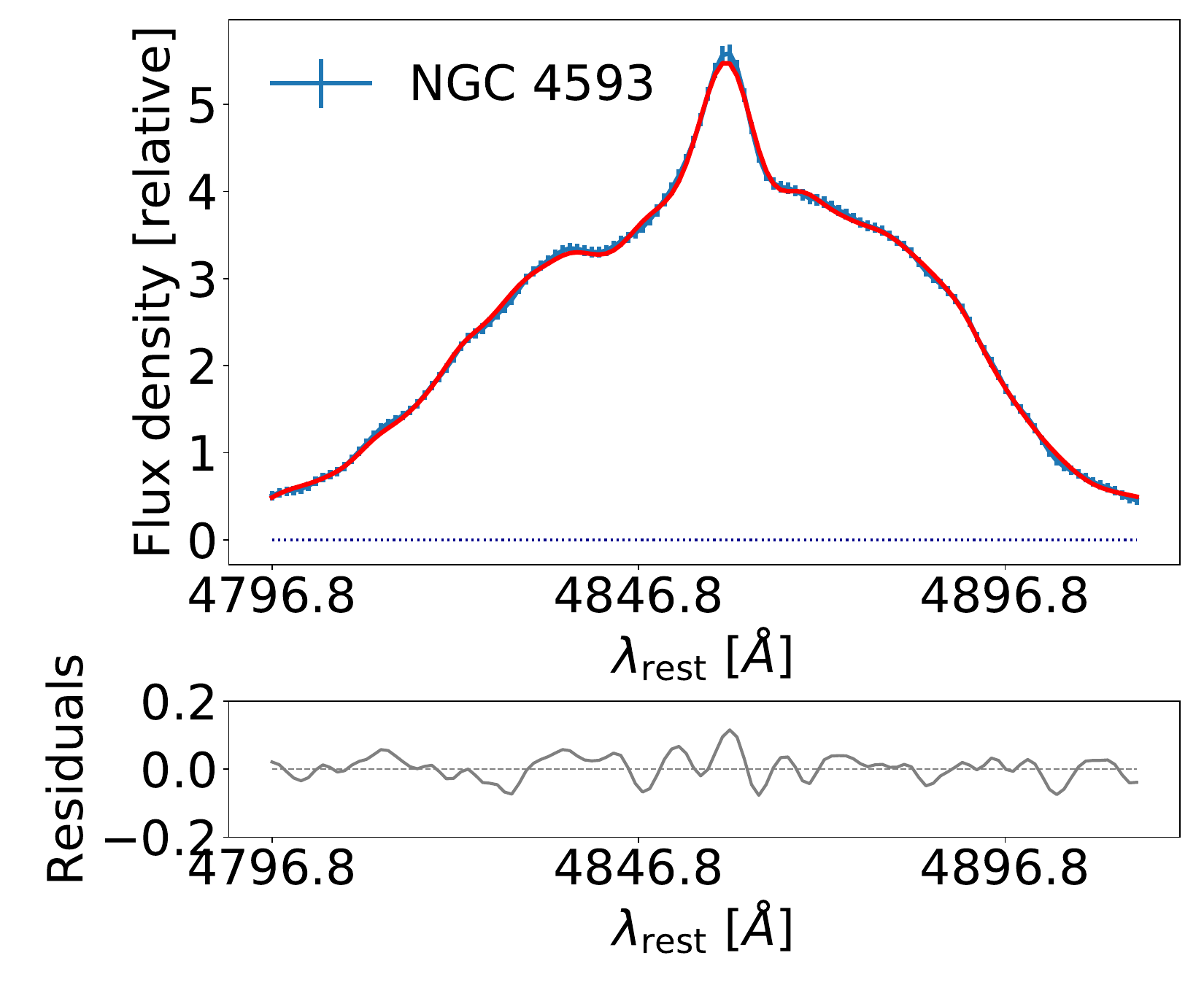}\\[0.2cm]
\includegraphics[width=0.33\textwidth]{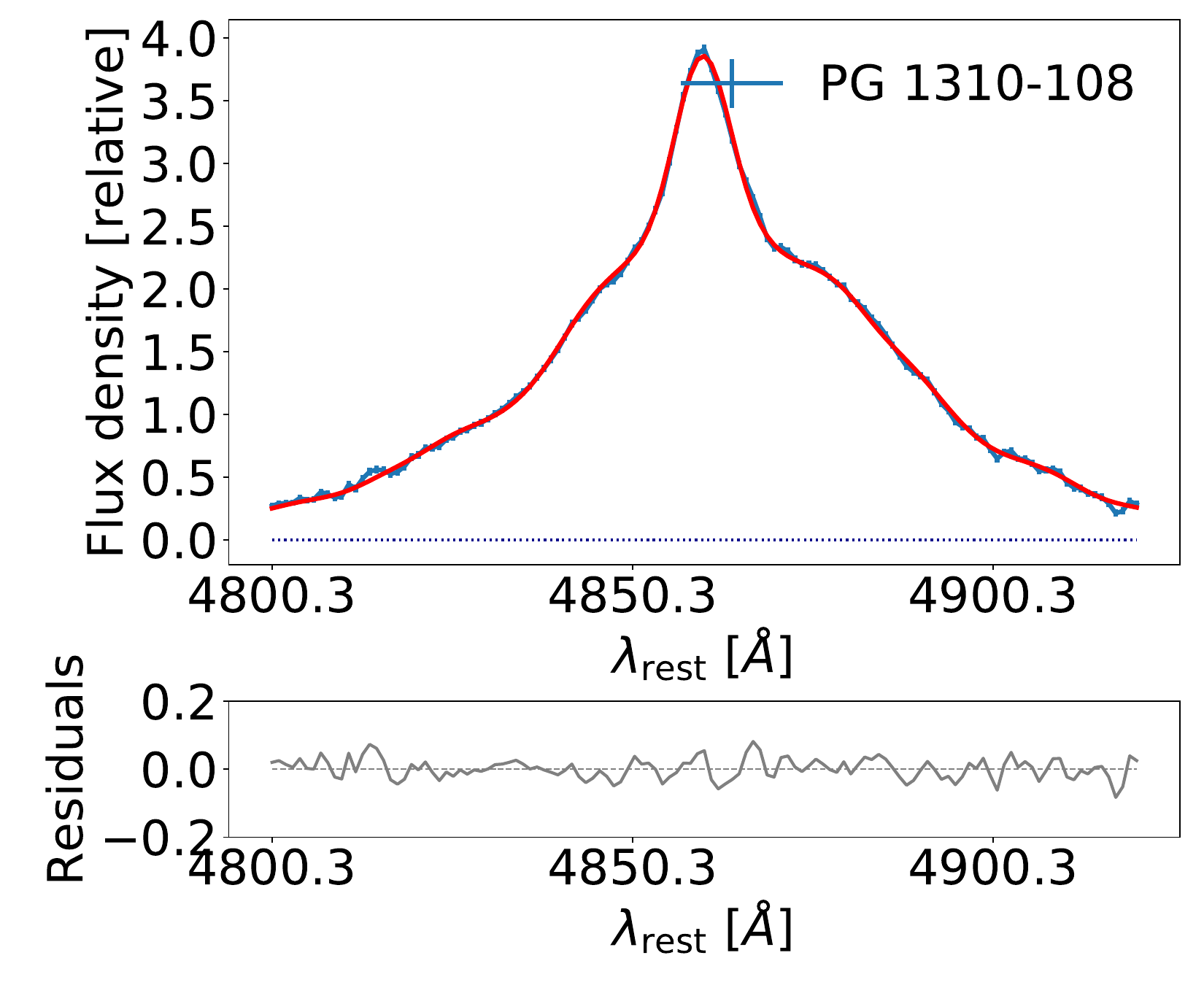}
\includegraphics[width=0.33\textwidth]{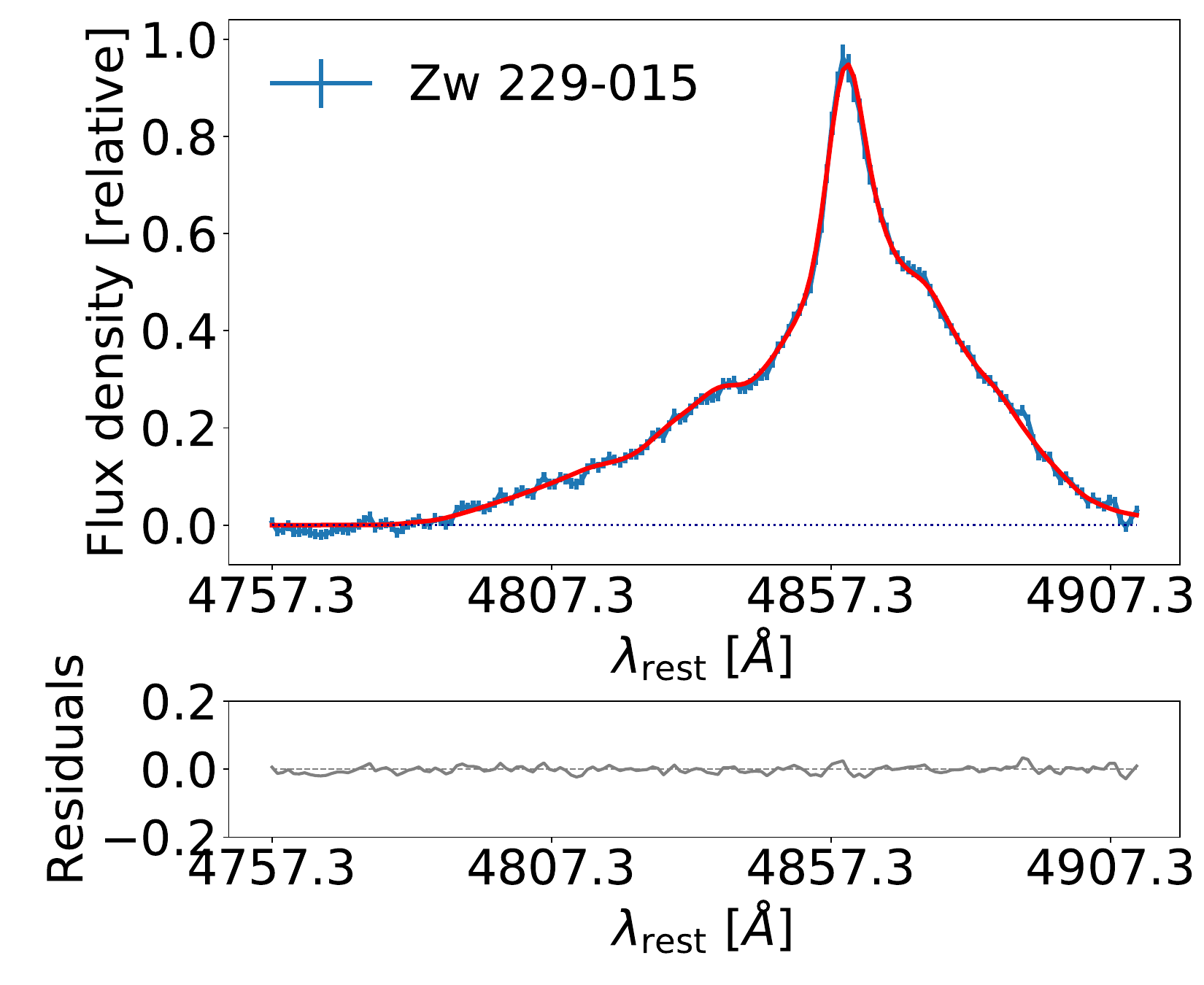}
\caption{Sample line profile generated by the model for each of the AGN. The blue line is the observed line profile and the red line is a line profile generated by the model and randomly selected from the posterior probability distribution. The residuals are shown in grey in the bottom panel.}
\label{spec_fit}
\end{figure*}

\begin{figure*}
\centering
\includegraphics[width=0.33\textwidth]{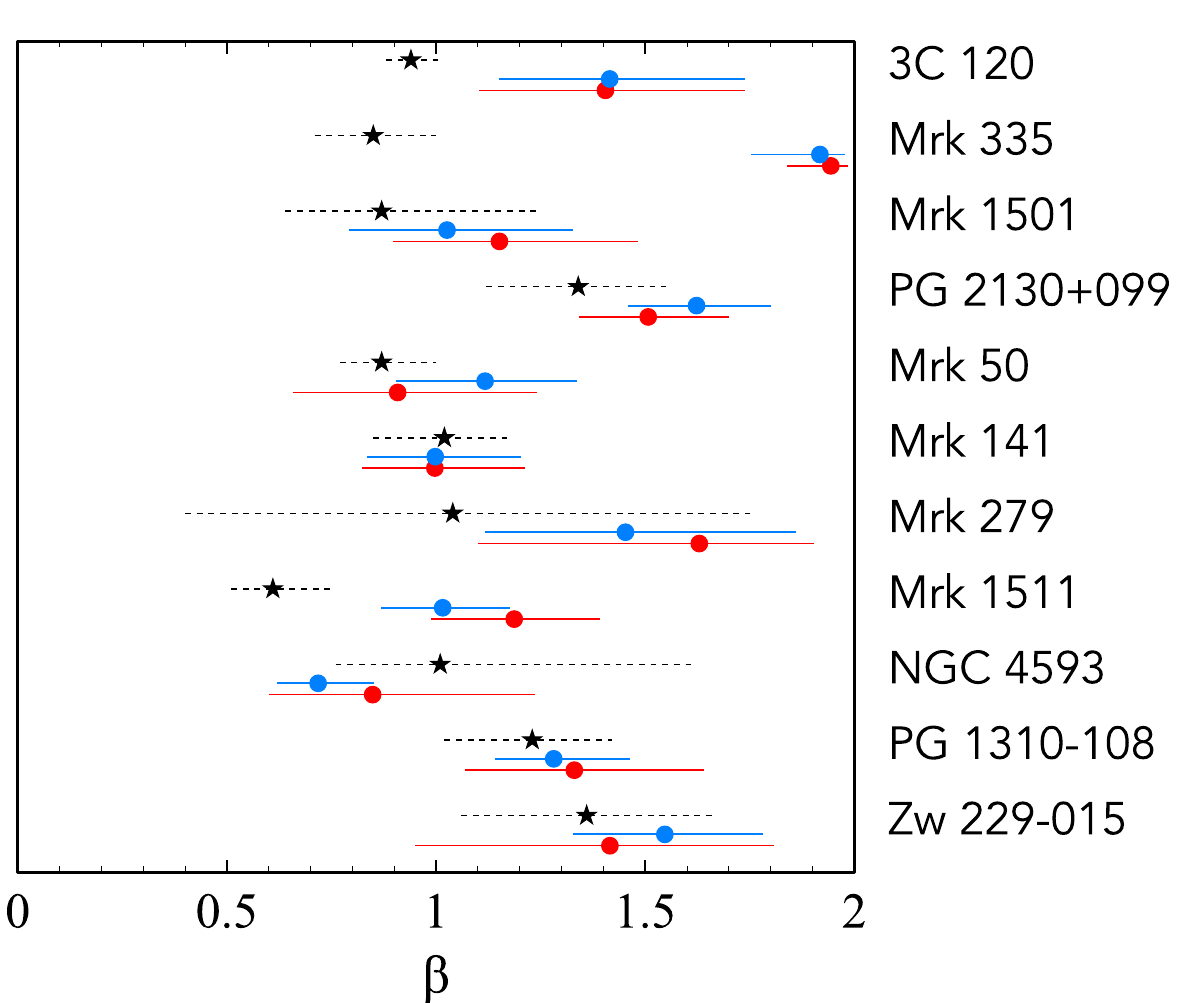}
\includegraphics[width=0.33\textwidth]{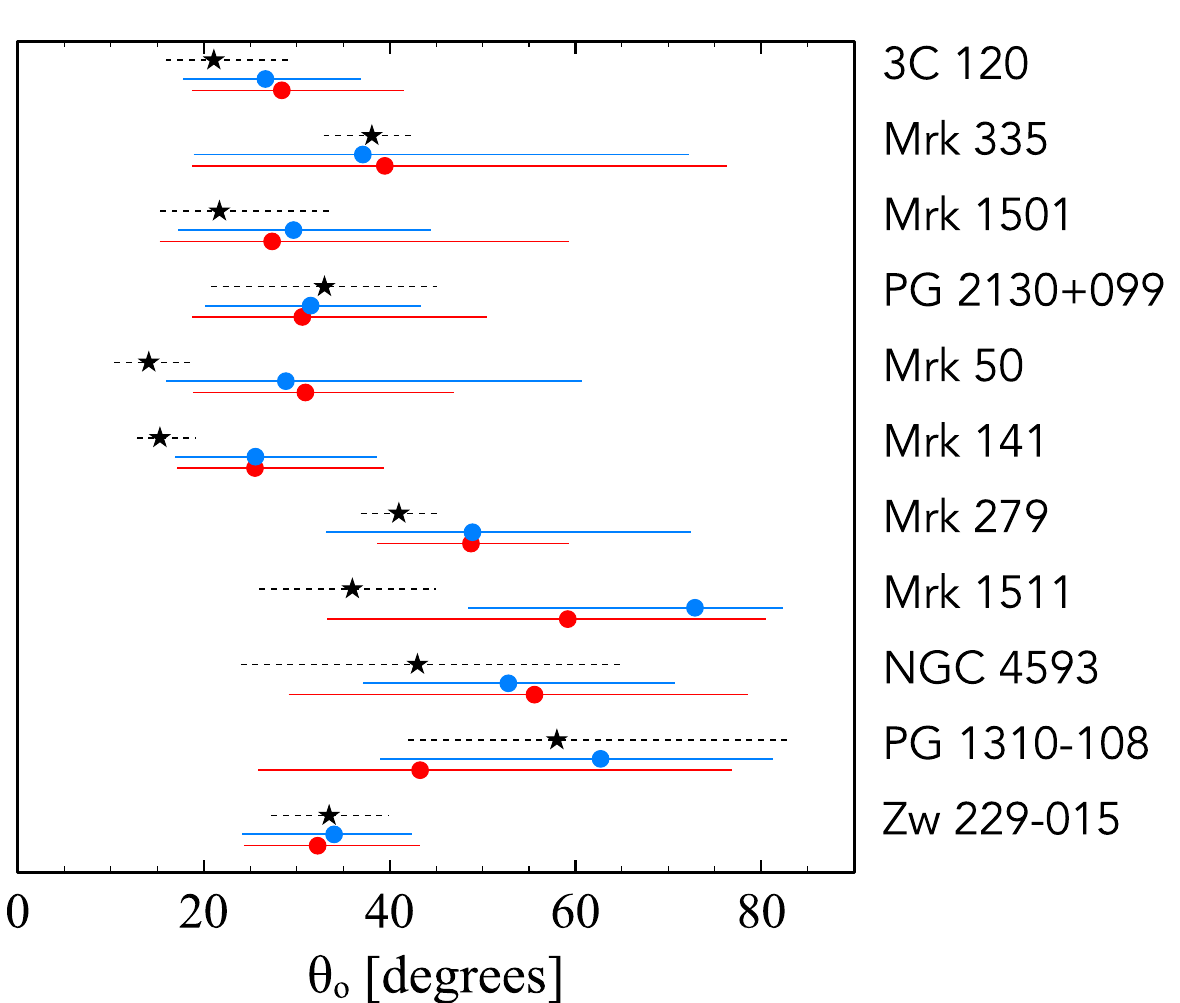}
\includegraphics[width=0.33\textwidth]{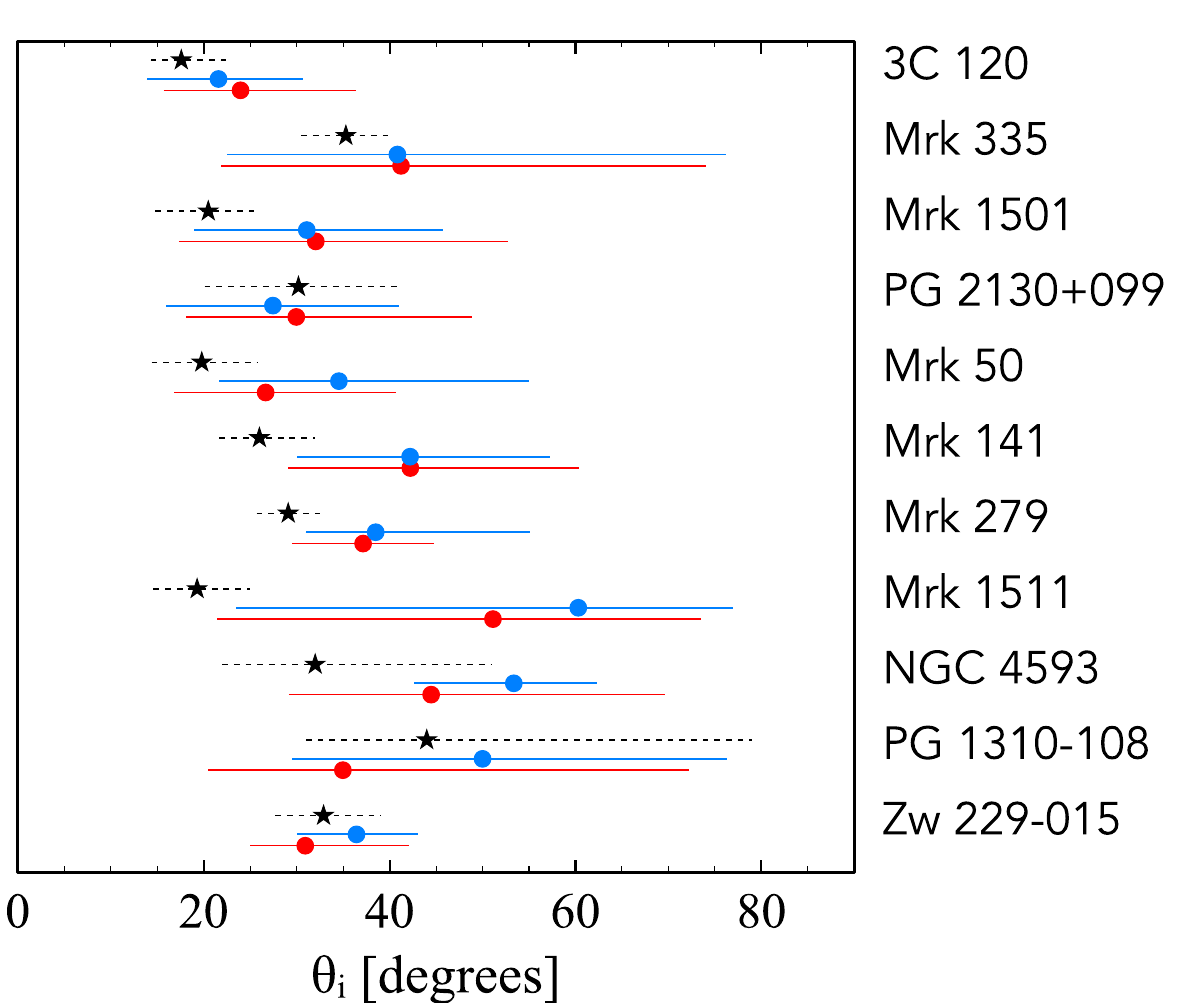}\\[0.1cm]
\includegraphics[width=0.33\textwidth]{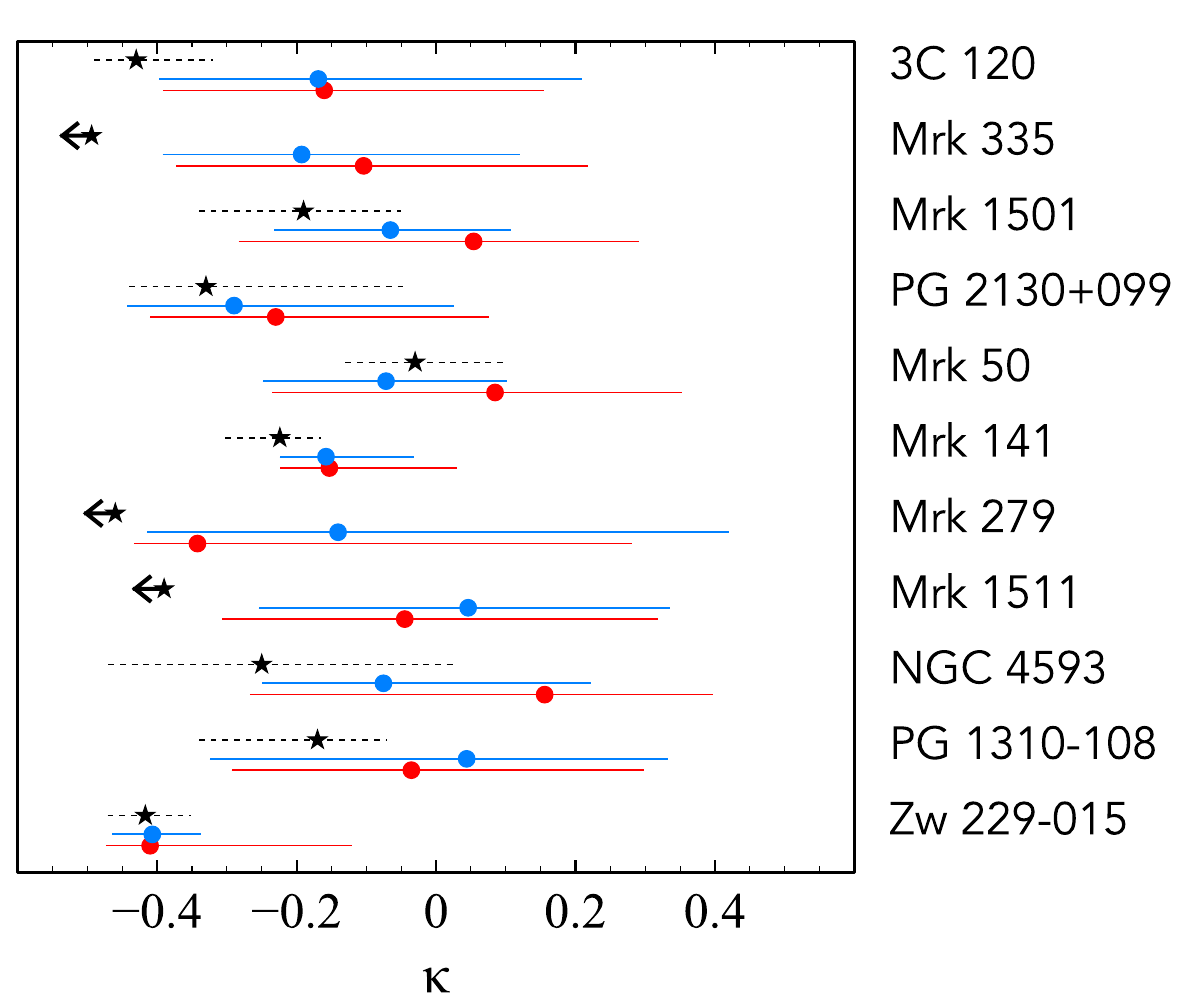}
\includegraphics[width=0.33\textwidth]{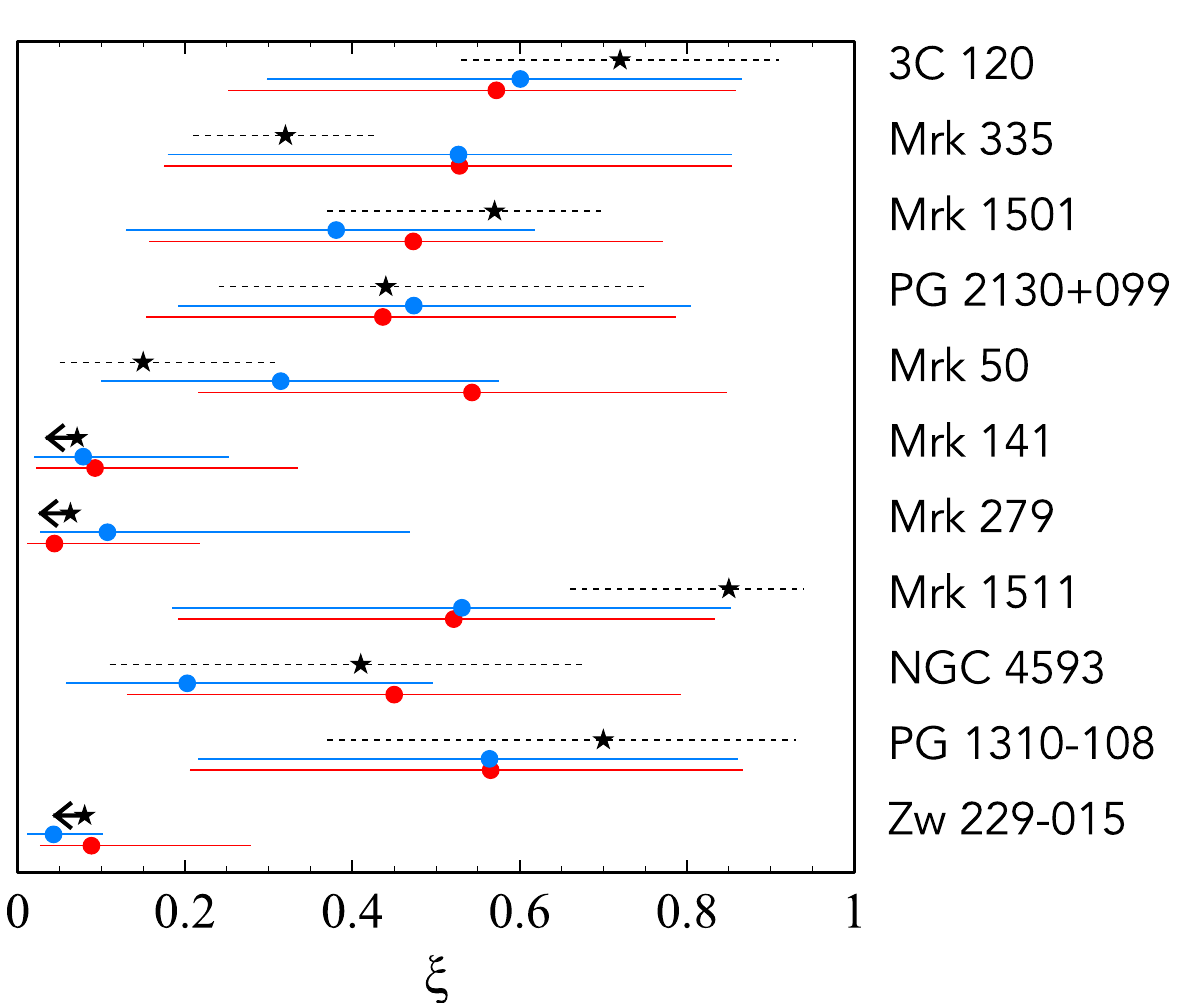}
\includegraphics[width=0.33\textwidth]{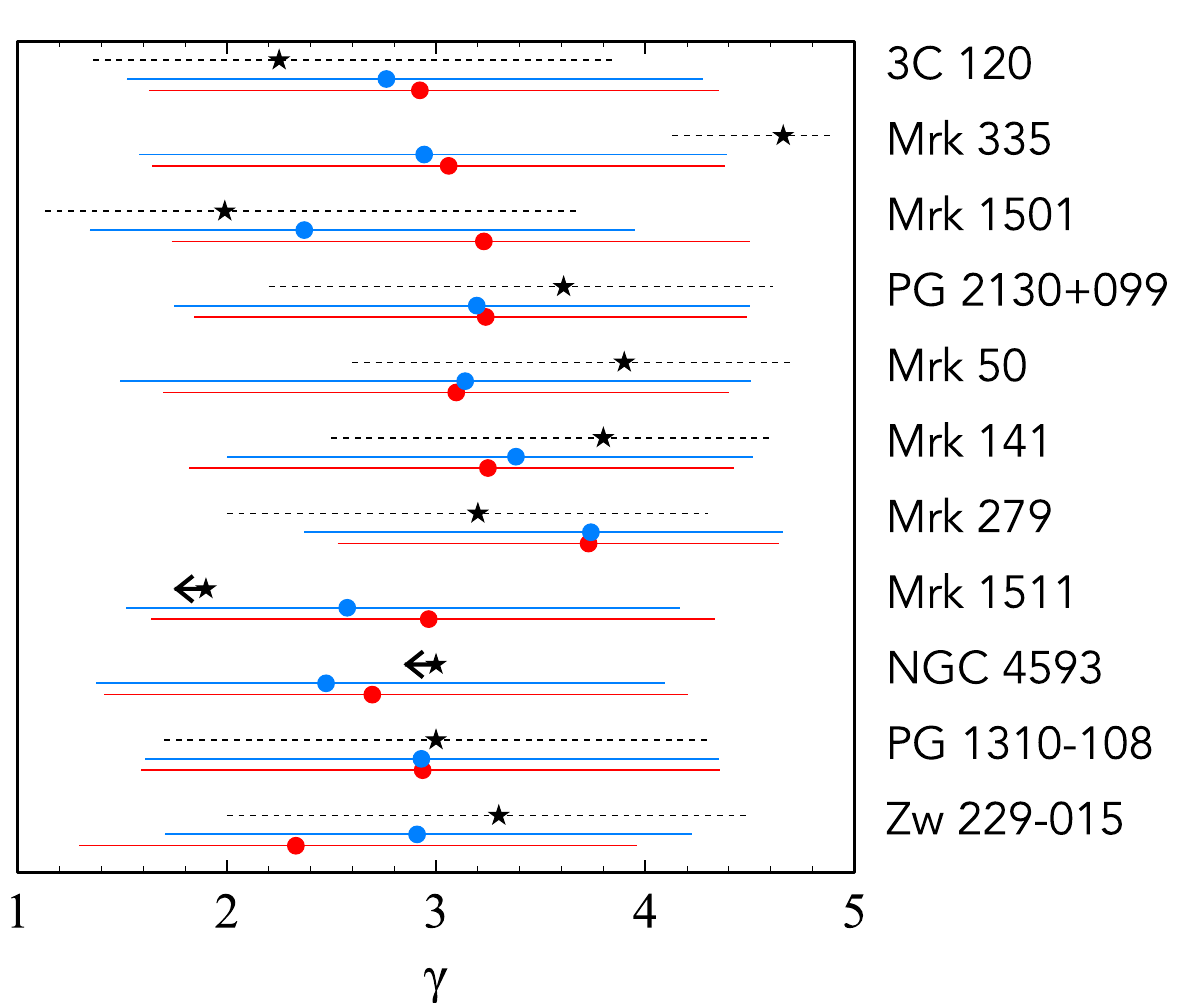}\\[0.1cm]
\includegraphics[width=0.33\textwidth]{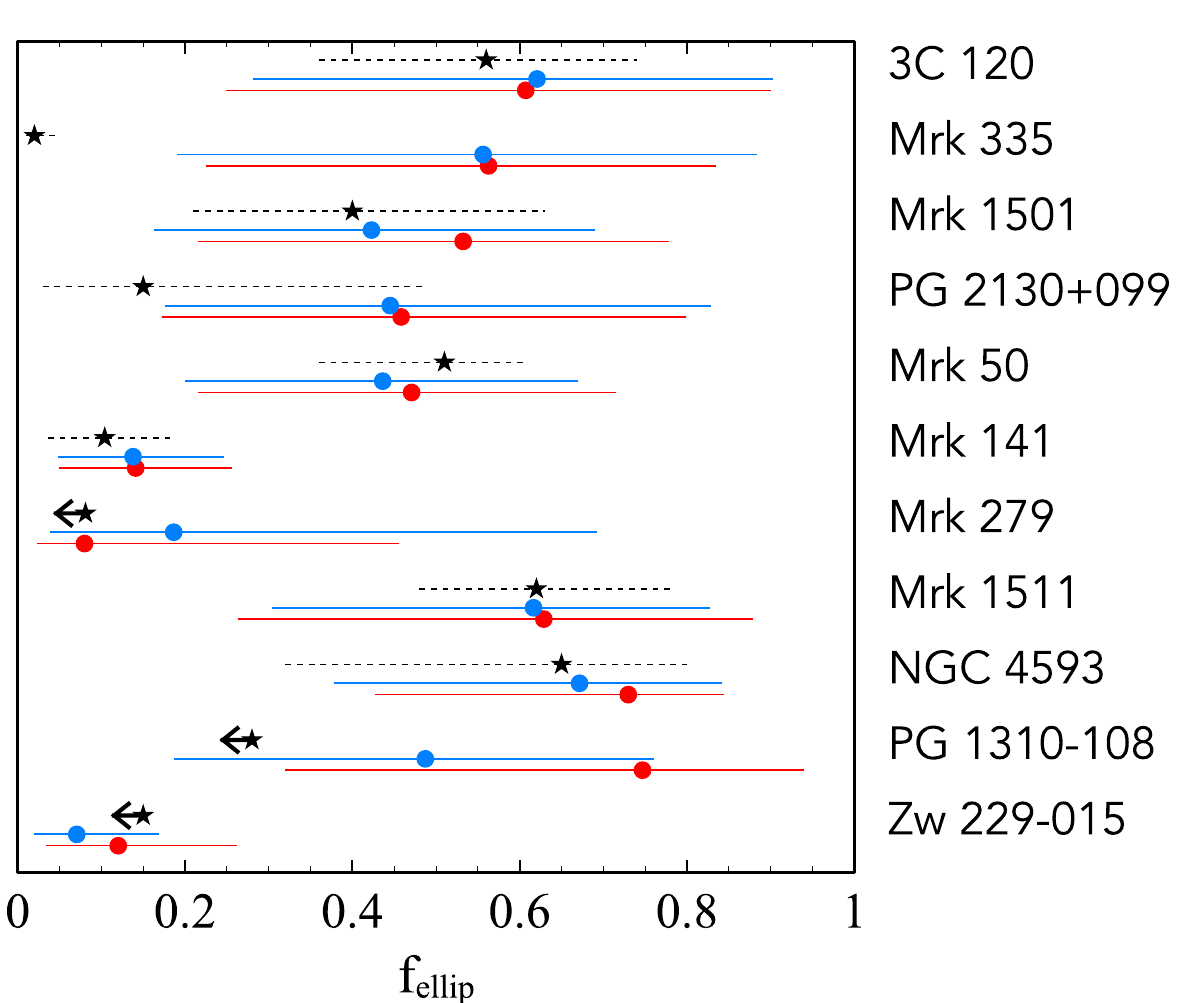}
\includegraphics[width=0.33\textwidth]{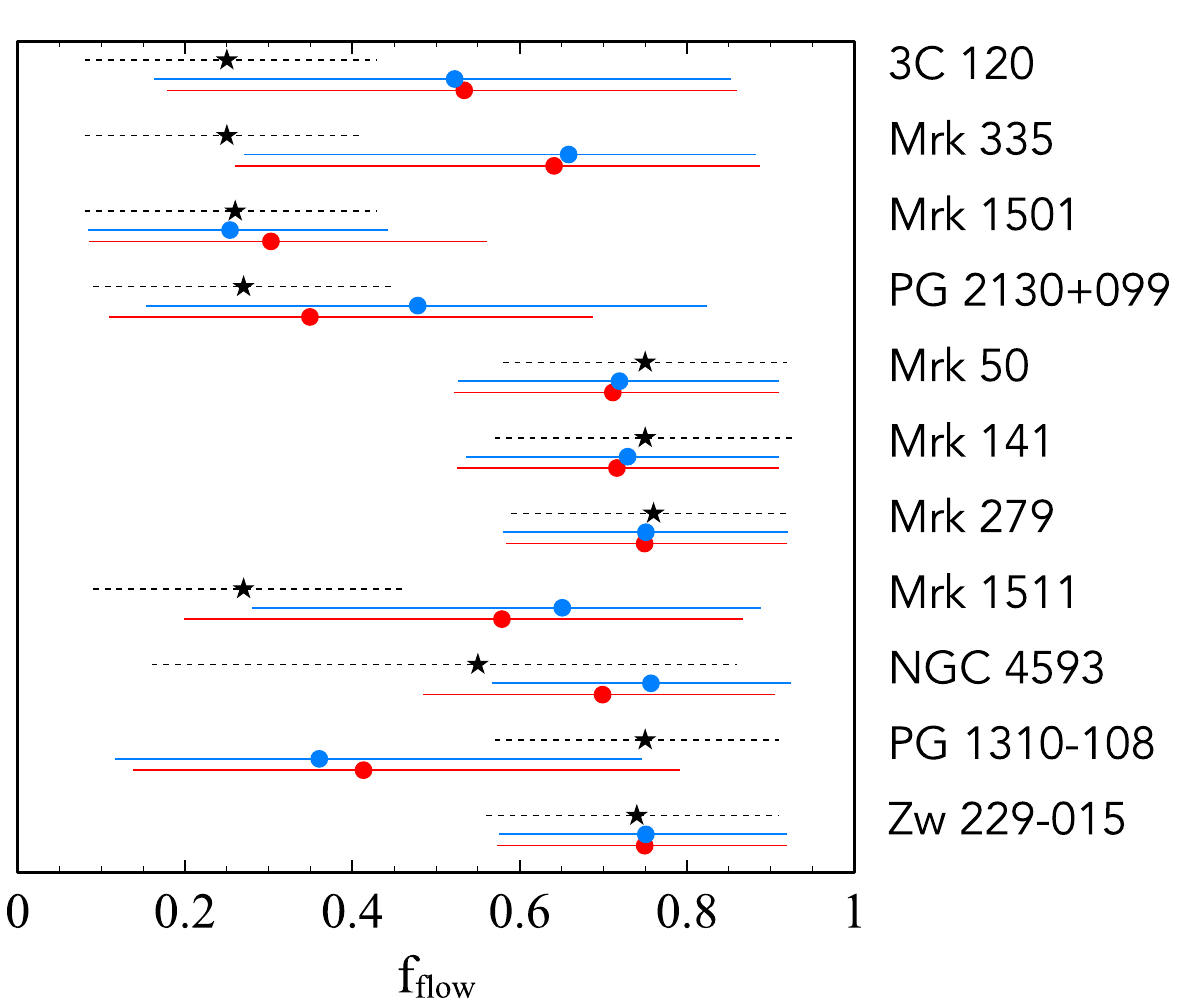}
\includegraphics[width=0.33\textwidth]{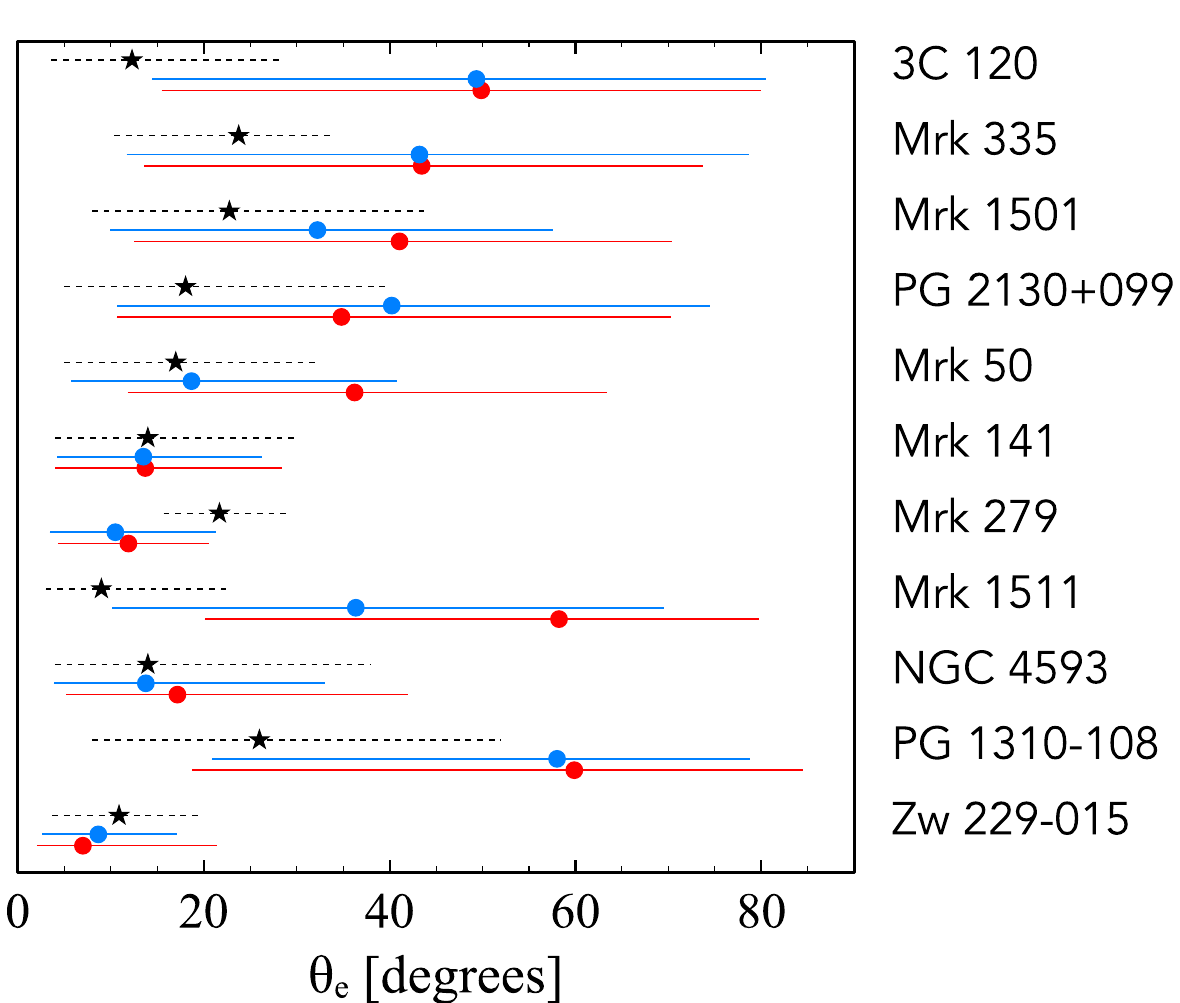}\\[0.1cm]
\includegraphics[width=0.33\textwidth]{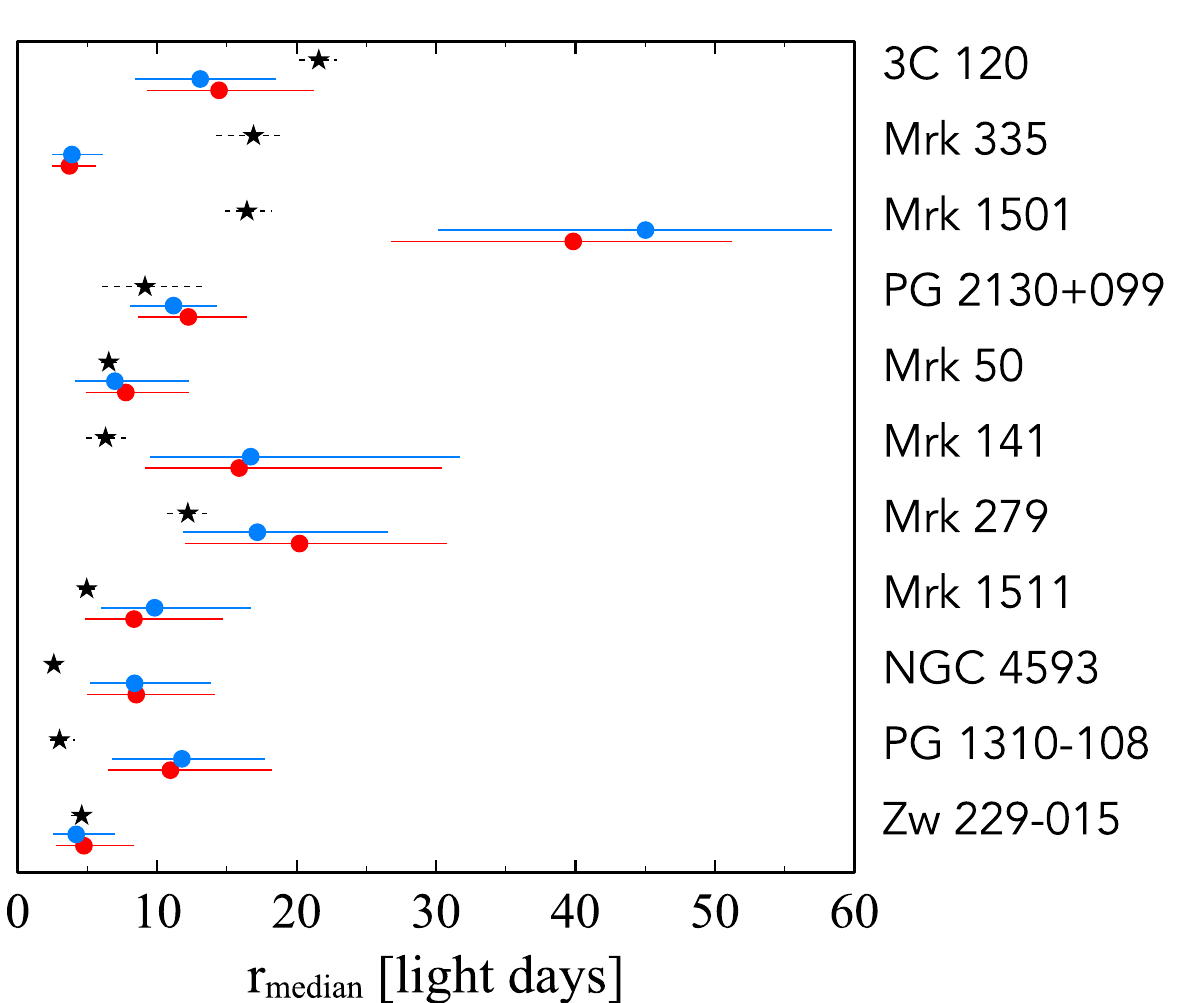}
\includegraphics[width=0.33\textwidth]{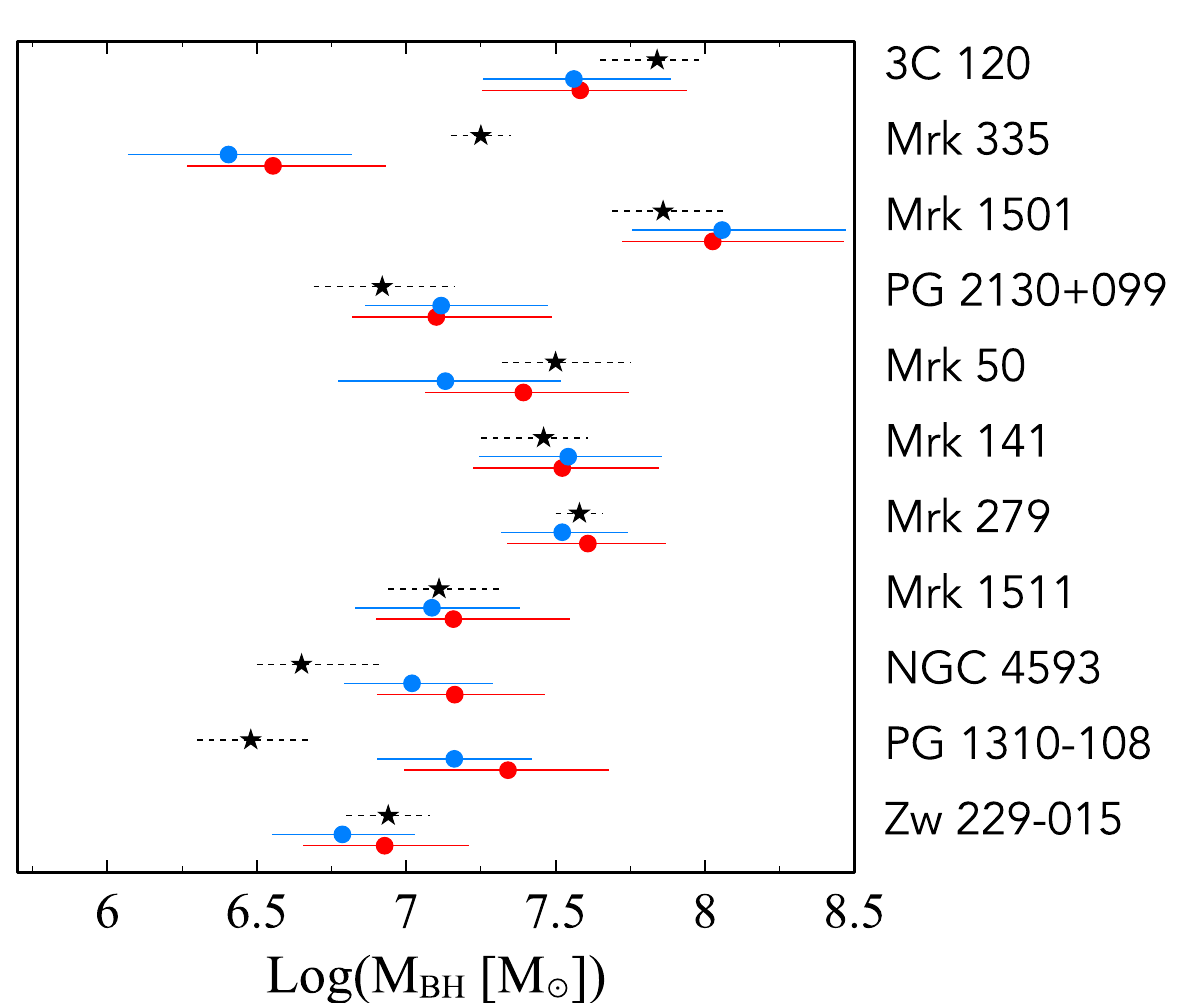}
\includegraphics[width=0.33\textwidth]{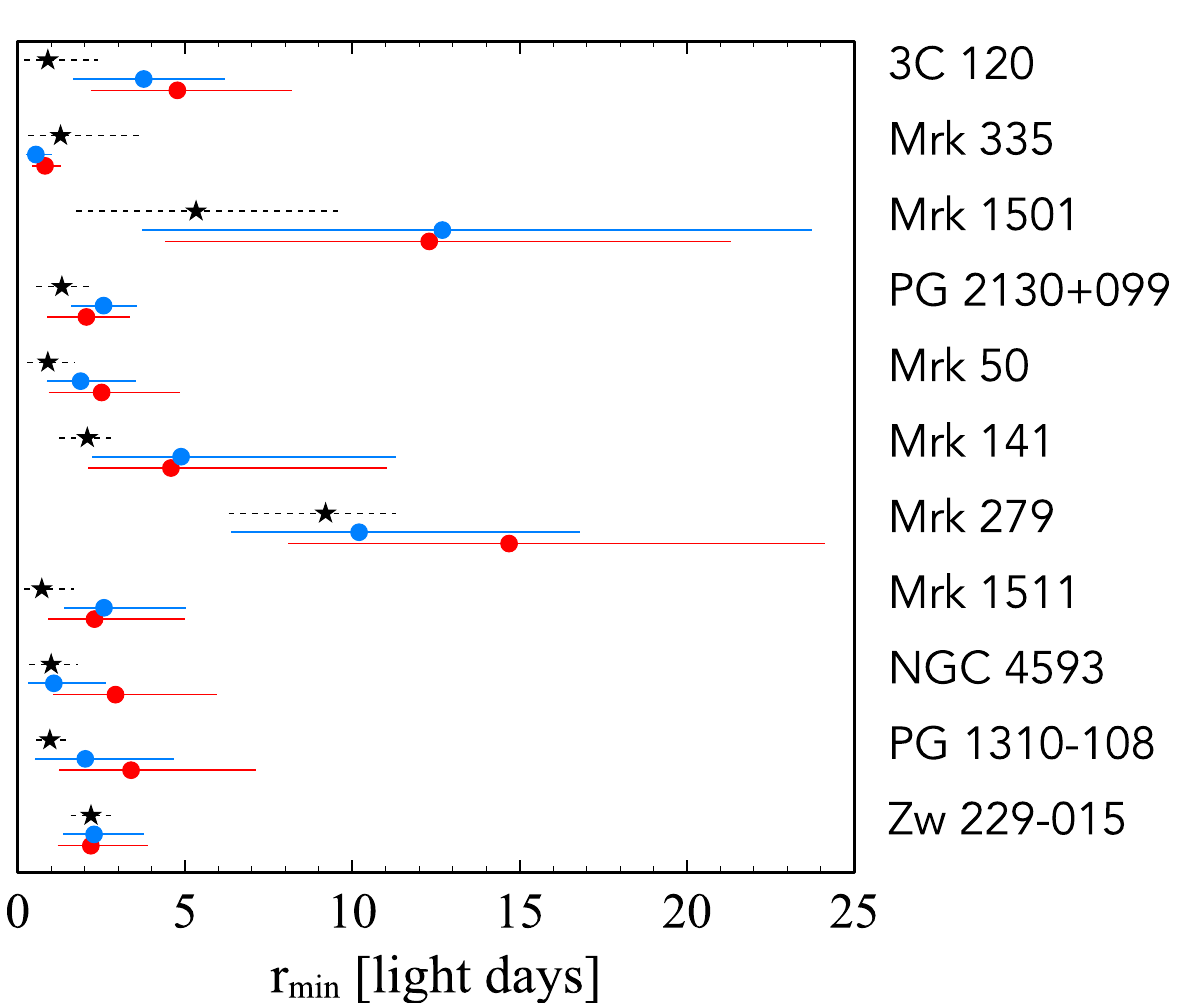}
\caption{Inferred values for the parameters and their respective 68\% confidence regions for each of the objects analysed. The first 4 objects from the top are part of the AGN10 monitoring campaign while the remaining 7 objects are from LAMP 2011. The filled red circles are the inferred values using the single-epoch analysis and the additional lower S/N epoch. The filled blue circles are the inferred values using the single-epoch analysis for the maximum S/N epochs and the black star symbols are the inferred values found by previous work using the full light-curve for the modelling of the BLR (\citealt{grier17} for AGN10 and \citealt{williams18} for LAMP 2011). The values inferred by \citealt{williams18} refer to their combined posterior probability distribution. We consider that a parameter cannot be constrained when the 68\% confidence range covers more than 50\% of the parameter space. Scale in the y-axis is arbitrary for visualisation purposes.}
\label{inferred_parameters_SN20}
\end{figure*}

\renewcommand{\arraystretch}{1.5}

\begin{landscape}
\begin{table}
\vspace{3cm}
\begin{tabular}{c | c | c | c | c | c | c | c | c | c | c | c }
\hline
\hline
Parameter & 3C 120 & Mrk 335 & Mrk 1501 & PG 2130+099 & Mrk 50 & Mrk 141 & Mrk 279 & Mrk 1511 & NGC 4593 & PG 1310-108 & Zw 229-015\\
 \hline
$\beta$ &   $1.42^{+0.32}_{-0.27}$ &   $1.92^{+0.06}_{-0.17}$ &   $1.03^{+0.30}_{-0.24}$ &   $1.62^{+0.18}_{-0.16}$ &   $1.12^{+0.22}_{-0.21}$ &   $1.00^{+0.21}_{-0.16}$ &   $1.45^{+0.41}_{-0.34}$ &   $1.02^{+0.16}_{-0.15}$ &   $0.72^{+0.13}_{-0.10}$ &   $1.28^{+0.18}_{-0.14}$ &   $1.55^{+0.23}_{-0.22}$ \\ 
$\theta_o$ (degrees) &   $27^{+10}_{-9}$ &   $37^{+35}_{-18}$ &   $30^{+15}_{-12}$ &   $32^{+12}_{-11}$ &   $29^{+32}_{-13}$ &   $26^{+13}_{-9}$ &   $49^{+24}_{-16}$ &   $73^{+10}_{-24}$ &   $53^{+18}_{-16}$ &   $63^{+19}_{-24}$ &   $34.0^{+ 8.4}_{- 9.9}$ \\ 
$\theta_i$ (degrees) &   $21.6^{+ 9.1}_{- 7.7}$ &   $41^{+35}_{-18}$ &   $31^{+15}_{-12}$ &   $28^{+14}_{-12}$ &   $35^{+20}_{-13}$ &   $42^{+15}_{-12}$ &   $38^{+17}_{-8}$ &   $60^{+17}_{-37}$ &   $53^{+9}_{-11}$ &   $50^{+26}_{-20}$ &   $36.4^{+6.7}_{- 6.3}$ \\ 
$\kappa$ &   $-0.17^{+0.38}_{-0.23}$ &   $-0.19^{+0.31}_{-0.20}$ &   $-0.07^{+0.17}_{-0.17}$ &   $-0.29^{+0.32}_{-0.15}$ &   $-0.07^{+0.17}_{-0.18}$ &   $-0.16^{+0.13}_{-0.07}$ &   $-0.14^{+0.56}_{-0.27}$ &   $0.05^{+0.29}_{-0.30}$ &   $-0.08^{+0.30}_{-0.17}$ &   $0.04^{+0.29}_{-0.37}$ &   $-0.407^{+0.070}_{-0.058}$ \\ 
$\gamma$ &   $2.8^{+1.5}_{-1.2}$ &   $2.9^{+1.4}_{-1.4}$ &   $2.4^{+1.6}_{-1.0}$ &   $3.2^{+1.3}_{-1.4}$ &   $3.1^{+1.4}_{-1.6}$ &   $3.4^{+1.1}_{-1.4}$ &   $3.7^{+0.9}_{-1.4}$ &   $2.6^{+1.6}_{-1.1}$ &   $2.5^{+1.6}_{-1.1}$ &   $2.9^{+1.4}_{-1.3}$ &   $2.9^{+1.3}_{-1.2}$ \\ 
$\xi$ &   $0.60^{+0.26}_{-0.30}$ &   $0.53^{+0.33}_{-0.35}$ &   $0.38^{+0.24}_{-0.25}$ &   $0.47^{+0.33}_{-0.28}$ &   $0.31^{+0.26}_{-0.21}$ &   $0.08^{+0.17}_{-0.06}$ &   $0.11^{+0.36}_{-0.08}$ &   $0.53^{+0.32}_{-0.35}$ &   $0.20^{+0.29}_{-0.15}$ &   $0.56^{+0.30}_{-0.35}$ &   $0.043^{+0.059}_{-0.032}$ \\ 

$\log_{10}(M_{\rm BH}/M_\odot)$ &   $7.56^{+0.33}_{-0.31}$ &   $6.41^{+0.41}_{-0.34}$ &   $8.06^{+0.42}_{-0.30}$ &   $7.12^{+0.36}_{-0.25}$ &   $7.13^{+0.39}_{-0.36}$ &   $7.54^{+0.31}_{-0.30}$ &   $7.52^{+0.22}_{-0.20}$ &   $7.09^{+0.30}_{-0.26}$ &   $7.02^{+0.27}_{-0.23}$ &   $7.16^{+0.26}_{-0.26}$ &   $6.79^{+0.24}_{-0.24}$ \\ 
$f_{\rm ellip}$ &   $0.62^{+0.28}_{-0.34}$ &   $0.56^{+0.33}_{-0.37}$ &   $0.42^{+0.27}_{-0.26}$ &   $0.45^{+0.38}_{-0.27}$ &   $0.44^{+0.23}_{-0.24}$ &   $0.14^{+0.11}_{-0.09}$ &   $0.19^{+0.51}_{-0.15}$ &   $0.62^{+0.21}_{-0.31}$ &   $0.67^{+0.17}_{-0.29}$ &   $0.49^{+0.27}_{-0.30}$ &   $0.07^{+0.10}_{-0.05}$ \\ 
$f_{\rm flow}$ &   $0.52^{+0.33}_{-0.36}$ &   $0.66^{+0.22}_{-0.39}$ &   $0.25^{+0.19}_{-0.17}$ &   $0.48^{+0.35}_{-0.32}$ &   $0.72^{+0.19}_{-0.19}$ &   $0.73^{+0.18}_{-0.19}$ &   $0.75^{+0.17}_{-0.17}$ &   $0.65^{+0.24}_{-0.37}$ &   $0.76^{+0.17}_{-0.19}$ &   $0.36^{+0.39}_{-0.24}$ &   $0.75^{+0.17}_{-0.18}$ \\ 
$\theta_e$ (degrees) &   $49^{+31}_{-35}$ &   $43^{+36}_{-32}$ &   $32^{+25}_{-22}$ &   $40^{+34}_{-30}$ &   $19^{+22}_{-13}$ &   $14^{+13}_{-9}$ &   $10^{+11}_{-7}$ &   $36^{+33}_{-26}$ &   $14^{+19}_{-10}$ &   $58^{+21}_{-37}$ &   $ 8.7^{+8.5}_{-6.0}$ \\ 
$r_{\rm min}$ (light days) &   $3.8^{+2.4}_{-2.1}$ &   $0.54^{+0.47}_{-0.30}$ &   $13^{+11}_{-9}$ &   $2.6^{+1.0}_{-1.0}$ &   $1.9^{+1.7}_{-1.0}$ &   $4.9^{+6.4}_{-2.6}$ &   $10.2^{+6.6}_{-3.8}$ &   $2.6^{+2.4}_{-1.2}$ &   $1.1^{+1.6}_{-0.8}$ &   $2.0^{+2.6}_{-1.5}$ &   $2.3^{+1.5}_{-0.9}$ \\ 
$r_{\rm mean}$ (light days) &   $21.7^{+6.2}_{-6.5}$ &   $13.6^{+4.7}_{-4.0}$ &   $55^{+12}_{-16}$ &   $22.4^{+3.0}_{-4.6}$ &   $9.9^{+7.2}_{-4.0}$ &   $22^{+17}_{-10}$ &   $24^{+11}_{-7}$ &   $13.0^{+9.1}_{-5.1}$ &   $9.8^{+6.4}_{-3.7}$ &   $18.9^{+8.2}_{-7.6}$ &   $7.3^{+4.1}_{-2.6}$ \\ 
$r_{\rm median}$ (light days) &   $13.1^{+5.4}_{-4.7}$ &   $3.9^{+2.2}_{-1.4}$ &   $45^{+13}_{-15}$ &   $11.2^{+3.1}_{-3.1}$ &   $7.0^{+5.3}_{-2.8}$ &   $17^{+15}_{-7}$ &   $17.2^{+9.4}_{-5.3}$ &   $9.8^{+6.9}_{-3.8}$ &   $8.4^{+5.5}_{-3.2}$ &   $11.8^{+6.0}_{-5.0}$ &   $4.2^{+2.8}_{-1.6}$ \\ 
$\tau_{\rm mean}$  (days)  &   $20.3^{+6.6}_{-5.8}$ &   $13.2^{+4.9}_{-3.7}$ &   $49^{+12}_{-15}$ &   $21.5^{+3.5}_{-4.4}$ &   $9.1^{+6.3}_{-3.4}$ &   $20^{+16}_{-8}$ &   $19.3^{+8.1}_{-6.0}$ &   $11.1^{+6.8}_{-4.1}$ &   $8.5^{+5.3}_{-3.3}$ &   $16.0^{+7.1}_{-5.4}$ &   $7.0^{+4.0}_{-2.5}$ \\ 
$\tau_{\rm median}$  (days)  &   $11.4^{+4.9}_{-4.1}$ &   $3.1^{+1.8}_{-1.0}$ &   $36^{+13}_{-12}$ &   $9.6^{+2.9}_{-2.7}$ &   $5.6^{+4.1}_{-2.2}$ &   $13^{+12}_{-6}$ &   $13.2^{+7.1}_{-4.4}$ &   $7.0^{+4.8}_{-2.6}$ &   $6.1^{+3.9}_{-2.3}$ &   $7.9^{+4.5}_{-2.9}$ &   $3.9^{+2.7}_{-1.4}$ \\ 
$r_{\rm out}$ (light-days) & 104 & 104 & 156 & 104 & 104 & 104 & 104 & 104 & 104 & 104 & 104\\
 $T$ & 1 & 1 & 1 & 2 & 1 & 1 & 1 & 1 & 1 & 1 & 1\\
\end{tabular}
\caption{Table of inferred broad line region geometry and dynamics parameters for the AGN modelled in this work. The parameters are determined from BLR modelling using single epoch spectra. The inferred parameter value quoted is the median of the posterior probability distribution and the uncertainties quoted are the 68\% confidence intervals. The last row quotes the temperature ($T$) used for each test.  \citetalias{raimundo19} discuss in detail how temperatures are selected in their Appendix.} 
\label{table_results}
\end{table}
\end{landscape}

\end{document}